\documentclass[11pt,a4paper]{article}
\usepackage{jheppub}
    
\usepackage{amsmath, amsfonts, amssymb, hyperref, xcolor, graphicx, mathtools, empheq, academicons}
\definecolor{nicered}{rgb}{.7,.1,.1}
\definecolor{nicegreen}{rgb}{.1,.5,.1}
\definecolor{darkblue}{rgb}{0,0,.5}
\hypersetup{colorlinks, citecolor=nicegreen, linkcolor=nicered, urlcolor=darkblue}

\newcommand{\ddV}{V^{(2)}}
\newcommand{\ddVt}{\tilde V^{(2)}}

\newcommand{\ddrho}{\frac{\text{d}^2}{\text{d} \rho^2}}
\newcommand{\drho}{\frac{\text{d}}{\text{d} \rho}}

\newcommand{\dz}{\frac{\text{d}}{\text{d} z}}
\newcommand{\ddx}{\frac{\text{d}^2}{\text{d} x^2}}
\newcommand{\dx}{\frac{\text{d}}{\text{d} x}}
\newcommand{\anynom}{\genfrac{\lbrace}{\rbrace}{0pt}{}}
\newcommand{\orcid}[1]{\href{https://orcid.org/#1}{\textcolor[HTML]{A6CE39}{\aiOrcid}}}

\arxivnumber{2202.04498}

\begin{document}

\title{Analytic thin wall false vacuum decay rate}

\author[a]{Aleksandar Ivanov 
\href{https://orcid.org/0000-0002-6198-8118}{\includegraphics[scale=0.3]{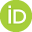}},}

\author[a,b]{Marco Matteini
\href{https://orcid.org/0000-0001-6481-3025}{\includegraphics[scale=0.3]{figures/orcid_32x32.png}},}

\author[b, a]{Miha Nemev\v{s}ek
\href{https://orcid.org/0000-0003-1110-342X}{\includegraphics[scale=0.3]{figures/orcid_32x32.png}},}

\author[c, d,e]{Lorenzo Ubaldi
\href{https://orcid.org/0000-0002-9567-9719}{\includegraphics[scale=0.3]{figures/orcid_32x32.png}}}

\affiliation[a]{Faculty of Mathematics and Physics, University of Ljubljana,\\ Jadranska 19, 1000 Ljubljana, Slovenia}
\affiliation[b]{Jo\v{z}ef Stefan Institute, Jamova 39, 1000 Ljubljana, Slovenia}
\affiliation[c]{INFN Sezione di Trieste, \\ Via Bonomea 265, 34136 Trieste, Italy}
\affiliation[d]{Institute for Fundamental Physics of the Universe, \\ Via Beirut 2, 34014 Trieste, Italy}
\affiliation[e]{Dipartimento di Fisica, Universit\`a di Genova and INFN, Sezione di Genova \\ Via Dodecaneso 33, 16146 Genova, Italy}

\emailAdd{ace.aleksandar0@gmail.com}
\emailAdd{marco.matteini@ijs.si}
\emailAdd{miha.nemevsek@ijs.si}
\emailAdd{ubaldi.physics@gmail.com}

\date{\today}

\abstract{
We derive a closed-form false vacuum decay rate at one loop in the thin wall limit, where the true and false 
vacua are nearly degenerate.
We obtain the bounce configuration in $D$ dimensions, together with the Euclidean action with a higher order 
correction, counter-terms and renormalization group running.
We extract the functional determinant via the Gel'fand-Yaglom theorem for low and generic orbital multipoles.
The negative and zero eigenvalues appear for low multipoles and the translational zeroes are removed.
We compute the fluctuations for generic multipoles, multiply and regulate the orbital modes.
We find an explicit finite renormalized decay rate in $D = 3, 4$ and give a closed-form expression 
for the finite functional determinant in any dimension.
}

\maketitle

\newpage

\tableofcontents

%
%
\section{Introduction} \label{secIntro}
Local ground states in physical systems, described by a field theory, may not be stable.
A deeper minimum may exist, or appear with varying temperature, and quantum/thermal tunneling can trigger
a first-order phase transition.
Such a transition takes the system from the metastable false vacuum (FV) to a deeper minimum, 
called the true vacuum (TV).
It is initiated by a sudden appearance of a bubble of TV upon the homogenous configuration of FV
and the probability of such events is given by the FV decay rate.

The theory of bubble nucleation was pioneered by the works of Langer~\cite{Langer:1967ax}, 
and applied to field theory by Kobzarev et al.~\cite{Kobzarev:1974cp}.
These initial results were rigorously derived and extended in seminal papers by Coleman on the 
bounce~\cite{Coleman:1977py} (see also~\cite{Coleman:1978ae, Coleman:1985}), the first quantum 
corrections with Callan~\cite{Callan:1977pt} and gravitational effects with de Luccia~\cite{Coleman:1980aw}.
Derivation of the rate was revisited in~\cite{Hammer:1978xu} and recently 
in~\cite{Andreassen:2016cff, Andreassen:2016cvx}, elucidating and confirming previous results.

Developing a theoretical understanding of bubble nucleation and subsequent expansion is important
for several reasons.
The stability of the potential in the Standard Model (SM) depends largely on the masses of the SM 
Higgs and the top.
The bound on the Higgs mass was historically considered in~\cite{Weinberg:1976mh, Linde:1976ds},
based on absolute stability, and in~\cite{Frampton:1976pb} from metastability/longevity.
There has been tremendous progress on quantifying the SM rate, for example in recent
works~\cite{Isidori:2001bm, Andreassen:2017rzq} and references therein.
In most theories beyond the SM, the presence of new particles may affect the stability of the Higgs
vacuum or even alter the landscape of the vacua altogether.
From these, only the physical vacua need to be selected, which requires them to be either globally
stable or at least sufficiently long-lived when metastable.
Thus, a precise knowledge of the decay rate puts specific constraints on the parameter space of
such models.

Furthermore, the shape of the potential may change significantly at high temperatures in the early
universe, leading to a first-order phase transition~\cite{Quiros:1999jp, Kapusta:2006pm, Laine:2016hma}.
At a sufficiently high temperature, the time dimension compactifies and gets replaced by 
temperature~\cite{Linde:1980tt}.
The FV decay rate at finite $T$ then depends on the three-dimensional bounce and the fluctuations 
come from the $D=3$ determinant as well~\cite{Affleck:1980ac}, see also the recent 
works~\cite{Croon:2020cgk, Ekstedt:2021kyx, Gould:2021ccf, Lofgren:2021ogg, Hirvonen:2021zej} regarding 
theoretical progress on the uncertainties of thermal rates.


Colliding bubbles in the early universe create gravitational waves~\cite{Witten:1984rs, Hogan:1986qda, 
Kosowsky:1992rz, Grojean:2006bp, Hindmarsh:2013xza, Cutting:2018tjt}.
These may be strong enough to be detected by the present observatories aLIGO~\cite{TheLIGOScientific:2014jea}
and aVIRGO~\cite{TheVirgo:2014hva}, although most TeV scale phase transitions typically predict
signals in reach of upcoming detectors, such as LISA~\cite{Audley:2017drz, Caprini:2019egz},
DECIGO~\cite{Kawamura:2011zz} and BBO~\cite{Crowder:2005nr, Corbin:2005ny}.
Moreover, the expanding bubble walls may carry CP-violating interactions and be responsible for the
creation of matter over anti-matter, for example in electroweak baryogenesis~\cite{Cline:2006ts} or in extended
scalar sectors.
Other examples in which phase transitions are relevant include the creation of primordial $B$-fields~\cite{Vachaspati:1991nm, 
Sigl:1996dm, DeSimone:2011ek, Tevzadze:2012kk, Ellis:2019tjf}, relation to~\cite{Chowdhury:2011ga, Fabian:2020hny}
and creation of dark matter~\cite{Baker:2019ndr}, neutrino physics~\cite{Brdar:2018num, Brdar:2019fur}
%
%
and primordial black holes~\cite{Baker:2021sno}, among others.

The usual approach to the calculation of the bubble nucleation rate is based on the semi-classical saddle 
point approximation in Euclidean spacetime.
This requires finding the bounce solution, which is an unstable~\cite{Derrick:1964ww} instanton configuration 
that extremizes the action and interpolates between the two minima.
An important simplification is that the bounce is proven to be $O(D)$ symmetric~\cite{Coleman:1977th}
for single field theories in flat spacetime with mild restrictions on the potential. 
This proof was extended to multi-fields in~\cite{Blum:2016ipp}.

Currently, various ways of finding the bounce configuration are available.
The first solution was found already in the original work~\cite{Coleman:1977py} and was derived in the
so-called thin wall (TW) limit, where the two minima are nearly degenerate.
In such a setting, the size of the instanton becomes large, and therefore the transition region (or the wall
of the bounce) from FV to TV becomes narrow, hence thin wall.
This limit may be useful as an order of magnitude estimate, e.g. in thermal field theory when the nucleation
temperature is not far away from the critical one.
However, its applicability is limited and deteriorates~\cite{Adams:1993zs, Sarid:1998sn} when minima 
become more separated.

It is obviously desirable to have analytic control over the bounce part of the rate and some examples
of exact solutions exist, for instance the Fubini-Lipatov instanton~\cite{Fubini:1976jm, Lipatov:1976ny} and its 
generalization~\cite{Loran:2006sf}, pure quartic~\cite{Lee:1985uv}, 
logarithmic~\cite{FerrazdeCamargo:1982sk, Aravind:2014pva} and quartic-quartic 
potentials~\cite{Dutta:2011rc, Guada:2020ihz}.
Also notable is the triangular potential with two piecewise linear segments that was solved exactly in  
$D=4$~\cite{Duncan:1992ai}, see also~\cite{Amariti:2020ntv}. 
The validity of such kink solutions was studied in~\cite{Dutta:2012qt} for a single field and 
in~\cite{Masoumi:2017trx} for multi-field theories.
The linear-quadratic solution was found in~\cite{Dutta:2011rc}, while~\cite{Pastras:2011zr} performed
the analytic continuation to Minkowski space.

Based upon the piecewise linear solution~\cite{Duncan:1992ai}, a general construction was developed 
in~\cite{Guada:2018jek}.
An arbitrary number of linear segments were joined into a polygonal shape, which was solved
semi-analytically in $D = 2, 3, 4, 6, 8$ dimensions.
The piecewise linear was perturbatively expanded to second-order corrections in the potential
and most significantly to any number of fields.
This approach was implemented in the \texttt{FindBounce} package~\cite{Guada:2020xnz}, 
which allows for a fast and arbitrarily precise evaluation of the bounce.
In recent years, new approaches to the bounce calculation were proposed, based on the tunneling 
potential~\cite{Espinosa:2018hue, Espinosa:2018szu}, machine learning techniques~\cite{Jinno:2018dek} and real time 
formalism~\cite{Andreassen:2016cff, Ai:2019fri, Hertzberg:2019wgx, Braden:2018tky}.
Nowadays, several publicly available tools exist that tackle the issue, even with 
multiple fields~\cite{Wainwright:2011kj, Masoumi:2016wot, Athron:2019nbd, Sato:2019wpo}.

Once the bounce is found, one can include one quantum corrections~\cite{Callan:1977pt}, which 
is a rather more involved calculation.
It includes fluctuations from all the fields that couple to the bounce and thus enter into the second 
variation of the action, which ought to be evaluated in the presence of the bounce. 
This is similar to the path integral in quantum mechanics~\cite{Feynman:1965, Kleinert:2004ev}, where 
one computes the (semi)classical particle trajectory and then performs the path integral around it.
To compute the path integral in field theory, we have to find the eigenvalues of the fluctuation
operator, which is the second functional derivative of the action, and then integrate over the resulting 
Gaussian coefficients.
We are thus left with a functional determinant~\cite{Callan:1977pt}, a product of all the eigenvalues 
of the operator.
Because the bounce is $O(D)$ symmetric, the fluctuations also come from a radially symmetric 
operator, which can therefore be decomposed into orbital multipoles.
The product over all the multipoles is divergent and has to be properly regulated, e.g. with
dimensional regularization, where poles of $\Gamma$ functions and the renormalization scale enter.
When the rate is written in terms of renormalized couplings, these artifacts go away, and one
is left with a finite result.
This much was stated already in the original work, however, an explicit solution is missing.

Following the original derivation of the rate~\cite{Callan:1977pt}, a couple of works found solutions
in the TW, in particular~\cite{Konoplich:1987yd} in $D=4$ and~\cite{Munster:1999hr} in $D=3$.
In more recent years, another calculation~\cite{Garbrecht:2015oea} was done using Green's functions,
claiming a numerical discrepancy with~\cite{Konoplich:1987yd}.
These works used explicit eigenfunctions with fixed (Dirichlet) boundary conditions.
Instead of having to find all of these, one can use the Gel'fand-Yaglom~\cite{Gelfand:1959nq}
theorem\footnote{
The paper~\cite{Gelfand:1959nq} is a survey article, where the first instance of the actual theorem is
attributed to Cameron \& Martin~\cite{Cameron:1945} and the derivation due to Montroll~\cite{Montroll:1952}, 
see also~\cite{Coleman:1985}.
 } 
and consider solving a semi-open boundary condition, which
automatically gives the product of eigenvalues without having to compute them\footnote{
For pedagogical examples involving functional determinants and physical applications,
see~\cite{Kirsten:2000ad, Kirsten:2001wz, Kleinert:2004ev, Dunne:2007rt}.
}.
This approach was adopted when deriving the renormalized rate using the WKB~\cite{Dunne:2005rt}
and $\zeta$ function~\cite{Dunne:2006ct} formalism, using a finite orbital cutoff~\cite{Dunne:2005rt}.
It is well suited for numerically evaluating the rate~\cite{Hur:2008yg} and was used to obtain an analytic result 
for the complete parameter space of the quartic-quartic potential~\cite{Guada:2020ihz} in $D=4$.
Progress has also been made regarding the precise treatment of gauge bosons, leading to
gauge invariant rates~\cite{Endo:2017gal, Endo:2017tsz} in single and multi-field theories~\cite{Chigusa:2020jbn}.

The aim of the present paper is to work out the TW decay rate analytically, including the bounce and quantum 
fluctuations via the Gel'fand-Yaglom theorem, while keeping the number of dimensions arbitrary.
The paper is organized in the following way.
We start by giving the expressions for the decay rate in $D = 3, 4$ in~\S\ref{secSummary}.
This is followed by the general setup for the calculation of the rate at one loop in~\S\ref{secSetup}.
The entire~\S\ref{secBounce} is devoted to the Euclidean action, where we define the thin wall
expansion and give the bounce field configuration in~\S\ref{subsecTWExp}, followed by the 
counter-terms and the running of parameters in~\S\ref{subsecBounceRGE}.
Section~\ref{secFluctuations} focuses on the quantum fluctuations.
After defining the problem and setting up the expectations for the behavior of the
fluctuations, we first derive the results for low multipoles in~\S\ref{subsecFluctLowL}.
These contain the single negative unstable eigenvalue and the $D$ translational
zero modes, which are removed perturbatively in~\S\ref{subsecFluctZero}.
We then move on to high orbital modes in~\S\ref{subsecFlucGenL} and find an expression,
which is valid in the UV.
In section~\ref{secRenorm} we regularize the functional determinant and show that the final 
decay rate is free of divergences and independent of the renormalization scale.
We conclude and give an outlook in~\S\ref{secConclusion}.
Technicalities are delegated to the appendices.
The Euclidean action, valid to higher orders can be found in the appendix~\ref{appAction}, 
details on the functional determinants are given in~\ref{appFluct} and the UV integrals that enter in
the renormalization procedure are collected in~\ref{appUVint}.
In appendix~\ref{appQuartic} we compute the renormalized functional determinant for the 
single real quartic using the $\zeta$ formalism and show that it matches the one using Feynman 
diagrams. 
The final appendix~\ref{appFinSumGenD} contains a general result for the functional determinant
in any dimension.

%
%
\section{Summary of the results} \label{secSummary}

\begin{figure}[!ht]
  \centering
  \includegraphics[width=.66 \linewidth]{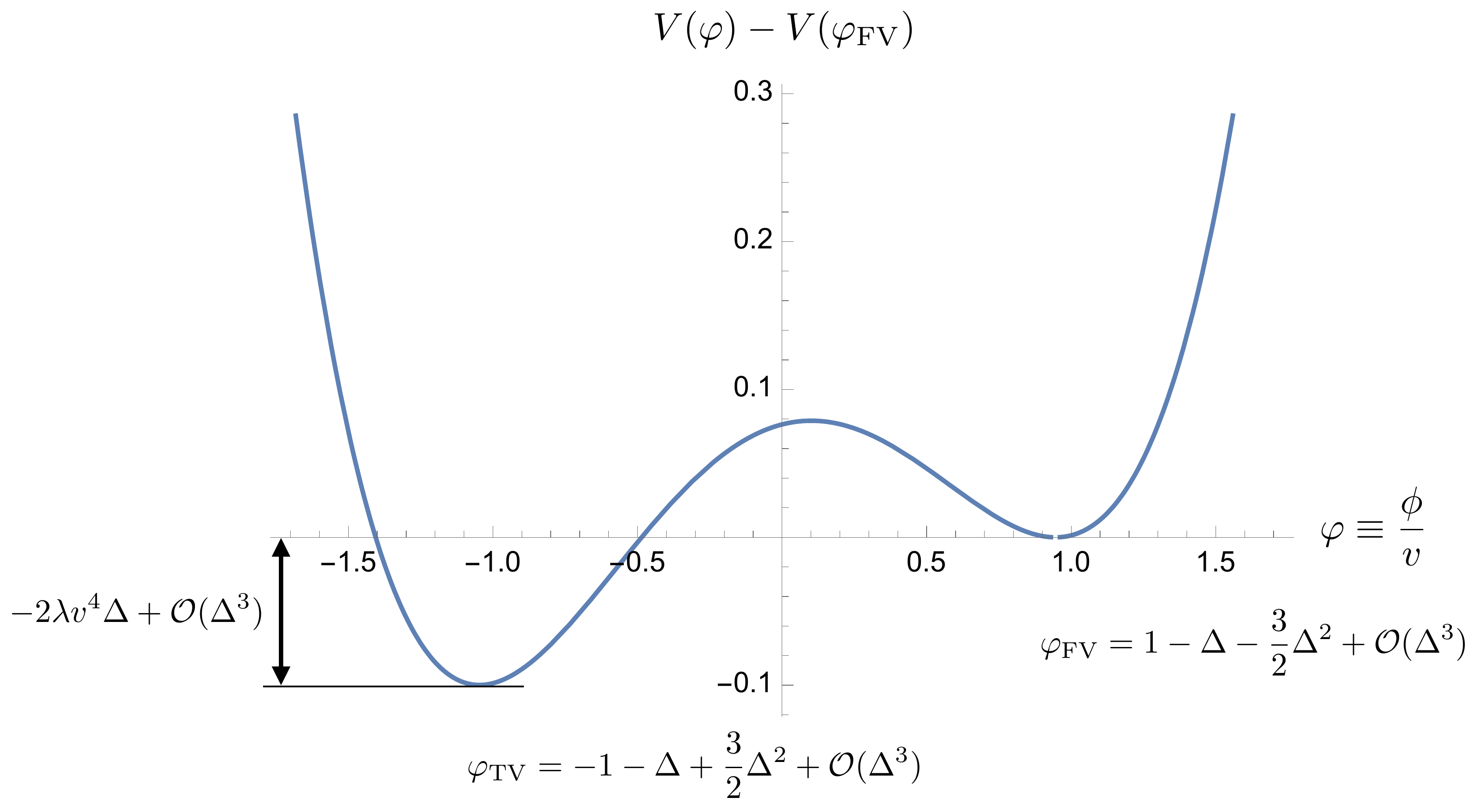}
  \caption{
  The potential~\eqref{eq:potential} as a function of $\varphi \equiv \phi / v$ with $V_{\rm FV}$ subtracted, 
  such that the difference vanishes at the FV.
  We set $\lambda v^4 = 1$, $\Delta = 0.05$ in the plot.
  }
  \label{fig:potential}
\end{figure}

We consider the following potential for a real scalar field $\phi$,
\begin{align} \label{eq:potential}
  V  = \frac{\lambda}{8} \left( \phi^2 - v^2 \right)^2 + \lambda \, \Delta \, v^3 \left(\phi - v \right)  \, ,
\end{align}
which is a slightly rewritten form of what was considered in~\cite{Coleman:1977py, Callan:1977pt}.
This notation introduces a dimensionless coupling $\Delta$, which governs the tunneling
and defines the TW limit when $\Delta \to 0$.
The linear term in $\phi$ is proportional to $\Delta$ and tilts the mexican hat potential, which breaks the 
degeneracy of the two minima, as seen in Fig.~\ref{fig:potential}.

We assume that $0 < \lambda \ll 1$ and $0 < \Delta \ll 1$ are small enough for the theory to remain perturbative 
and for the TW expansion to be applicable.
We chose $\Delta > 0$ to keep the FV on the right (see Fig.~\ref{fig:potential}), while an upper bound
from having the second minimum is given by $\Delta < 1/(3 \sqrt 3) \simeq 0.19$.
In this regime we compute the FV decay rate analytically and find
\begin{align} \label{eqGamVSum}
  \frac{\Gamma}{\mathcal V} \simeq \left(  \left(\frac{S}{2 \pi}\right) \frac{12}{e^{D-1}}
  \lambda v^2 \right)^{D/2} \exp \left [ -S - \frac{1}{\Delta^{D-1}}
  \begin{cases}
    \frac{20 + 9 \ln 3}{54} \, , & D = 3 \, ,
    \\
    \frac{27 - 2 \pi \sqrt 3}{96}  \, , & D = 4 \, ,
  \end{cases} \right]
\end{align}
with the Euclidean bounce action given by
\begin{align} \label{eqSDelta2}
  S &= \frac{1}{\Delta^{D-1}}
  \begin{cases}
    \frac{2^5  \pi v}{3^4 \sqrt{\lambda}} \left( 1 - \left( \frac{9 \pi^2}{4} - 1 \right) \Delta^2 \right) \, , 	& D = 3 \, ,
    \\
    \frac{\pi^2}{3 \lambda} \left( 1 - \left( 2 \pi^2 + \frac{9}{2} \right) \Delta^2 \right) \, , & D = 4  \, .
  \end{cases}
\end{align}
Here the $\Delta^3$ corrections vanish, hence $S$ is valid up to $\mathcal O(\Delta^4)$.
The rate in~\eqref{eqGamVSum} consists of the prefactor and the exponent. 
Both $\lambda$ and $v$ are dimensional parameters in general $D$\footnote{
In particular, at tree level $[\lambda] = 4-D$, $[\phi] = [v] = D/2-1$, and $[\Delta]$ = 0.
}, but $\lambda v^2$ always has mass dimension 2, or equivalently (length)$^{-2}$, and $\Delta$ is 
dimensionless.
The argument of the exponent in~\eqref{eqGamVSum} is dimensionless and has two parts.
The first comes from the renormalized bounce action in~\eqref{eqSDelta2}, given in terms of renormalized
couplings, defined at the scale $\mu_0 = \sqrt \lambda v$.
The second one comes from the functional determinant~\eqref{eqFuncDetD3} in $D=3$,  
and~\eqref{eqFuncDetD4full} in $D=4$. 
We omit the $\mathcal O(\Delta^2/ \Delta^{D-1})$ and higher corrections in such a term.
As expected, the rate goes to zero either when $\lambda \to 0$, in which case the potential vanishes, 
or when $\Delta \to 0$, in which case the vacua are degenerate.
Depending on the relative size of the couplings, the $\Delta^2$ correction of the action in~\eqref{eqSDelta2}
and the one loop contribution from the functional determinant in~\eqref{eqGamVSum} may be comparable
and they should both be included.

%
%
\section{Setup} \label{secSetup}

The problem of computing the FV decay rate was lucidly set up in~\cite{Coleman:1977py, Callan:1977pt}. 
The rate at one loop is given by~(3.11) of \cite{Callan:1977pt}, which we write for generic $D$ dimensions in 
Euclidean space as
\begin{equation} \label{CCGamma}
 \frac{\Gamma}{\mathcal V} = \left( \frac{S_R}{2\pi \hbar} \right)^\frac{D}{2} \left\vert  
 \frac{\det' \mathcal O}{\det \mathcal O_\text{FV} }
 \right\vert^{- \frac{1}{2}} e^{-\frac{S_R}{\hbar} - S_{\rm ct} } \left( 1 + {\cal O}(\hbar) \right) \, ,
\end{equation}
where $\mathcal O = -\partial_\mu \partial^\mu + V^{(2)}$, $S_R$ is the renormalized bounce action and $S_{\rm ct}$ 
is its one loop counterterm.
The second derivative of the potential with respect to $\phi$ is denoted by $V^{(2)}$ and is calculated on the bounce
field configuration, while $V^{(2)}_{\rm FV}$ is evaluated at the FV.
The $\det^\prime$ signifies the omission of zero eigenvalues from the determinant.
There are $D$ such eigenvalues associated with the translational symmetry in $D$ dimensions. 
The removal of each zero mode brings about the insertion of $\sqrt{S/(2\pi \hbar)}$, hence we get the first factor in the 
equation above.
In order to keep the tree level action finite, we subtract the constant FV part of the potential, as shown in
Fig.~\ref{fig:potential}.
Likewise, the FV of the action counter-term must be subtracted to keep $S_{\rm ct}$ 
finite at one loop.

The rate is thus a properly renormalized expression, free of UV divergences and independent
of the renormalization scale $\mu$ at a given loop level. 
As we will see, the $\ln \mu$ and $1/\epsilon$ terms\footnote{
The $\ln \mu$ is only present in even $D$ dimensions. 
Using dimensional regularization, we define $\epsilon = 4 - D$ in four dimensions, and analogously in generic $D$.} 
coming from $S_R$ and $S_{\rm ct}$ in the exponent, will be compensated by the terms coming 
from the regularized functional determinant.

We kept factors of $\hbar$ explicit in~\eqref{CCGamma}, but from here on we set $\hbar = 1$. 
One can recover the $\hbar$ counting from the dimensionless action by considering the kinetic term to 
be $(\partial_\mu \phi)^2 = v^2 (\partial_\mu \varphi)^2$, while the potential term has an overall 
$\lambda v^4$ factor when expressed in terms of $\varphi$. 
Rescaling $v \to \hbar^{-1/2} v$ and $\lambda \to \hbar \lambda$ gives $S \to S/\hbar$.
 
%
%
\section{The bounce} \label{secBounce}
Let us begin with the bounce field configuration, which enjoys a spherical $O(D)$ symmetry, as shown 
in~\cite{Coleman:1977py, Coleman:1977th}.
The bounce extremizes the Euclidean action
\begin{align} \label{eqSD0}
  S &= \int_D \, \left( \frac{1}{2} \dot \phi^2 + V - V_{\rm FV}  \right) \, ,
  &
  \int_D &= \Omega \int_0^\infty \text{d} \rho \, \rho^{D-1} \, ,
  &
  \Omega &= \frac{2 \pi^{D/2}}{\Gamma(D/2)} \, ,  
\end{align}
where $\rho^2 = t^2 + x_i^2$ is the Euclidean radius, $\Omega$ is the $D$-dimensional surface element and
the dot denotes a derivative over $\rho$.
We need to subtract the FV constant $V_{\rm FV}$, i.e. the value of the potential at the FV, to make the action finite.
The extremization of $S$ corresponds to solving the Euler-Lagrange equations of motion with 
appropriate boundary conditions:
\begin{align} \label{eqBounce}
\begin{split}
  \ddot \phi + \frac{D-1}{\rho} \dot \phi &= \frac{\text{d} V}{\text{d} \phi} \, ,
  \qquad
  \dot \phi(0) = \dot \phi(\infty) = 0\, ,
  \\
  \phi(0) &= \phi_{\text{in}} \, , 
  \qquad
  \phi(\infty) = \phi_{\text{FV}} \, .
\end{split}
\end{align}
These conditions ensure that the solution remains finite~\cite{Coleman:1977py}, while $\phi_{\text{in}}$ is in 
principle arbitrary.
As we shall see, the TW expansion automatically sets $\phi_{\text{in}}$ to the TV value $\phi_\text{TV}$.

%
%
\subsection{Thin wall expansion of the action} \label{subsecTWExp}
Throughout this work, we use $\Delta$ as an expansion parameter for the bounce action and the fluctuations.
The entire perturbation series is set in powers of $\Delta$: the expansion for the fields, the Euclidean action 
and the functional determinant.
It is convenient to work with the dimensionless field $\varphi$ and the dimensionless 
Euclidean coordinate $z$, defined as
\begin{align} \label{eqPhir}
  \frac{\phi}{v} \equiv  \varphi &= \sum \varphi_n \,  \Delta^n \, ,
  &
  z = \sqrt{\lambda} v \ \rho - r \, ,
\end{align}
where $n \geq 0$ counts the powers of $\Delta$, starting from the leading $n = 0$ term. 
The size of the bounce instanton (or the bubble wall) is set by the dimensionless Euclidean radius $r$,
which becomes infinite in the TW, when $\Delta \to 0$.
This justifies the expansion
\begin{align} 
  r = \frac{1}{\Delta} \sum r_n \, \Delta^n \, .
\end{align}

The bounce equation in~\eqref{eqBounce} is rewritten using dimensionless variables, expanded
in powers of $\Delta$ and solved in Appendix~\ref{appAction} at each order, such that
\begin{align}
  \varphi_0 &= \text{th} \frac{z}{2} \, ,  \label{bounce0} 
  \\  
   \varphi_1 &= -1 \, , \label{bounce1} 
   \\
  \begin{split}
    \varphi_2 & =  \frac{3}{4 (D-1) \text{ch}^2(z/2)} \bigl( 
    \left( 2 - D - 2 \left(4 + \text{ch} z \right) \ln(1 + e^z) \right) \text{sh} z
    \\
    & \qquad \qquad \qquad \qquad \qquad  - z \left(D - e^z \left(4 + \text{sh} z \right) \right) + 3 (\text{Li}_2(-e^z) - \text{Li}_2(-e^{-z}))
    \bigr) \, .
  \end{split}
  \label{bounce2}
\end{align}

\begin{figure}[t!]
  \centering
  \includegraphics[width=\linewidth]{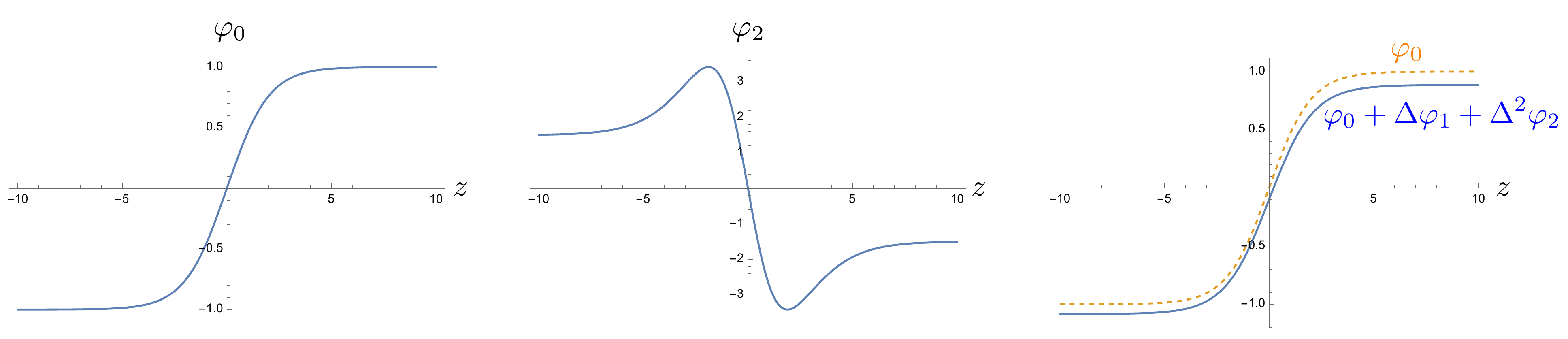}
  \caption{Bounce field configuration in $D = 4$ with $\Delta = 0.1$.
  Left: leading order $\varphi_0$, center: $\varphi_2$ from~\eqref{bounce2} 
  and on the right: $\varphi = \varphi_0 + \Delta \varphi_1 + \Delta^2 \varphi_2$.
  }
  \label{fig:bounce}
\end{figure}

At the leading order the bounce solution $\varphi_0$ interpolates between the two symmetric minima.
At next-to-leading order the bounce gets shifted by a constant $\varphi_1$, while at $n = 2$ the 
$\varphi_2$ describes a non-trivial deformation of the bubble profile, as illustrated on Fig.~\ref{fig:bounce}.
Examining the asymptotics at $\rho = 0, \infty$ or $z = \pm \infty$, we have
\begin{align}
  \varphi_0(z \to \pm \infty) &= \pm 1 \, ,
  &
  \varphi_1(z \to \pm \infty) &= -1 \, ,  
  &
  \varphi_2(z \to \pm \infty) &= \mp \frac{3}{2} \, ,
\end{align}
as in Fig.~\ref{fig:potential}.
This implies that the extremal ends of the bounce correspond to the two vacua $\phi(\rho = 0) = \phi_{\text{in}} = \phi_{\rm TV}$, 
$\phi(\rho \to \infty) = \phi_{\rm FV}$, with increasing level of precision at higher orders of $\Delta$.
While this construction does not allow for an arbitrary $\phi_{\text{in}}$ outside of the TV, one does get a more accurate
estimate of the action at higher powers of $\Delta$,  as reported in Appendix~\ref{appAction}, where we go up to $\varphi_3$.

Solving the bounce equation at the leading order gave us the shape of the bounce in $\varphi_0$, but not its position.
In order to complete the extremization, we need to fix the bubble radius $r$. 
To fix $r \sim r_0/\Delta$ we have to go to $n = 1$.
This can be done by either extremizing the action explicitly as in~\cite{Coleman:1977py} or implicitly 
by solving~\eqref{eqBounce} at $n = 1$.
These two procedures are equivalent and interchangeable at any order.
As explained in Appendix~\ref{appAction}, the radii up to second order, are given by
\begin{align}
  r_0 & = \frac{D-1}{3} \, , & 
  r_1 & = 0 \, , & 
  r_2 & = \frac{6 \pi^2 - 40 + D (26 - 4 D - 3 \pi^2)}{3(D-1)} \, .
\end{align}

With the shape and position of the bounce fixed, the leading order action is given by
\begin{equation} \label{eqS0}
  S_0 = \frac{\Omega \, v^{4-D}}{\lambda^{D/2-1} \Delta^{D-1}} \left(\frac{D-1}{3} \right)^{D-1} \frac{2}{3 D} \, .
\end{equation}
While $v$ and $\lambda$ are dimensional, the combination $v^{4-D} / \lambda^{D/2-1}$ that enters in $S_0$ 
is dimensionless in any $D$.
The action diverges as $1/\Delta^{D-1}$ and thereby suppresses the rate in the TW limit.
In $D = 4$ we have $S_0 = \pi^2/(3 \lambda \Delta^3)$, so the decay rate goes as $\Gamma \propto \exp{(-\pi^2/(3\lambda \Delta^3))}$. This is reminiscent of the 
$\exp(-1/g^2)$ behaviour of gauge instantons~\cite{tHooft:1976snw}, with $g$ the gauge coupling, since quartics like $\lambda$ act as $g^2$.
It is also closely related to the unstable quartic of the SM, where $\Gamma \propto \exp(-8\pi^2/(3\lambda))$.
The TW result in~\eqref{eqS0} multiplies it by $1/\Delta^3$, which further suppresses the rate\footnote{
As discussed above, the dimensions of $\lambda$ and $v$ are such that their ratio $v^{4-D}/\lambda^{D/2-1}$ 
in~\eqref{eqS0} is dimensionless.
However, in $D=4$ thermal field theory fields have dimension 1, while $[\lambda] = 0$, whereby the $S_3$ 
action from~\eqref{eqS0} has dimension 1.
In such an instance, the 3D action gets divided by $T$ to keep the exponent $S_3/T$ in the rate 
dimensionless.
}.

Proceeding to higher orders, as worked out in Appendix~\ref{appAction}, we get the following correction
\begin{equation} \label{Sbounce2}
 S = S_0 \left( 1 + \Delta^2 \left(\frac{1 + D \left(25 - 8 D - 3 \pi^2 \right)}{2 (D-1)} \right) \right) \, ,
\end{equation}
which is valid up to $\mathcal O\left( \Delta^4 \right)$, since the $\Delta^3$ term vanishes.
This result is of interest for two reasons: it provides a more precise estimate of the TW action in any $D$,
and serves as an upper bound for $\Delta$, below which the TW approximation is still valid.

%
%
\subsection{Renormalized bounce action} \label{subsecBounceRGE}

Because we are calculating a one loop quantity, we need to renormalize the bounce action,
which requires adding the action counter-terms and running the relevant couplings.
We follow the usual procedure with dimensional regularization, where one encounters
a pole near integer even dimensions. 
Here we discuss the $D = 4$ case, where the deviation from the integer $D$ is given by the 
usual $\varepsilon = 4 - D$.

{\bf Four dimensions.}
{\em Counterterms.}
We begin with the textbook derivation of counter-terms that renormalize the potential without 
the $\Delta$ term.
We then turn on the linear $\Delta$ term and show that it does not affect the UV structure of the theory at one loop.
In other words, the $\Delta$ counter-term is zero and this parameter does not run.

Let us define the counter-term potential $V_{\text{ct}}$ and a shorthand $V^{(n)}$ for the $n$-th derivative,
evaluated at $\phi = \langle \phi \rangle$,
\begin{align}
  V_{\text{ct}} &= \frac{\delta_{m^2}}{2} \phi^2 + \frac{\delta_\lambda}{4} \phi^4 \, ,
  &
  V^{(n)} \equiv \frac{\text{d}^n V}{\text{d} \phi^n} (\langle \phi \rangle) \, .
\end{align}
Removing the infinity from the four-point function fixes the quartic counter-term
\newlength{\feynheight}
\setlength{\feynheight}{10mm}
\begin{align} \label{eqDeltaLmbd}
  \raisebox{-5mm}{\includegraphics[height=\feynheight]{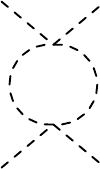}} \quad + \quad
  \raisebox{-5mm}{\includegraphics[height=\feynheight]{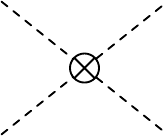}} &= 0 \, ,
  &
  &\Rightarrow
  &
  \delta_\lambda &= \frac{1}{\left( 4 \pi \right)^2 2 \varepsilon} V^{(4) 2} \, .
\end{align}
The $1/\varepsilon$ pole in the 3-point function vanishes under the condition
\begin{align} \label{eq3pCT}
  \raisebox{-5mm}{\includegraphics[height=\feynheight]{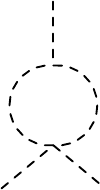}} \quad + \quad
  \raisebox{-5mm}{\includegraphics[height=\feynheight]{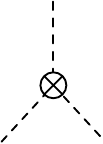}} &= 0 \, ,
  &
  &\text{ iff }
  &
  V^{(3)} &= \langle \phi \rangle V^{(4)} \, .
\end{align}
This relation is automatically satisfied when the cubic term, i.e. $V^{(3)}$, comes solely from 
expanding the quartic after spontaneous breaking.
Moving on to the tadpole, we remove the divergence with the quadratic counter-term
\begin{align} \label{eqDeltaMu}
  &\raisebox{-5mm}{\includegraphics[height=\feynheight]{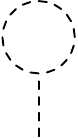}} \quad + \quad
  \raisebox{-5mm}{\includegraphics[height=\feynheight]{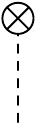}} = 0 \, ,
  &
  &\delta_{m^2} = \frac{1}{\left( 4 \pi \right)^2 \varepsilon} V^{(4)} \left( V^{(2)} - \frac{1}{2} V^{(4)} \langle \phi \rangle^2 \right)  \, .
\end{align}
Using this counter-term and the relation in~\eqref{eq3pCT}, the two-point infinities cancel automatically
\begin{align}
  \raisebox{-5mm}{\includegraphics[height=0.7\feynheight]{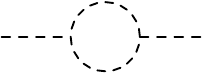}} \, + \,
  \raisebox{-5mm}{\includegraphics[height=0.7\feynheight]{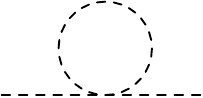}} \, + \,
  \raisebox{-0.7mm}{\includegraphics[width=2cm]{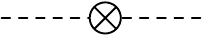}} &= 0 \, .
\end{align}
This procedure was set up for an arbitrary potential, now we can apply it to our $V$, given in \eqref{eq:potential}.
When the two minima are degenerate, the potential reduces to the simple quartic with 
$\langle \phi \rangle = v$, where all the consistency relations, derived above, are valid.

Now turn on the linear $\Delta$ term.
The fourth and third derivatives remain the same, such that the $\delta_\lambda$ in~\eqref{eqDeltaLmbd} and the 
running of $\lambda$ is unaffected.
On the other hand, the position of the minima shifts away from $v$, so $\langle \phi \rangle$ is $\Delta$-dependent.
However, the finiteness of tadpoles and 2-point functions was derived for arbitrary $\langle \phi \rangle$ and
remains valid for any $\Delta$.
Thus, even if we were to add a counter-term for $\Delta$, we would have to set it to zero and $\Delta$ does not run.

With the $\delta_\lambda$ and $\delta_{m^2}$ at hand, we can calculate the counter-term for the 
Euclidean action.
Plugging the $V$ from~\eqref{eq:potential} into~\eqref{eqDeltaLmbd} and~\eqref{eqDeltaMu}, we have
\begin{align} \label{eqCTs}
  \delta_\lambda &= \frac{9 \lambda^2}{\left( 4 \pi \right)^2 2 \varepsilon}\, ,
  &
  \delta_{m^2} &= - \frac{3 \lambda^2 v^2}{\left( 4 \pi \right)^2 2 \varepsilon}\, .
\end{align}
The $\langle \phi \rangle$ terms in~\eqref{eqDeltaMu} cancel away and $\delta_{m^2}$ does not
depend on $\Delta$ to all orders in $n$.
With \eqref{eqCTs} we can compute the one-loop counter-term action:
\begin{align} \label{eqSct}
\begin{split}
  S_{\rm ct} &= \int_D \left( V_{\text{ct}} - V_{\text{ctFV}} \right)
  = \frac{3 \lambda^2}{8 \left( 4 \pi \right)^2 \varepsilon} \int_D \left( 
  3 \left(\phi^4 - \phi_\text{FV}^4 \right) - 2 v^2 \left(\phi^2 - \phi^2_\text{FV} \right)\right) 
   \simeq -\frac{3}{16 \varepsilon \Delta^3} \, .
\end{split}
\end{align}
This integration was performed in the TW expansion up to $\mathcal O(\Delta^2)$, which requires including 
$\varphi_0$ and $\varphi_1$, see~\eqref{eqPhi2ct} and~\eqref{eqPhi4ct}.

{\em Running.} With the $\delta_\lambda$ computed in~\eqref{eqDeltaLmbd}, we get the usual one loop
$\beta_\lambda$ function that defines the running of $\lambda$
\begin{align}
  \beta_\lambda &= \frac{\text{d} \lambda}{\text{d} \ln \mu} = \frac{9 \lambda^2}{\left( 4 \pi \right)^2} \, ,
  &
  \lambda(\mu) &\simeq \lambda_0 + \frac{9 \lambda_0^2}{\left( 4 \pi \right)^2} \ln \frac{\mu}{\mu_0} \, .
\end{align}
Here $\mu_0^2 \sim V^{(2)}_{\text{FV}} \sim \lambda_0 v^2$ is the scale where the renormalized $\lambda_0$ 
is measured by an anxious observer in the FV. 
Plugging the running coupling $\lambda(\mu)$ into the action, and including the counter-terms in \eqref{eqSct}, we obtain
\begin{align} \label{eqSRun}
  S_R +  S_{\rm ct} &= S \left( 1 -  \frac{9 \lambda_0}{\left( 4 \pi \right)^2} 
  \left(\frac{1}{\varepsilon} + \ln \frac{\mu}{\mu_0} \right) \right) \, , 
\end{align}
with $S$ the tree-level bounce action given in \eqref{Sbounce2}.
This is the combination that enters at the exponent in \eqref{CCGamma}, and
as we shall see, these are precisely the terms that will cancel the divergence and the $\mu$-dependence
from the functional determinant.
Moreover, while the semi-classical $S$ in~\eqref{eqS0} goes as $\hbar^{-1}$, the running and counterterms are 
one loop suppressed, and thus one order higher in $\hbar$, i.e. $\hbar^0$.

{\bf Other dimensions.}
In the $\overline{\text{MS}}$ scheme the poles of $\Gamma$ functions vanish in odd dimensions and there are no 
associated infinities and no renormalization scale, such that $S_R +  S_{\rm ct} = S$.
In higher even dimensions, diagrams with more than two propagators become divergent, which means
one has to introduce further counterterms for $\phi^6$ in $D=6$ and so on.
These all get fixed, starting from the last divergent diagram with the highest number of external legs,
and then going down to the quartics and tadpoles.
%
%
In the end, all of them depend only on $\lambda$ and $v$, and no new parameters enter.
In case there is a clear physical motivation to consider a particular dimension $D \geq 6$, one can follow
through these steps.
However, the main focus of this work is on the functional determinant, therefore a general treatment of 
counterterms for any $D$ goes beyond our scope.

%
%
\section{Fluctuations} \label{secFluctuations}
With the Euclidean action and the associated bounce at hand, we are ready to move on to the main
objective, the calculation of the fluctuations.
We wish to evaluate the ratio of determinants in \eqref{CCGamma}
\begin{equation} \label{detratio}
 \left \vert \frac{\det' \mathcal O}{\det \mathcal O_\text{FV}}  \right\vert^{- \frac{1}{2}} = 
 \left\vert \prod_{l=0}^\infty \frac{\det'{\cal O}_l}{\det{\cal O}_{l{\rm FV}}}  \right\vert^{- \frac{1}{2}} \, .
\end{equation}
Exploiting the $O(D)$ symmetry of the problem, we perform the angular separation of the radial variable $\rho$,
and expand the fluctuations in hyper-spherical multipoles $l$:   
\begin{align} \label{eqFluctO0}
  {\mathcal O}_l &= -\frac{\text{d}^2}{\text{d} \rho^2} - \frac{D-1}{\rho} \frac{\text{d}}{\text{d} \rho} + 
  \frac{l \left( l + D - 2 \right)}{\rho^2} + V^{(2)} \, ,
\end{align}
where $V^{(2)} = \text{d}^2 V/\text{d} \phi^2$ is evaluated on the bounce. 
The operator ${\cal O}_{l{\rm FV}}$ is the same as ${\cal O}_l$, with $V^{(2)}$ replaced by $V^{(2)}_{\rm FV}$.
The prime in $\det'$ means that the zero eigenvalues are removed.

The functional determinant $\det \mathcal O_l$ is computed around the bounce configuration, i.e.
$\phi = \overline \phi + \psi$, where  $\overline \phi$ is the bounce solution computed in the previous section 
and $\psi$ are the fluctuations around it.
To compute the determinant, one needs to define the basis for $\psi$ and find the eigenvalues of the fluctuation 
operator with fixed Dirichlet boundary conditions at $\rho = 0$ and $\infty$, and then multiply them.
However, instead of solving the entire eigensystem, we take advantage of the Gel'fand-Yaglom 
theorem~\cite{Gelfand:1959nq}, according to which
\begin{equation} \label{GelYagth}
  \frac{\det {\cal O}_l}{\det {\cal O}_{l \rm{FV}}} = 
  \lim_{\rho \to \infty} \left( \frac{\psi_l(\rho)}{\psi_{l \rm{FV}}(\rho)} \right)^{d_l} \, ,
\end{equation}
where the degeneracy is given by~\cite{Kleinert:2004ev,Dunne:2006ct} 
\begin{align} \label{eqDegL}
  d_l &= \frac{(2l + D-2)(l+D-3)!}{l! (D-2)!} \, .
\end{align}
Setting $l=0$ we get $d_0 = 1$ regardless of $D$ and thereby a single $s$-wave eigenmode, whose eigenvalue 
should be negative~\cite{Coleman:1987rm}.
On the other hand, plugging $l=1$ into~\eqref{eqDegL} gives $d_1 = D$ and corresponds to translations 
of the nucleated bubble in all possible $D$ directions.
Since the bounce is $O(D)$ symmetric and can nucleate anywhere, we expect to get $D$ eigenmodes with 
zero eigenvalues due to translational symmetry.
We will take care of them in section~\ref{subsecFluctZero}.

The $\psi_l$ and $\psi_{l{\rm FV}}$ that enter in~\eqref{GelYagth} are now solutions of
\begin{align} \label{psileqs}
 \mathcal O_l \, \psi_l &= 0 \, , &
  \mathcal O_{l \text{FV}} \, \psi_{l\text{FV}} & = 0 \, ,
\end{align}
with the boundary conditions $\psi_l (\rho \sim 0) \sim \rho^l$, $\psi_{l\rm{FV}} (\rho \sim 0) \sim \rho^l$ fixed 
only at the origin.
The theorem allows to compute the ratio of the two determinants
\begin{align} \label{Rldef}
  R_l &\equiv \frac{\psi_l}{\psi_{l{\rm FV}}} \, ,
  &
  &\text{ and }
  &
  \frac{\det {\cal O}_l}{\det {\cal O}_{l \rm{FV}}} &= R_l(\infty)^{d_l} \, ,
\end{align}
without explicitly computing all the eigenvalues.
We can trade the $\psi_l$ in~\eqref{psileqs} for $R_l$ (see Appendix~\ref{appFluct}), and solve
\begin{align}
  \ddot \psi_{l{\rm FV}} +  \frac{D-1}{\rho} \dot \psi_{l{\rm FV}} & =  \left( \frac{l \left( l + D - 2 \right)}{\rho^2} + V_{\rm FV}^{(2)} \right) \psi_{l \rm{FV}}  \, , 
 \label{eqPsiFV}
  \\
  \ddot R_l + 2 \left( \frac{\dot \psi_{l\text{FV}}}{\psi_{l\text{FV}}} \right) \dot R_l &=
  \left(V^{(2)} - V^{(2)}_\text{FV}   \right) R_l \, , 
 \label{eqFluctR}
\end{align}
with boundary conditions
\begin{align} \label{eqRlBcs}
  \psi_{l\rm{FV}} (\rho \sim 0) & \sim \rho^l \, , &  R_l (\rho = 0) &= 1 \, ,  &  \dot R_l(\rho = 0) &= 0 \, .
\end{align}
The main advantage of analyzing $R_l$ directly~\cite{Baacke:2003uw} is that it is bounded on the entire $\rho$ interval, 
because $V^{(2)} - V^{(2)}_\text{FV} \xrightarrow{\rho \to \infty} 0$. 
This was not the case for $\psi_l$ and $\psi_{l\text{FV}}$, which diverge exponentially when $\rho \to \infty$, even
though their ratio stays finite.
Once we solve \eqref{eqFluctR} and find $R_l(\infty)$, we expect the following behavior
\begin{align} \label{eqRlExpect}
  R_l(\infty) =
  \begin{cases}
    < 0, & l = 0 \, ,
    \\
    0, & l = 1 \, ,
    \\
    1, & l \gg 1 \, .
  \end{cases}
\end{align}
The first two come from the instability of the bounce and translational invariance.
We will calculate the spectrum in this low multipole/IR limit in the upcoming section~\ref{subsecFluctLowL} 
and then show how to remove the zeroes from the determinant in~\ref{subsecFluctZero}.
The last limit in~\eqref{eqRlExpect}, when $l \gg 1$, becomes obvious upon examining~\eqref{eqFluctO0}.
In this UV limit, the orbital term dominates over the rest, and the $V^{(2)}$ term can be neglected.
Therefore, there is no significant difference between $\mathcal O_l$ and $\mathcal O_{l \text{FV}}$, 
so that $\psi_l \to \psi_{l \text{FV}}$ or $R_l \to 1$.
The behavior of $R_l$ in the UV will be examined separately in section~\ref{subsecFlucGenL}.

%
%
\subsection{Low multipoles} \label{subsecFluctLowL}
To study the low multipoles, let us start with the FV part in~\eqref{eqPsiFV}. 
Introducing
\begin{equation} \label{nudefl}
 \nu = l + \frac{D}{2} - 1 \, ,
\end{equation}
rescaling $\psi_{l{\rm FV}} \to \rho^{\frac{D-1}{2}} \psi_{l{\rm FV}}$ and switching to dimensionless variables,~\eqref{eqPsiFV} 
simplifies to 
\begin{equation} \label{chiFVzeq}
\frac{\text{d} \psi_{l\text{FV}}^2}{\text{d} z^2} = 
  \left(  \frac{\nu^2 - \frac{1}{4}}{(z+r)^2} + \ddVt_{\text{FV}}  \right) \psi_{l \rm{FV}}  \, ,
\end{equation}
with the FV constant given by
\begin{equation}  \label{Vappex}
 \ddVt_\text{FV} = \frac{1}{2} (3 \varphi_{\rm FV}^2 - 1) = 1 - 3 \Delta - 3\Delta^2 \, .
\end{equation}
We are only interested in terms up to $\Delta^2$ and will consistently be dropping the $\Delta^{p>2}$
powers in this section, such that
\begin{align} \label{eqPsiFVLowL}
  \frac{\text{d} \psi_{l\text{FV}}^2}{\text{d} z^2}
  &= \left( 1 - 3 \Delta - 3 \Delta^2 + \Delta^2 \left( \frac{\nu^2 - \frac{1}{4}}{r_0^2} \right) \right) \, \psi_{l\text{FV}} \, .
\end{align}
It becomes clear why we need to go to $\mathcal O(\Delta^2)$: there is no $l$ dependence at lower orders.
Moreover, up to this order, there is no $z$ dependence\footnote{The $z$ dependence enters from ${\cal O}(\Delta^3)$ onwards.} 
on the right hand side of \eqref{eqPsiFVLowL}, so the FV solution is simply
\begin{align} \label{eqPsiFVLowLSol}
  \psi_{l\text{FV}}(z) & \simeq c_{\rm FV} \exp \left[ \left( 1 - \frac{3}{2} \Delta + 
  \left(\frac{\nu^2 - \frac{1}{4}}{2 r_0^2} - \frac{21}{8} \right) \Delta^2 \right) z \right] \, .
\end{align}
We expanded the exponent in powers of $\Delta$ and dropped the term proportional to
$e^{-z}$ in order to satisfy the boundary condition \eqref{eqRlBcs} at $z \to -\infty$, corresponding to $\rho = 0$.
The FV solution has the form $\psi_{l \text{FV}} = \psi_{l \text{FV}0} \psi_{l \text{FV}1}^\Delta \psi_{l \text{FV}2}^{\Delta^2} \ldots$ 
that motivates taking a {\em multiplicative expansion} for $R_l$
\begin{align} \label{eqRlMult}
  R_l &= \prod_{n\geq 0} R_{l n}^{\Delta^n} \, ,
  &
  \ln R_l = \sum_{n \geq 0} \ln R_{l n} \Delta^n \, . 
\end{align}
One further simplifying technicality in the calculation of $R_l$ is to work with an exponentially compensating 
variable $x = e^z$,
which turns~\eqref{eqFluctR} into
\begin{equation} \label{Rlxall}
\frac{{\rm d}^2 R_l}{{\rm d}x^2} + \frac{1}{x} \left( 2 \frac{{\rm d}\psi_{l{\rm FV}} / {\rm d}z}{\psi_{l{\rm FV}}} + 1 \right) \frac{{\rm d} R_l}{{\rm d}x} = \frac{1}{x^2} \left( \tilde V^{(2)} - \tilde V^{(2)}_{\rm FV}  \right) R_l \, .
\end{equation}
Here $ \tilde V^{(2)} = \frac{1}{2} (3 \varphi^2 - 1)$,
with $\varphi$ the bounce solution from~\S\ref{subsecTWExp}.
We plug the bounce expansion from~\eqref{eqRlMult} into~\eqref{Rlxall} and solve it up to order $\Delta^2$.
Up to this order, the combination $\frac{{\rm d}\psi_{l{\rm FV}} / {\rm d}z}{\psi_{l{\rm FV}}}$ does not depend on $x$.
Further details of the calculation of $R_l$ are deferred to Appendix~\ref{appFluct}, here we only discuss the results.
From the first equation in~\eqref{eqRlMult}, we have
\begin{align} \label{eqRlMultExp}
  R_l = R_{l0} \left( 1 + \Delta \ln R_{l1} + \Delta^2 \left( \frac{1}{2} \ln^2 R_{l1} + \ln R_{l2} \right) \right) \, ,
\end{align}
and the solutions of~\eqref{Rlxall} up to $n=2$ are found to be
\begin{align} \label{eqRl0Lowlsol}
  R_{l 0} &= \frac{1}{\left(1 + x \right)^2 } \, ,
  \\
  \ln R_{l1} & = 3\left( r + \ln x \right) \, ,  \label{lnRl1} 
  \\
  \ln R_{l2}(x \to \infty) & = \frac{3}{4} \frac{\left( l - 1 \right) \left( l + D - 1 \right)}{ \left( D - 1 \right)^2} \, x^2 \, .
\end{align}
In the last equation, we only show the high-$x$ part of $\ln R_{l2}$, as prescribed by the Gel'fand Yaglom 
theorem~\eqref{GelYagth}, which grows as $x^2$.
In this limit, the $R_{l0}$ goes as $x^{-2}$, implying that $R_l \to 0$ for the first two orders
and a non-vanishing result comes in at $\mathcal O(\Delta^2)$.
Indeed, $\ln R_{l2} \propto x^2$, which compensates exactly the large-$x$ behavior of $R_{l0}$. 
As anticipated, the first non-trivial $l$-dependent term appears at second order in $\Delta$ and
is obtained by plugging the solutions into \eqref{eqRlMultExp}, such that
\begin{equation} \label{eqRlinfLowl}
  R_l(\infty) =  \Delta^2 e^{D-1} \frac{3}{4} 
  \frac{\left( l - 1 \right) \left( l + D - 1 \right)}{ \left( D - 1 \right)^2} \, .
\end{equation}

Some comments on the low-$l$ fluctuations are in order.
\begin{itemize}
  \item The final expression in~\eqref{eqRlinfLowl} conforms to expectations from~\eqref{eqRlExpect}: 
  the ratio of multiplied eigenvalues is negative for $l = 0$ and vanishes for $l = 1$.
  \item  The $R_l(\infty)$ does not go to 1 when $l \gg 1$, as argued in~\eqref{eqRlExpect},
  because the entire setup applies only for low $l$.
  Looking back at our starting point in~\eqref{eqPsiFVLowL}, the $l$ dependence enters via the 
  $\Delta^2 \nu^2$ term, which was counted as $\mathcal O(\Delta^2)$.
  Such power counting only makes sense as long as $\nu < 1 / \Delta$.
  In the following section, we will have to adapt it in order to access the higher multipoles.
  \item The factor $e^{D-1}$ in \eqref{eqRlinfLowl} comes from $R_{l1}^\Delta = e^{(3 r_0 / \Delta + 3 \ln x) \Delta}$,
  tracing back to the $3 r$ term in \eqref{lnRl1}.
  This in turn is needed to implement the boundary condition at the origin $\rho = 0$, corresponding to $z = -r$.
  As explained in Appendix~\ref{appFluct}, proper implementation of the boundary conditions is crucial
  in order to keep the $\Delta$ power counting valid.
\end{itemize}

%
%
\subsection{Zero removal} \label{subsecFluctZero}

The $\det'{\cal O}_l$ that enters \eqref{detratio} and the total rate in~\eqref{CCGamma}, contains the truncated product 
of eigenvalues with the removed $l = 1$ zeroes due to the translational invariance of the bounce.
In the Gel'fand-Yaglom approach, we cannot simply handpick and remove any individual eigenvalue, 
because all the eigenvalues with the same $l$ are multiplied together.
Instead, we subtract the zeroes perturbatively~\cite{Jevicki:1976kd, Endo:2017gal, Endo:2017tsz, 
Andreassen:2017rzq} by first off-setting the fluctuation potential with a small dimensionful parameter 
$\mu^2_\varepsilon$ and then finding the corresponding solution of
\begin{equation} \label{eqOeps}
  \left( \mathcal O_1 + \mu^2_\varepsilon \right) \psi_1^\varepsilon = 0 \, .
\end{equation}
By introducing such a dimensional regulator in the off-set, the ratio of determinants does not approach
zero as in~\eqref{eqRlinfLowl}, but instead goes to
\begin{align} \label{eqR1eps}
  R^\varepsilon_1(\infty) &= \frac{\psi_1^\varepsilon(\infty)}{\psi^\text{FV}_1(\infty)} 
   \simeq \frac{ \left(\mu^2_\varepsilon + \gamma_1 \right) 
  \prod_{n=2}^\infty  \gamma_{\textbf n}}{\prod_{n=1}^\infty \gamma^{\text{FV}}_{\textbf n}}
  = \mu^2_\varepsilon  R^\prime_1(\infty) \, .
\end{align}
Here, the $\textbf n$ refers to the collective index over all the eigenvalues and $\gamma_1$ corresponds to
the $D$-fold degenerate $l=1$ eigenvalue.
It also becomes clear from~\eqref{eqR1eps} that the dimension of $R^\prime_1$ is reduced by 2 with respect to $R_l$.
To calculate the reduced determinant, we thus have to compute
\begin{equation} \label{eqR1pDef}
  R^\prime_1(\infty) = \lim_{\mu^2_\varepsilon \to 0} \frac{1}{\mu^2_\varepsilon} R^\varepsilon_1(\infty) \, .
\end{equation}

In the TW power counting scheme, the $l$ dependence in the determinant appears at the $\Delta^2$ order.
Therefore, it is enough to lift the fluctuation potential at the same order, i.e. 
$V^{(2)} \to V^{(2)} + \Delta^2 \mu_\varepsilon^2$ in the $R_l$ equation \eqref{eqFluctR}.
This does not affect the FV, which has no zeroes to begin with.
It also does not affect the $l=1$ fluctuations at the leading and next-to-leading order,
i.e. $R_{l 0}$ and $R_{l 1}$ remain the same.
However, it does impact the $R_{l 2}$ fluctuations at the second order, where the solution for $l = 1$ is found 
to be
\begin{align} \label{eqR21prime0}
  \ln R_{1 2}^\varepsilon & = \frac{1}{12}  \frac{ \mu^2_\varepsilon}{\lambda v^2} x^2 \, .
\end{align}
The technical details of this derivation are left to the appendix~\ref{appFluct}, 
see~\eqref{eqV2tldeps}-\eqref{eqR21primeDer}.
Plugging the result back into the multiplicative expansion defined in~\eqref{eqRlMultExp} and 
using~\eqref{eqR1pDef}, we have
\begin{align} \label{eqR21prime1}
\begin{split}
  R^\prime_1(\infty) &=  R_{l 0}(\infty) e^{D-1} \lim_{\mu^2_\varepsilon \to 0} 
  \frac{1}{\Delta^2 \mu^2_\varepsilon} \Delta^2 \ln R_{2 1}^\varepsilon 
  = \frac{e^{D-1}}{12} \frac{1}{\lambda v^2} \, .
\end{split}
\end{align}
This is exactly what we would expect on dimensional grounds because each removal of zero gives us an 
additional power of $1/(\lambda v^2)$ and $[\lambda v^2] = 2$.
The first part of the determinant is thus
\begin{equation} \label{eqFuncDetDecon}
 \left \vert \frac{\det' \mathcal O}{\det \mathcal O_\text{FV}}  \right\vert^{- \frac{1}{2}} = 
 \left( \left\vert R_0 \right \vert R_1^{\prime D} \prod_{l=2}^\infty \frac{\det{\cal O}_l}{\det{\cal O}_{l{\rm FV}}} \right)^{- \frac{1}{2}} \, ,
\end{equation}
which is how the dimensional part enters in~\eqref{eqGamVSum}.

It may be insightful to derive this result in an alternative way.
We can look for the explicit $l=1$ eigenvalue $\gamma_1$ with Dirichlet boundary conditions, and then 
divide $R_l$ by it.
To this end, consider the ansatz $\psi_l = Y_{lm} \dot \Phi$, inspired by the fact that $\partial_\mu \Phi$ 
satisfies the equations of motion with zero eigenvalues~\cite{Lee:2014yud}.
Plugging it into $\mathcal O_l \psi_l = \gamma_1 \psi_l$ with $\mathcal O_l$ from~\eqref{eqFluctO0}, we have
\begin{align} \label{eqFluctl1}
  - \dddot \Phi - \frac{D-1}{\rho} \ddot \Phi + V^{(2)} \dot \Phi + 
  \frac{l \left( l + D - 2 \right)}{\rho^2} \dot \Phi = \gamma_1 \dot \Phi \, .
\end{align}
Taking the derivative of the bounce equation \eqref{eqBounce} with respect to $\rho$ and rearranging the terms, 
we get to the following identity
\begin{align}
  - \dddot \Phi - \frac{D-1}{\rho} \ddot \Phi + V^{(2)} \dot \Phi = 
-  \frac{D - 1}{\rho^2} \dot \Phi \, .
\end{align}
Plugging it into~\eqref{eqFluctl1}, using $\rho = (r + z)/(\sqrt{\lambda}v)$, $r = r_0 / \Delta$, 
with $r_0 = (D-1)/3$, and expanding in $\Delta$, 
we obtain the desired Dirichlet eigenvalue
\begin{align}
  \gamma_1 &= 9 \lambda v^2 \Delta^2 \frac{(l-1)(l+D-1)}{(D-1)^2} \, .
\end{align}
After dividing $R_l$ in \eqref{eqRlinfLowl} by $\gamma_1$, the reduced determinant comes about
\begin{align}
  R_l^\prime(\infty) &= \frac{R_l(\infty)}{\gamma_1} = \frac{e^{D-1}}{12} \frac{1}{\lambda v^2} \, ,
\end{align}
which is in agreement with~\eqref{eqR21prime1}.

%
%
\subsection{Generic multipoles} \label{subsecFlucGenL}
The functional determinant in~\eqref{detratio} involves a product of orbital multipoles up to infinity.
However, the ratio of eigenvalues $R_l$, given in section~\ref{subsecFluctLowL}, only applies for low $l$.
We need to modify the $\Delta$ power counting such that we can access the UV part of the 
spectrum with arbitrary high $\nu$, given by the UV-valid ratio $R_\nu$.

The orbital numbers $\nu$ enter in the $R_l$ equation~\eqref{eqFluctR} via the FV solution 
$\psi_{l\text{FV}}$ that comes from solving~\eqref{chiFVzeq}.
The key change here is to treat the combination $\Delta \nu$, which enters into the FV equation~\eqref{chiFVzeq}
in the TW expansion, as an order one parameter.
With such modification, the $\nu$ multipoles enter the iterative expansion already at the leading 
order (instead of $\Delta^2$ as in section~\ref{subsecFluctLowL}).
The leading order FV solution is thus given by
\begin{align}
  \psi_{\nu \text{FV}} &\simeq e^{k_\nu z} \, ,
  &
  k_\nu^2 = 1 + \frac{\Delta^2 \nu^2}{r_0^2} \, ,
  \label{knudefin}
\end{align}
where the multipoles are characterized by $k_\nu$, which goes from 1 for low multipoles (in 
agreement with the low multipole solution) and up to $\infty$.
Next, we consider the UV-valid equation for $R_\nu$ at the leading order, written in terms of the 
FV fluctuations
 \begin{equation} \label{Rnuxeqmain}
  \ddx R_{\nu 0} + \frac{2 k_\nu + 1}{x} \dx R_{\nu 0} - \frac{1}{x^2} \left( 
  \ddVt - \ddVt_{\rm FV} \right) R_{\nu 0} = 0 \, ,
\end{equation}
which is solved by
\begin{align}
  R_{\nu 0} &= \frac{1}{\left(1 + x \right)^2} \left( 1 + 4 \frac{k_\nu - 1}{2 k_\nu + 1} x + 
  \frac{\left( k_\nu - 1 \right)\left(2 k_\nu - 1 \right)}{\left( k_\nu + 1 \right)\left(2 k_\nu + 1 \right)} x^2 \right) \, .
\end{align}
For low multipoles we have $k_\nu \to 1$, which reduces the above to $R_{\nu 0} \simeq 1/(1+x)^2$ and 
matches with~\eqref{eqRl0Lowlsol}.
On the other hand, at large $x$ we have 
\begin{align} \label{eqRnu0inf}
  R_{\nu 0}(\infty) &= \frac{\left( k_\nu - 1 \right)\left(2 k_\nu - 1 \right)}{\left( k_\nu + 1 \right)\left(2 k_\nu + 1 \right)} \, .
\end{align}

However, the leading term alone is not enough, one needs to proceed to all orders in $n$, to
recuperate and sum all the $\mathcal O(\Delta^0)$ terms.
A systematic procedure of isolating all such terms is explained in detail in Appendix~\ref{appFluct}.
While getting the individual terms is rather complicated, the final result after all the orders are summed
up is remarkably simple
\begin{equation}
  U = 3 r_0 \left( k_\nu - \sqrt{k_\nu^2 -1} \right) \, .
\end{equation}
Together with $R_{\nu 0}$ in \eqref{eqRnu0inf}, one gets the complete leading-order result
\begin{align} \label{eqRnuUFact}
  R_\nu(\infty) &= R_{\nu 0}(\infty) e^U \, ,
  \\
  \label{lnRnuU}
  \ln R_\nu(\infty) &= \ln \frac{(k_\nu -1)(2k_\nu -1)}{(k_\nu+1)(2k_\nu+1)} + 3 r_0 \left( k_\nu - \sqrt{k_\nu^2 -1} \right) \, .
\end{align}
Before proceeding, let us comment on the salient features of this elegant result.
\begin{itemize}
  \item The final result in~\eqref{eqRnuUFact} is a {\em complete} expression for the product of eigenvalues
  at a given multipole, which is non-zero at the leading order in $\Delta$.
  This is in contrast to the low-$l$ result in~\eqref{eqRlinfLowl}, where the first non-vanishing term appeared 
  at ${\cal O}(\Delta^2)$. 
  If we expand~\eqref{eqRnuUFact} for low multipoles ($\nu \ll 1/\Delta$)
   up to ${\cal O}(\Delta^2)$, using the definition of $k_\nu$ in~\eqref{knudefin},
  we do not recover~\eqref{eqRlinfLowl}. This is not a surprise: the result~\eqref{eqRnuUFact}
  is only valid at the leading order, ${\cal O}(\Delta^0)$, does not include all the ${\cal O}(\Delta^2)$
  corrections, so it cannot reproduce~\eqref{eqRlinfLowl}. That is why we needed a separate 
  calculation for the low multipoles. 
  Note when $k_\nu \to 1$ one gets
  $U(k_\nu \to 1) = 3 r_0$, confirming the $e^{D-1}$ factor which appears in~\eqref{eqRlinfLowl},
  and traces back to~\eqref{lnRl1}.   
  As we shall see in the following section, the large multipoles are the ones that dominate in the determinant,
  and at the leading order those are fully accounted for by~\eqref{lnRnuU}. 
  \item At large multipoles when $k_\nu \to \infty$, the determinant goes to unity, in agreement with the expectation 
  from \eqref{eqRlExpect}. 
  Higher order corrections in $U$ behave as $\sim 1/k_\nu$ and vanish for large $k_\nu$, such that they do
  not affect the correct behavior of $R_{\nu 0}$.
  However, they do provide sub-leading powers of $k_\nu$, which are important in the sum we have to perform
  to obtain the determinant, as we will see in the next section.
\end{itemize}

%
%
\section{Renormalized determinant} \label{secRenorm}
In the previous section we computed the negative $l=0$ eigenmodes and removed the zero modes with $l=1$,
now we wish to evaluate the infinite product in~\eqref{eqFuncDetDecon}.
Taking the log and using Gel'fand-Yaglom \eqref{GelYagth}, one has
\begin{equation} \label{logdetRnumain} 
  \ln \left(\frac{\det \mathcal O}{\det \mathcal O_\text{FV}} \right) = 
  \sum_{\nu = D/2-1}^\infty d_\nu \ln R_\nu \, ,
\end{equation}
where in this section we denote the asymptotic value $R_\nu(\infty)$ with $R_\nu$, to ease the notation.
The sum in~\eqref{logdetRnumain} formally starts from $l = 0$, corresponding to $\nu = D/2 -1$,
 and goes to $\infty$, but as we shall shortly see, it 
is dominated by large multipoles in the TW limit\footnote{
The sum starting at $l=0$ is in contrast with the product left in~\eqref{eqFuncDetDecon} starting at $l=2$.
However it is convenient to define the starting point at $l=0$ here, in order to deal properly with the regularization,
as we will see shortly.
}.
The low multipoles are suppressed by powers of $\Delta$ and we can use the generic multipole fluctuations 
from~\eqref{lnRnuU} to obtain the dominant contribution to the sum.

Let us examine the structure of divergences in the UV limit when $\nu \to \infty$.
The degeneracy factor in terms of $\nu$ is given by
\begin{equation} \label{dnugene}
  d_\nu = \frac{2\nu \left( \nu + \frac{D}{2} -2 \right)!}{(D-2)! \left( \nu - \frac{D}{2} + 1 \right)!} \simeq 
  \frac{2}{(D-2)!}\nu^{D-2} \, ,
\end{equation}
where the last approximation is valid in general $D$ for large $\nu$, but is exact for any $\nu$ in $D=2,3,4$.
Then we expand~\eqref{lnRnuU} up to including ${\cal O}(\nu^{-3})$
\begin{equation} \label{divsumpart}
  \sum_{\nu \gg 1} d_\nu  \ln R_{\nu \gg 1} \sim -\frac{3 r_0 (2 - r_0)}{(D-2)! \Delta} \,
  \sum_{\nu \gg 1} \nu^{D-2} \left(\frac{1}{\nu} - \frac{1}{\nu^3} \left(\frac{r_0}{2 \Delta} \right)^2 \right) \, .
\end{equation}
This sum clearly diverges and the number of divergent terms depends on the dimension $D$.
In $D = 2, 3$, only the first term diverges: it produces a log divergence in $D = 2$ and a linear one in $D = 3$.
In $D = 4$, the first term gives a quadratic divergence and the second a logarithmic one, and so on to
higher dimensions. 

This is hardly surprising - the functional determinant is a one-loop quantity, which may diverge and
must be regularized. 
To obtain a consistent finite result, we have to use the same regularization scheme that was adopted to compute 
the renormalized bounce action in Section~\ref{subsecBounceRGE}, i.e. the $\overline{\rm MS}$ scheme.
That can be done either by computing the one-loop effective action using Feynman diagrams~\cite{Baacke:2003uw},
or with the $\zeta$-function regularization scheme developed in \cite{Dunne:2006ct}, which 
was shown to be equivalent to $\overline{\rm MS}$\footnote{
As a further check that the two schemes are indeed equivalent we provide an explicit example in 
Appendix~\ref{appQuartic}, where we use the bounce solution for a purely quartic potential \cite{Andreassen:2017rzq}, 
which has a simpler form compared to the TW bounce.}.
In the rest of this section we compute the renormalized determinant explicitly in three and four dimensions.
We begin with some specifications useful in generic dimensions.

{\bf Finite sum.}
To perform the sum over the multipoles in~\eqref{logdetRnumain} we use the Euler-Maclaurin (EuMac) approximation.
For later convenience it is useful to define $\Sigma_D$ as the log of the determinant for generic $D$, 
with individual terms in the sum $\sigma_D$ given by
\begin{align} \label{eqDefSigmaD}
  \Sigma_D &= \sum_{\nu = \nu_0}^\infty\, \sigma_D = \sum_{\nu = \nu_0}^\infty d_\nu
  \left( \ln R_\nu - \ln R_\nu^a \right) \, .
\end{align}  
The lower boundary is given by $\nu_0 = D/2 -1$ 
and the $\ln R_\nu^a$ term stands for the generic asymptotic subtraction, needed to regulate 
the UV part ($\nu \to \infty$) and make the sum finite.
The EuMac formula approximates the sum as an integral, $\Sigma_D^{\int}$, plus corrections from the boundaries,
$\Sigma_D^\text{bnd}$, which in our case consist of the summand $\sigma_D$ 
and its derivatives $\sigma_D^{(j)}$ evaluated at $\nu_0$, 
\begin{align} \label{eqDefEM}
  \Sigma_D &\simeq \Sigma_D^{\int} + \Sigma_D^{\text{bnd}} + R_p \, ,
  \\
  \Sigma_D^{\int} &= \int_{\nu_0}^\infty \text{d} \nu \, \sigma_D \, ,
  &
  \Sigma_D^{\text{bnd}} = \frac{1}{2} \sigma_D(\nu_0)   - \sum_{j = 1}^{\left \lfloor {\frac{p}{2}}\right \rfloor} \frac{B_{2 j}}{(2 j)!}
  \sigma_D^{(2 j - 1)}(\nu_0) \, .
  \label{EuMacorr}
\end{align}
Here, $B_{2 j}$ are the Bernoulli numbers and $R_p$ is the remainder, dependent on $p$, which is the
desired order of approximation.
In general $\Sigma_D^\text{bnd}$ also contains $\sigma_D$ evaluated at the upper boundary. Here, such 
terms $\sigma_D^{(p \geq 0)}(\infty)$, vanish because the subtraction of the asymptotic part guarantees 
that the sum is finite.

{\bf Three dimensions.}
The renormalized functional determinant~\cite{Dunne:2005rt, Dunne:2006ct} requires the subtraction of 
a single term
\begin{align} \label{eqLnDetRenD3}
  \ln \left( \frac{\det \mathcal O}{\det \mathcal O_{\text{FV}}} \right) &= 
  \sum_\nu d_\nu \left(\ln R_\nu - \frac{1}{2 \nu} I_1 \right) \, .
\end{align}  
Here $d_\nu = 2 \nu$. 
The first term in the parenthesis is the ratio of determinants that we computed in~\eqref{lnRnuU}.
The second term is the asymptotic subtraction $\ln R_\nu^a$, which makes the sum finite.
Note we only have to remove terms proportional to $1/\nu$.
The $I_1$ coefficient is given by~\cite{Dunne:2005rt, Dunne:2006ct}
\begin{align} \label{eqI1}
  I_1 &= \int_0^\infty \text{d} \rho \, \rho \left( V^{(2)} - V^{(2)}_\text{FV} \right) 
  \simeq - 3 \left( 2 - r_0 \right) \left(\frac{r_0}{\Delta}\right) \, .
\end{align}
It is convenient to introduce $y = \Delta \nu/r_0$, which counts as $\mathcal O(\Delta^0)$
and thus helps to clarify the overall $\Delta$ dependence. The variable $k_\nu$, which enters
in $\ln R_\nu$, see~\eqref{lnRnuU}, is related to $y$ as $k_\nu = \sqrt{1+y^2}$.
Using~\eqref{eqI1}, $r_0 = 2/3$, and comparing~\eqref{eqLnDetRenD3} to~\eqref{eqDefSigmaD}, 
we get the asymptotic form $\ln R_\nu^a = -2/y$.
The integral piece of the EuMac approximation, after performing the change of variable $\nu \to y$, is given by  
\begin{align} \label{eqFuncDetD3}
  \Sigma_3^{\int} &\simeq 2 \left(\frac{r_0}{\Delta} \right)^2 \int_{y_0}^\infty 
  \text{d} y \, y \left( \ln R_{\nu} + \frac{2}{y} \right)
  = \frac{1}{\Delta^2} \frac{20 + 9 \ln 3}{27} \, ,
\end{align}
and does not depend on the precise value of the lower boundary of
$y$, which can be extended to zero. 
Here it becomes clearer why the low-multipole result~\eqref{eqRlinfLowl} is not important when we sum
(integrate) over multipoles. Were we to use~\eqref{eqRlinfLowl}, instead of $R_\nu$ 
from~\eqref{eqRnuUFact}, in~\eqref{eqFuncDetD3}
to integrate over low multipoles, we would get a contribution suppressed by $\Delta^2$ 
compared to~\eqref{eqFuncDetD3}. That is subdominant, and we are only interested in the leading order terms
in this section. We learn that the integral in~\eqref{eqFuncDetD3} is manifestly dominated by the high multipoles.

For the remaining terms in the EuMac, we need to evaluate $\sigma_3(\nu_0)$ by treating $\nu = \mathcal O(1)$ and 
expanding in small $\Delta$.
The leading term comes from the asymptotic part and goes as $\sigma_3 \simeq 8/(3 \Delta) + \mathcal O(\Delta^0)$,
and is therefore suppressed compared to the leading $\Delta^{-2}$ dependence in~\eqref{eqFuncDetD3}.
When taking higher derivatives over $\nu$, we get corrections that
are further suppressed by powers of $\Delta$;
the Bernoulli terms are thus irrelevant. 

The final result for the determinant
is $ \ln \left( \frac{\det \mathcal O}{\det \mathcal O_{\text{FV}}} \right) =   \Sigma_3^{\int}$,
with $\Sigma_3^{\int}$ given in~\eqref{eqFuncDetD3}. This enters directly in the final expression
for the decay rate~\eqref{eqGamVSum}.
Recall that we are working in the $\overline{\rm MS}$ scheme, so there are no counter-terms
and no running in $D=3$.  

{\em Thermal field theory.}
At high temperatures, when the periodic boundary conditions go below the size of the instanton, the time
coordinate in 4D quantum field theory gets compactified and we end up with a 3D theory.
In such case~\cite{Linde:1980tt, Affleck:1980ac} (see also~\cite{Laine:2016hma}), the rate is given by
\begin{align} \label{eqGamVT}
  \frac{\Gamma}{\mathcal V} &\simeq \left( \frac{\lambda_-}{2 \pi} \right) 
  \left( \frac{S_3}{2 \pi T} \right)^{\frac{3}{2}}
  \sqrt{\frac{\det \mathcal O_{\text{FV}}}{\det' \mathcal O}} \, e^{-S_3/T} \, ,
\end{align}
where $(-\lambda_-^2)$ is the single negative eigenvalue of the 3D fluctuation operator.
The Euclidean action is exactly the one given as $S_3$ in~\eqref{eqS0}, but with the usual 4D counting
of dimensions.
Likewise, the determinant in~\eqref{eqGamVT} has to be calculated in $D = 3$, precisely what we
got in~\eqref{eqFuncDetD3} above, and is given by
\begin{align}
  \ln \sqrt{\frac{\det \mathcal O_{\text{FV}}}{\det' \mathcal O}} \simeq -\frac{1}{2} \Sigma_3 \, .
\end{align}

{\bf Four dimensions.}
In this case the renormalized determinant needs the following subtractions~\cite{Dunne:2005rt, Dunne:2006ct},
\begin{align} \label{eqLnDetRenD4}
\begin{split}
  \ln \left( \frac{\det \mathcal O}{\det \mathcal O_{\text{FV}}} \right) &= 
  \sum_\nu d_\nu \left(\ln R_\nu - \frac{1}{2 \nu} I_1 + \frac{1}{8 \nu^3} I_2 \right) - \frac{1}{8} \tilde I_2 \, .
\end{split}
\end{align}
Here we have $d_\nu = \nu^2$.
The $I_1$ coefficient in~\eqref{eqI1} is also valid for $D=4$, while $I_2$ and $\tilde I_2$ are 
obtained by inserting two powers of $V^{(2)}$. In the TW limit one has
\begin{align} \label{eqI2}
  I_2 &= \int_0^\infty \text{d} \rho \, \rho^3 \left( V^{(2) 2} - V^{(2) 2}_\text{FV} \right)
  \simeq - 3 \left( 2 - r_0 \right) \left(\frac{r_0}{\Delta}\right)^3 \, ,
  \\ \label{eqI2tld}
  \begin{split}
  \tilde I_2 &= \int_0^\infty \text{d} \rho \, \rho^3 \left( V^{(2) 2} - V^{(2) 2}_\text{FV} \right)
  \left( \frac{1}{\varepsilon} + \gamma_E + 1 + \ln \left(\frac{\mu \rho}{2}\right) \right) 
  \\
  &\simeq I_2 \left( \frac{1}{\varepsilon} + \gamma_E + \frac{5}{4} + 
  \ln \left(\frac{\mu }{2 \sqrt \lambda v \Delta}\right) \right) \, . \qquad \qquad (D=4)
  \end{split}
\end{align}
We anticipate that the $I_2$ will be valid for the asymptotic subtractions in higher $D$.
Note that the $\tilde I_2$ part of the renormalized determinant is outside the sum, and in the last line we have
written the result valid for $D=4$. 
It contains the $1/\varepsilon$ pole and depends on the renormalization scale $\mu$. 
The role of $\ln R_\nu^a = I_1/(2 \nu) -  I_2/(8 \nu^3)$, inside the sum, is to remove the divergent 
parts of $\ln R_\nu$:
the $I_1$ term regulates the quadratic divergence,
while $I_2$ removes the logarithmic one.

Using $r_0 = 1$ and the variable $y = \Delta \nu / r_0$, we have $\ln R_\nu^a = - 3/(2 y) + 3/(8 y^3)$
that enters in~\eqref{eqDefSigmaD}. The integral part of the EuMac approximation is then
\begin{align} \label{eqSigma4int}
  \Sigma_4^{\int} &\simeq \frac{1}{\Delta^3} \int_{y_0}^\infty \text{d} y \, y^2 \left( 
  \ln R_\nu + \frac{3}{2 y} - \frac{3}{8 y^3} \right)
  = \frac{3}{8 \Delta^3} \left( \frac{9 - 4 \sqrt 3 \pi}{36} + \ln 2 y_0 \right)\, .
\end{align}
Here, the finite piece $ \frac{3}{8 \Delta^3} \left( \frac{9 - 4 \sqrt 3 \pi}{36} \right)$ comes from intermediate multipoles 
with $\Delta \nu \sim 1$ and behaves as $1/\Delta^3$.
As in the $D=3$ case discussed above, the low-multipole contribution from~\eqref{eqRlinfLowl}
would be suppressed, so we do not need to account for it in the integral. 
%
%
On the other hand, we get a $\ln y_0$ term that seems to depend on the low multipoles. 
It comes solely from the term in the asymptotic subtraction $\ln R_\nu^a$
 which goes as $1/y^3$ and integrates into $\ln y_0$, 
due to lower boundary $y_0 = \Delta \nu_0/r_0$ in~\eqref{eqSigma4int}.
However, this is an artifact of the $\zeta$-regularization scheme.
As we shall see, $y_0$ is just a regulator that will disappear from the final result.

Let us move on to the EuMac corrections given by $\Sigma_D^\text{bnd}$ in~\eqref{eqDefEM}.
Here the lower boundary is at $\nu_0 \sim 1$, so we can expand the summands $\sigma_D$ from~\eqref{eqDefSigmaD}
in small $\Delta$ and keep only the leading part, which is of order $\Delta^{-3}$.
The only terms that contribute come from the $3/8y^3$ piece of $\ln R_\nu^a$, and are of the form 
$\sigma_4^{(j)}(\nu_0) = 3 (-)^{j+1} j!/(8 \Delta^3 \nu_0^{j+1})$, with $j \geq 0$. Recall that
the superscript $(j)$ here denotes the $j$-th derivative with respect to $\nu$.
Since $\nu_0\sim 1$, when we plug $\sigma_4^{(j)}(\nu_0)$ into the EuMac approximation in~\eqref{eqDefEM}, 
the Bernoulli terms eventually grow large and the sum starts to diverge.

We can delay the onset of this divergence by raising the $\nu_0$ in the Bernoulli terms. 
To accomplish this, we split the original sum in~\eqref{eqDefSigmaD} as follows
\begin{equation} \label{SumSplit}
\Sigma_D = \Sigma_D^{\rm low} + \Sigma_D^{\rm high} = \sum_{\nu = \nu_0}^{\nu_1} \sigma_D + \sum_{\nu = \nu_1+1}^\infty \sigma_D \, .
\end{equation}
Here the splitting point is such that ${\cal O}(1) = \nu_0 \ll \nu_1 < 1/\Delta$. 
To evaluate $\Sigma_D^{\rm high}$ we use the EuMac formula and replace $\nu_0$ in~\eqref{EuMacorr} with 
$\nu_1 + 1 \simeq \nu_1$. Following the reasoning above, we now find that the EuMac boundary 
corrections $\Sigma^{\rm bnd}$ containing $\sigma_4^{(j)}(\nu_1)$ are formally suppressed, because $\nu_1 \gg 1$, 
so we can neglect them.
Then the high sum is well approximated by the integral~\eqref{eqSigma4int}, with $y_0$ replaced by 
$y_1 = \Delta \nu_1 / r_0$, such that
\begin{align} \label{Sig4high}
  \Sigma_4^{\rm high} &
  = \frac{3}{8 \Delta^3} \left( \frac{9 - 4 \sqrt 3 \pi}{36} + \ln 2 y_1 \right)\, .
\end{align}

The low sum $\Sigma_4^{\rm low}$ can be performed explicitly. In the range up to $\nu_1 < 1/\Delta$
the leading piece in the summand $\sigma_4$ comes from the $3/(8y^3)$ term in $\ln R_\nu^a$, that
is $3/(8 \Delta^3 \nu)$ once multiplied by $d_\nu = \nu^2$.
Compared to it, all the other terms in $\sigma_4$  
are suppressed by powers of $\Delta$, so we neglect them. Then $\Sigma_4^{\rm low}$
 is given by the Harmonic number 
$H_{\nu_1}$, which can be expanded for $\nu_1 \gg 1$
\begin{align} \label{Sig4low}
  \Sigma_4^\text{low} = - \frac{3}{8 \Delta^3} \sum_{\nu = 1}^{\nu_1} \frac{1}{\nu} 
  = - \frac{3}{8 \Delta^3} H_{\nu_1} \simeq - \frac{3}{8 \Delta^3} \left( \ln \nu_1 + \gamma_E \right) \, .
\end{align}
The low-multipole sum of the fluctuations from~\eqref{eqRlinfLowl} still do not play any role. 
They would give terms suppressed by $\Delta^2$ compared to~\eqref{Sig4low}, which we neglect.
Adding up \eqref{Sig4low} and \eqref{Sig4high}, we get
\begin{align} \label{eqFuncDetD4}
  \Sigma_4 &  = \frac{3}{8 \Delta^3} \left( \frac{9 - 4 \sqrt 3 \pi}{36} - \gamma_E  + \ln 2 \Delta \right) \, .
\end{align}
It is reassuring that the dependence on the arbitrary point of separation $\nu_1$ is gone.

{\em Renormalized rate.}
To complete the calculation of the decay rate, we need to sum up the renormalized functional 
determinant, consisting of the finite sum $\Sigma_2$ above and the $\tilde I_2$ term in~\eqref{eqI2tld},
\begin{align}  \label{eqFuncDetD4full}
  \ln \left( \frac{\det \mathcal O}{\det \mathcal O_{\text{FV}}} \right) = \Sigma_4 - \frac{\tilde I_2}{8} \, ,
\end{align}  
with the renormalized Euclidean action $S_R + S_{\rm ct}$ in~\eqref{eqSRun}, that gives
\begin{align} \label{lndetS4}
\begin{split}
  \ln \frac{\Gamma}{\mathcal V} &\ni - S_R - S_{\rm ct}
  - \frac{1}{2} \left( \Sigma_4 - \frac{\tilde I_2}{8} \right)
  = - S - \frac{1}{\Delta^3} \frac{27 - 2 \pi \sqrt 3}{96} \, .
\end{split}
\end{align}
The action $S$ is given in~\eqref{Sbounce2} in terms of renormalized couplings.
The finite expression~\eqref{lndetS4} enters in the final result of~\eqref{eqGamVSum}.
Both, the $1/\varepsilon$ pole and the $\ln \mu$ from the running of $\lambda$ have canceled out.
The remaining $\mu_0 \simeq \sqrt \lambda v$ in~\eqref{eqSRun} cancels with $\sqrt \lambda v$ in the $\ln$ 
part of~\eqref{eqI2tld}, and the combination $-\gamma_E + \ln 2\Delta$ in~\eqref{eqFuncDetD4} also
cancels against the analogous terms in~\eqref{eqI2tld}.

The procedure of obtaining a finite functional determinant can be generalized to any even or odd dimension
in general.
This is done in Appendix~\ref{appFinSumGenD}.

%
%
\section{Conclusions and outlook} \label{secConclusion}

In this work we presented a dimensionally unified treatment of the false vacuum decay for a 
single real scalar in the benchmark TW limit.
We started by reviewing the bounce solution, which is valid for any $D$, including the second order 
correction in the expansion parameter $\Delta$, as well as the counter-terms and RGE running.
Employing the Gel'fand-Yaglom theorem, we found the orbital multipoles of the functional 
determinant around the bounce in two regions.
The first region was the low multipole one, where the negative eigenvalues and the translational 
zeroes appear.
The zeroes are removed perturbatively and that is how we get the correct physical dimension
of the rate.
The second region is computed by modifying the $\Delta$ counting such that $\Delta l = \mathcal O(1)$ 
and is therefore valid for higher multipoles, going up to infinity.

The renormalization section shows how the infinities in the orbital summation get regulated
in the TW limit.
In particular, we review how the subtractions in inverse powers of multipoles, 
$1/\varepsilon$ poles and the $\mu$
dependence are derived and evaluated in the TW limit, which gives us the asymptotic
part of the determinant that subtracts the leading divergencies.
Finally, we show how the multipoles are added up using the Euler-Maclaurin approximation.

It turns out that the summations are somewhat different for even and odd dimensions.
In odd dimensions, the dominant contribution to the determinant comes from higher multipoles
and no renormalization scale dependence appears.
In even dimensions, the single term from the asymptotic subtraction brings in a dependence
on low multipoles, which has to be treated carefully.
We resolve this issue by splitting the sum into two parts and deriving a consistent result for lower
and larger multipoles, while showing that the exact point of separation does not enter in the final
result.

A natural extension of this work would be to include additional degrees of freedom,
such as a globally symmetric potential with additional scalars, would-be-Goldstones, gauge 
bosons and fermions~\cite{wip:2022}, analytically in the TW limit.
This would require a somewhat general approach to renormalization, similar to the simplification 
that was done for the running of couplings~\cite{Jones:1981we}.
It may also be desirable to develop a framework for evaluating the total rate at one loop for 
general potentials.
Such a formalism may be developed from a semi-analytical polygonal bounce~\cite{Guada:2018jek} 
setup, where the shape of the bounce is given explicitly on each segment.
The upshot of this construction is that it applies to a broad class of models and matches
exactly to the known Euclidean action in the TW limit.
Most significantly, the bounce part was already suited for multi-fields, which would make it
a good starting point for developing the theory of one-loop fluctuations for generic BSM setups.

\smallskip
{\bf Note added:} After the publication of this work, a question regarding Coleman's
original thin wall action was brought to our attention thanks to Alonso Rodrigo and Adam Pluciennik.
We found a couple of typos in~\cite{Coleman:1977py}: equation (4.12) is missing a factor
of 2 and should read $S_1 = 2 \mu^3/(3 \lambda)$.
Equation (4.15) is missing a factor of 2 in the second term and should read 
$S_E = -1/2 \ \pi^2 R^4 \epsilon + 2 \pi^2 R^3 S_1$.
Equation (4.19) is correct and using the correct $S_1 = 2 \mu^3/(3 \lambda)$, one arrives at 
$B = 2^4 \, (\pi^2 \mu^{12})/(6 \epsilon^3 \lambda^4)$, a bounce action 
in the thin-wall limit, which is larger by a factor of 16 compared to the one given in equation (4.21)
of~\cite{Coleman:1977py}.
This agrees with our leading order result for the action in $D=4$ when using the translation
$\Delta = \epsilon/(2 \lambda v^4)$ and $\mu^2 = \lambda v^2$.

\acknowledgments 
We are very grateful to Andreas Ekstedt, Oliver Gould, and Joonas Hirvonen, for pointing out a contribution
to the $\tilde I_2$ UV integral we had missed in the first version of the manuscript.
We would like to thank Borut Bajc for discussions on finite temperature field theory and Victor Guada
for discussions at the initial stage of the collaboration.
MN was supported by the Slovenian Research Agency under the research core funding No. P1-0035 and 
in part by the research grant J1-8137 and J1-3013.
MM was supported by the Slovenian Research Agency's young researcher program 
under grant No. PR-11241.
LU and MN acknowledge the support of the COST action CA16201 - ``Unraveling new physics at the LHC 
through the precision frontier'', in particular for sponsoring the two short-term scientific missions at the IJS.
LU was supported by the MIUR grant PRIN 2017FNJFMW.

%
%
\appendix

%
%
\section{Calculation of the bounce and the Euclidean action} \label{appAction}
Here we provide a detailed derivation of the bounce field configuration and the bounce action. 
The starting point is eq.~\eqref{eqBounce}, which in terms of the dimensionless variables reads
\begin{equation} \label{eqBouncedimless}
  \frac{\text{d}^2 \varphi}{\text{d} z^2} + \frac{D-1}{z+r}  \frac{\text{d} \varphi}{\text{d} z} - \frac{1}{2} \varphi(\varphi^2 - 1) - \Delta = 0 \, .
\end{equation}
This is subject to the boundary conditions
\begin{align}
  \frac{\text{d} \varphi}{\text{d} z} (z = \infty) & =  \frac{\text{d} \varphi}{\text{d} z} (z = -r \to -\infty) =  0 \, , \label{bcder} \\
  \varphi(z = \infty) & = \varphi_{\rm FV} = 1 - \Delta - \frac{3}{2} \Delta^2 - 4 \Delta^3 + {\cal O}(\Delta^4)  \label{bcphi} \, .
\end{align}
It is useful to recall that at the true vacuum
\begin{equation}
  \varphi_{\rm TV} = - 1 - \Delta + \frac{3}{2} \Delta^2 - 4 \Delta^3 + {\cal O}(\Delta^4) \, .
\end{equation}

Using the expansions
\begin{equation} \label{expandApp}
\varphi = \sum_{n\geq 0} \Delta^n \varphi_n \, , \qquad \qquad r = \frac{1}{\Delta} \sum_{n \geq 0} \Delta^n r_n \, ,
\end{equation}
in \eqref{eqBouncedimless} we get
\begin{align}
& \frac{\text{d}^2 \varphi_0}{\text{d} z^2} - \frac{1}{2} \varphi_0(\varphi_0^2 - 1) \label{eqBEom0} \\
 & + \Delta \left[   \frac{\text{d}^2 \varphi_1}{\text{d} z^2} + \frac{1}{2}  \varphi_1 \left(1 - 3 \varphi_0^2 \right) - 1 + (D-1)\frac{1}{r_0} \frac{\text{d} \varphi_0}{\text{d} z}  \right]  \label{eqBEom1} \\
 & + \Delta^2 \left[  \frac{\text{d}^2 \varphi_2}{\text{d} z^2}  + \frac{1}{2} \varphi_2 \left(1 - 3 \varphi_0^2 \right) -\frac{3}{2} \varphi_0 \varphi_1^2  + (D-1)\left( \frac{1}{r_0} \frac{\text{d} \varphi_1}{\text{d} z} - \frac{z+r_1}{r_0^2}  \frac{\text{d} \varphi_0}{\text{d} z} \right)\right]  \label{eqBEom2} \\
 \begin{split}
 & +\Delta^3 \left[  \frac{\text{d}^2 \varphi_3}{\text{d} z^2} +\frac{1}{2} \varphi_3  \left(1 - 3 \varphi_0^2 \right) - 3 \varphi_0 \varphi_1 \varphi_2 - \frac{1}{2} \varphi_1^3   \right. \\
& \left. \qquad \qquad  + (D-1)\left(  \frac{1}{r_0} \frac{\text{d} \varphi_2}{\text{d} z} - \frac{z+r_1}{r_0^2}  \frac{\text{d} \varphi_1}{\text{d} z} + \frac{z^2 + 2z r_1 + r_1^2 - r_0 r_2}{r_0^3} \frac{\text{d} \varphi_0}{\text{d} z} \right) \right] \label{eqBEom3}
 \end{split} \\
&  + {\cal O}(\Delta^4) = 0 \, .
\end{align}
The strategy is to first solve the $n=0$ part, \eqref{eqBEom0}, plug the $\varphi_0$ solution into the $n=1$ equation, \eqref{eqBEom1} solve for $\varphi_1$, and so on iterating to higher orders. We will clarify how solving
the equations for $n \geq 1$ fixes the $r_n$ coefficients.

{\bf The leading order.} 
By noting that 
\begin{equation}
  \frac{\text{d}^2 \varphi_0}{\text{d} z^2} = \frac{{\rm d}({\rm d}\varphi_0 / {\rm d}z)}{{\rm d}\varphi_0} \frac{\text{d} \varphi_0}{\text{d} z} \, ,
\end{equation}
we can write \eqref{eqBEom0} as
\begin{equation}
  {\rm d}({\rm d}\varphi_0 / {\rm d}z) \frac{\text{d} \varphi_0}{\text{d} z}  = \frac{1}{2} \varphi_0(\varphi_0^2-1) {\rm d}\varphi_0 \, .
\end{equation}
Integrating both sides, requiring the derivative to vanish at $z \to \infty$, gives
\begin{equation} \label{derphi0}
  \frac{\text{d} \varphi_0}{\text{d} z}  = - \frac{1}{2} (\varphi_0^2 - 1) \, ,
\end{equation}
where we chose the minus sign in order to satisfy the boundary conditions.

Integrating once more with the boundary condition \eqref{bcphi}, that is $\varphi_0(z = \infty) = 1$,
we get $\int \text{d} \varphi_0/(1 - \varphi_0^2) = \text{ath} \varphi_0 = z/2$.
The bounce configuration at the leading order is thus
\begin{align} \label{eqBounce0}
  \varphi_0 &= \text{th} \frac{z}{2} \, .
\end{align}
This is odd under $z \to -z$, and interpolates between the FV at $z = \infty$ and the TV at $z = - \infty$.

Plugging the leading order solution into the action~\eqref{eqSD0}, we have
\begin{align} \label{eqSdn0}
  S_0 = \frac{\Omega v^{4-D}}{\lambda^{D/2-1} \Delta^{D-1}}
  \int_{-r}^\infty  \text{d} z \,
   \left( r_0 + \Delta z \right)^{D-1} \left( \frac{1}{2} \left(\frac{\text{d} \varphi_0}{\text{d} z}\right)^2 + 
  \frac{1}{8} \left( \varphi_0^2 - 1 \right)^2 + \Delta \left( \varphi_0 - 1 \right) \right) \, .
\end{align}
Here the index $n$ in $S_n$ refers to the order of the expansion in 
$\Delta^n$.
The first two terms of the integral are even under parity and vanish exponentially as $z \to \pm \infty$,
because $(\text{d} \varphi_0/\text{d} z)^2 = 1/4 (\varphi_0^2 -1)^2 \propto \exp(\mp 2z)$.
For such terms we can extend the lower limit of integration from $-r \sim -r_0/\Delta \xrightarrow{\Delta \to 0} -\infty$,
which amounts to neglecting terms\footnote{
Terms with $e^{-1/\Delta}$ go to zero faster than any power of $\Delta$ when $\Delta \to 0$. That is why it is 
consistent in our approach, based on counting powers of $\Delta$, to always drop such exponentially suppressed terms.
} of the form $e^{-1/\Delta}$.
The integral then becomes, upon using \eqref{derphi0},
\begin{align} 
  &\int_{-r}^\infty \text{d} z \, r_0^{D-1} \left( \frac{1}{2} \left(\frac{\text{d} \varphi_0}{\text{d} z}\right)^2 + 
  \frac{1}{8} \left( \varphi_0^2 - 1 \right)^2 \right) 
  = r_0^{D-1} \int_{-\infty}^\infty \left(\frac{\text{d} \varphi_0}{\text{d} z}\right)^2 \text{d} z 
  \\
  &= r_0^{D-1} \int_{-1}^1 \text{d} \varphi_0 \, \frac{\text{d} \varphi_0}{\text{d} z}
  = \frac{2}{3} r_0^{D-1} \, ,
   \label{surfint}
\end{align}
where we omitted the sub-leading $\Delta z$ terms, which will be added once we get to the $\Delta^2$ 
corrections below. 
The last term in~\eqref{eqSdn0}
looks as if it should not belong to $n=0$, because it is proportional to $\Delta$. 
However, this is not the case. The combination $(\varphi_0 -1)$ goes to a constant, $-2$, as $z \to - r \to -\infty$,
implying that the lower limit of integration cannot be extended to $-\infty$.
We can either integrate the term directly, keeping $-r$ as the lower boundary, which brings in the $1/\Delta$ power to
bring down the term to $n=0$, or equivalently integrate by parts (using
$\Delta z$ as the integration variable):
\begin{align}
  &\int_{-r}^\infty  \text{d} \Delta z \, \left( r_0 + \Delta z \right)^{D-1} \left( \varphi_0 - 1 \right) 
  \\ \label{eqSEn1}
  &= \frac{1}{D} \left( r_0 + \Delta z \right)^D \left( \varphi_0 - 1 \right) \Bigr|_{-r_0}^\infty
  - \frac{1}{D} \int_{-\infty}^\infty \text{d} z\, \left( r_0 + \Delta z \right)^D \frac{\text{d} \varphi_0}{\text{d} z} 
  \\
  &= - \frac{1}{D} r_0^D \int_{-\infty}^\infty \text{d} z \, \frac{\text{d} \varphi_0}{\text{d} z} 
  = - \frac{1}{D} r_0^D \varphi_0 \bigr|_{-1}^1 
  = - \frac{2}{D} r_0^D \, .
\end{align}
In~\eqref{eqSEn1} the boundary term vanishes, while in the integral with ${\rm d}\varphi_0 / {\rm d} z$ we can extend
the lower limit to $-\infty$, committing only an exponentially small error. We keep only the $r_0^D$ term, since the 
rest is sub-leading in $\Delta$. We refer to this as a volume term, since it is proportional to $r_0^D$, while \eqref{surfint}
is a surface term, proportional to $r^{D-1}$.
Combining them we get
\begin{align} 
  S_0 &= \frac{\Omega v^{4-D}}{\lambda^{D/2-1} \Delta^{D-1}} \left( \frac{2}{3} r_0^{D-1} - \frac{2}{D} r_0^D \right) \, ,
\end{align}
that can be extremized over $r_0$
\begin{align} \label{eqS0xtr0}
  \frac{\text{d} S_0}{\text{d} r_0} = 0 \quad \Rightarrow \quad  r_0 = \frac{D-1}{3} \, .
\end{align}  
We end up with the Euclidean bounce action \eqref{eqS0} at the leading order.
A sanity check can be made by splitting the action into the kinetic $\mathcal T$ term, which is $1/2$ of the surface 
term in~\eqref{surfint}, and the potential piece $\mathcal V$, coming from the other half of the surface plus the volume term:
\begin{align} \label{eqS0TV}
  S_0 &= \frac{\Omega v^{4-D}}{\lambda^{D/2-1} \Delta^{D-1}} \frac{r_0^{D-1}}{3}  
  \left( 1 + \frac{D - 6 r_0}{D} \right) = \mathcal T + \mathcal V \, .
\end{align}
This verifies that the two terms of the action are indeed related by 
\begin{align} \label{eqDerrick}
  \left(D - 2 \right) \mathcal T = -D \mathcal V \, ,
\end{align}
as stated by Derrick's~\cite{Derrick:1964ww} theorem, once the action is properly extremized by 
$r_0$ in~\eqref{eqS0xtr0}.

{\bf Higher order corrections.}
The equations for $n \geq 1$, as we see in \eqref{eqBEom1}, \eqref{eqBEom2}, \eqref{eqBEom3}, all have the form
\begin{equation}
 \frac{\text{d}^2 \varphi_n}{\text{d} z^2} + \frac{1}{2}  \varphi_n \left(1 - 3 \varphi_0^2 \right) = f^{\rm nhom}(\varphi_{j<n}(z),r_{j<n}) \, ,
\end{equation}
with the homogeneous part on the left hand side, the non-homogeneous part on the right hand side. 
With $\varphi_0 =  \text{th}(z/2)$, the solution to the homogeneous part,
\begin{equation} \label{phinhom}
\varphi_n^{\rm hom} = \frac{c_{1n}}{4 {\rm ch}^2(z/2)} + \frac{c_{2n}}{4 {\rm ch}^2(z/2)} [6 z + 8 {\rm sh}(z) + {\rm sh}(2z)] \, ,
\end{equation}
is the sum of an even function in $z$, multiplied by $c_{1n}$, and an odd function, multiplied by $c_{2n}$.

At $n =1$ we have \eqref{eqBEom1}:
\begin{equation} \label{n1boun}
 \frac{\text{d}^2 \varphi_1}{\text{d} z^2} + \frac{1}{2}  \varphi_1 \left(1 - 3 \varphi_0^2 \right)  = 1 - (D-1)\frac{1}{r_0} \frac{\text{d} \varphi_0}{\text{d} z} \, .
\end{equation}
We can understand some features by inspection. As $\varphi_0$ is an odd function of $z$, $\varphi_0^2$ and 
${\rm d}\varphi_0 / {\rm d}z$ are even functions. Hence the equation admits a solution $\varphi_1$ which is even. Moreover,
as $\varphi_0^2(z = \pm \infty) = 1$ and the derivatives of the field must vanish at $z = \pm \infty$, we have 
$\varphi_1(z = \pm \infty) = -1$. Therefore we know immediately that $\varphi_1$ is forced to interpolate between
the FV and the TV at order $\Delta$.
The solution to \eqref{n1boun} is
\begin{align}
\varphi_1 & =  \frac{c_{11}}{4 {\rm ch}^2(z/2)} + \frac{c_{21}}{4 {\rm ch}^2(z/2)} [6 z + 8 {\rm sh}(z) + {\rm sh}(2z)] \\
\begin{split}
&+  \frac{1} {3 r_0 \left(e^z+1\right)^2}  \left[ 
{\rm ch} (z) \left((D-1) (3 z+1)-9 r_0 (z+1)\right) +(D-1) (3 z+2) {\rm sh} (z)  \right.  \\ 
& \left. \qquad \qquad \qquad -4 D -3 r_0 ({\rm ch} (2
   z)+{\rm sh} (z) (3 z+2 {\rm ch} (z)+4))+9 r_0+4 \right] \, .
\end{split}
\end{align}
The boundary condition $\varphi_1(z = \infty) = -1$ requires $c_{21} = 0$. This sets ${\rm d}\varphi_1 / {\rm d}z = 0$ at 
$z = \infty$. Requiring ${\rm d}\varphi_1 / {\rm d}z = 0$ also at $z = -\infty$ fixes $r_0 = (D-1)/3$. With these, 
the solution simplifies to
\begin{equation}
\varphi_1 = -1 + \frac{c_{11}}{4 {\rm ch}^2(z/2)} \, .
\end{equation}
This is an even function, as anticipated, and we see that the boundary conditions do not fix the constant $c_{11}$. 
We are free to set it to zero, so that
\begin{equation}
\varphi_1 = -1 \, .
\end{equation}
Note that solving the bounce equation at this order gives us the same $r_0$ as we found in \eqref{eqS0xtr0} by
extremizing the action, and provides a further consistency check.

At $n=2$, using $\varphi_1$ and $r_0$ from above into \eqref{eqBEom2}, we get
\begin{align} \label{eqBEom2simp}
  \frac{\text{d}^2 \varphi_2}{\text{d} z^2} + \frac{1}{2}\varphi_2 \left( 1 - 3 \varphi_0^2 \right) 
  &= \frac{3}{2} \varphi_0 + \frac{9}{D-1} \frac{\text{d} \varphi_0}{\text{d} z} \left( z + r_1 \right) \, .
\end{align}
With the derivatives vanishing at $z = \pm \infty$, this equation implies $\varphi_2(z = \pm \infty) = \mp \frac{3}{2}$. 
Hence, $\varphi_2$ is also forced to interpolate between true and false vacua. Note that setting $r_1 = 0$ at this stage
would make it apparent that \eqref{eqBEom2simp} admits a solution $\varphi_2$ odd under $z$. Let us keep $r_1$
in the game for a moment. The solution then is
\begin{equation}
\varphi_2  =  \frac{c_{12}}{4 {\rm ch}^2(z/2)} + \frac{c_{22}}{4 {\rm ch}^2(z/2)} [6 z + 8 {\rm sh}(z) + {\rm sh}(2z)] + \varphi_2^{\rm nhom}(z, r_1) \, ,
\end{equation}
with $\varphi_2^{\rm nhom}(z, r_1)$ a function of $z$ and $r_1$ computed from the non-homogeneous part 
on the right hand side of \eqref{eqBEom2simp}. 
The boundary condition $\varphi_2(z = \infty) = - \frac{3}{2}$ requires $c_{22} = 0$, 
which also makes ${\rm d}\varphi_2 / {\rm d}z = 0$ at $z = \infty$. Asking that the derivative vanish also at $z = -\infty$
requires $r_1 = 0$. Thus, the boundary condition forces this parameter to vanish.
We can fix the remaining parameter by demanding that the solution be an odd function of $z$,
\begin{equation}
c_{12} =  \frac{3(\pi^2 + 3D -9)}{2(D-1)} \, ,
\end{equation}
so that
\begin{align} \label{eqPhi2}
\begin{split}
  \varphi_2 = \frac{1}{4 r_0 \text{ch}^2(z/2)} \bigl( 
    & \left( 2 - D - 2 \left(4 + \text{ch} z \right) \ln(1 + e^z) \right) \text{sh} z
    \\
    & - z \left(D - e^z \left(4 + \text{sh} z \right) \right) + 3 (\text{Li}_2(-e^z) - \text{Li}_2(-e^{-z}))
  \bigr)  \,,
\end{split}  
\end{align}
with the asymptotic behavior $\varphi_2(\pm\infty) = \mp 3/2$.

At $n=3$, using $\varphi_1 = -1$, $r_1 = 0$ into \eqref{eqBEom3}, we have
\begin{align} \label{eqBEom3simp}
  \frac{\text{d}^2 \varphi_3}{\text{d} z^2} + \frac{1}{2}\varphi_3 \left( 1 - 3 \varphi_0^2 \right) 
  &= - 3 \varphi_0 \varphi_2 - \frac{1}{2} - (D-1) \left( \frac{1}{r_0} \frac{\text{d} \varphi_2}{\text{d} z}  + \frac{z^2 - r_0 r_2}{r_0^3} \frac{\text{d} \varphi_0}{\text{d} z} \right) \, .
\end{align}
As both $\varphi_0$ and $\varphi_2$ are odd functions of $z$, this equation admits an even solution $\varphi_3$. Again,
by inspection we see that the asymptotics are such that $\varphi_3$ interpolates between true and false vacua.
The solution, after plugging in $\varphi_0, \varphi_2, r_0$, is
\begin{equation}
\varphi_3  =  \frac{c_{13}}{4 {\rm ch}^2(z/2)} + \frac{c_{23}}{4 {\rm ch}^2(z/2)} [6 z + 8 {\rm sh}(z) + {\rm sh}(2z)] + \varphi_3^{\rm nhom}(z, r_2) \, .
\end{equation}
The boundary condition $\varphi_3(z = \infty) = -4$ fixes 
\begin{equation}
c_{23} = - \frac{3\pi^2(D-2)}{(D-1)^2} \, ,
\end{equation}
which results in ${\rm d}\varphi_3 / {\rm d}z = 0$ at $z = \infty$. Asking that the derivative vanish at $z = -\infty$ fixes
\begin{align} \label{r2App}
  r_2 &= \frac{6 \pi^2 - 40 + D (26 - 4 D - 3 \pi^2)}{3(D-1)} \, .
\end{align}
The parameter $c_{13}$, analogously to $c_{11}$ at $n=1$, remains unfixed, and we can choose to set it to zero. 
Finally we have
\begin{align}
\varphi_3 & = - \frac{1}{16 {\rm ch}^2(z/2) (D-1)^2} \Big[ 2 {\rm ch} (z) \left(16 D^2+D \left(-72 z+24 \pi ^2-77\right)  \right.  \nonumber \\ 
& \qquad  \left.  +144 (D-1) \log \left(e^z+1\right) +72 z  (z+1)-48 \pi ^2+97\right) \nonumber \\
& \qquad +216 (3 D-7) \text{Li}_3\left(-e^z\right)+72 \text{Li}_2\left(-e^z\right)
   (-3 (D-3) z+8 (D-2) {\rm sh} (z)  \nonumber \\ 
   &\qquad +(D-2) {\rm sh} (2 z))+27 (D-3) z^2+9 \left(4 \pi ^2 (D-2)-15 (D-1)\right) z  \nonumber \\ 
   &\qquad +270 D \log \left(e^z+1\right)+12 {\rm sh} (z) \left(3 z (-5 D+4 z+7) \right. \nonumber \\ 
   &\qquad \left. +24 (D-3) z \log
   \left(e^z+1\right)+4 \pi ^2 (D-2)\right) \nonumber \\ 
   &\qquad +3 {\rm sh} (2 z) \left(3 z (-D+2 z+1)+12 (D-3) z \log
   \left(e^z+1\right)+2 \pi ^2 (D-2)\right)\nonumber  \\ 
   &\qquad +3 {\rm ch} (2 z) \left(3 z (-D+2 z+1)+6 (D-1) \log
   \left(e^z+1\right)+2 \pi ^2 (D-2)\right) \nonumber \\ 
   &\qquad +16 (D-4) (2 D-5)-18 \pi ^2 (D-2)+36 z^3-270 \log
   \left(e^z+1\right) \Big] \, .
\end{align}
We did not manage to compute analytically the bounce solution beyond $n=3$.

{\bf Comments on the bounce solution.}
We have seen that with this iterative setup, the bounce interpolates between the FV and the TV
 at any order $n$. 
This means that, with the $\Delta$ expansion initiated here, the field will never go beyond the TW in the 
sense that $\phi_{\text{in}} = \phi_{\text{TV}}$ for any power\footnote{In order to have 
$\phi_{\text{in}} \neq \phi_{\text{TV}}$, one should reconsider the bounce equation and include the linear 
$\Delta$ term at the leading order and expand the friction term only.} of $\Delta$.
By going to higher orders we are simply describing the TW instanton between the TV 
and the FV more accurately.

Note that a shift of the radius in the leading order bounce solution produces
\begin{equation}
\varphi_0(z + c^{\rm shift}_n \Delta^n) = \varphi_0(z) + \frac{c^{\rm shift}_n}{2 {\rm ch}^2(z/2)} \Delta^n + {\cal O}(\Delta^{2n}) \, .
\end{equation}
Comparing to \eqref{phinhom} and to the expansion of $r$ \eqref{expandApp}, we see that a nonzero coefficient $c_{1n}$ 
in \eqref{phinhom} results in a shift 
$r_{n+1} \to r_{n+1} + c_{1n}/2$. At $n=1$ we chose to set $c_{11} = 0$. 
An alternative choice would be to keep the $c_{11}$ term in $\varphi_1$ and set $r_2 = 0$; in this case the derivative
boundary condition for the $n=3$ equation would fix $c_{11}$ to twice the value of \eqref{r2App}. 
There is no consistent solution with both $c_{11} = 0$ and $r_2 = 0$, but this becomes clear only when one 
gets to the $n=3$ order. 
In conclusion, we learn that the $\Delta$ expansion of $r$ is useful because by introducing more redundant parameters
it allows for the bounce solution $\varphi$ to take a simpler form.

{\bf The bounce action at $n=2$.}
As we shall see, both $r_2$ and $\varphi_2$ enter in the kinetic and potential part of the
action at $\Delta^2$, but eventually cancel out.
So let us examine the several second-order sources of the Euclidean action.
We found that the leading contribution to the action in~\eqref{eqS0TV} consists of the kinetic $\mathcal T$
and potential $\mathcal V$ terms at zero order
\begin{align}
  \mathcal T_0 &= \frac{\Omega v^{4-D}}{\lambda^{D/2-1} \Delta^{D-1}} \frac{r_0^{D-1}}{3} \, ,
  &
  \mathcal V_0 &= \mathcal T_0 \left( 1 - \frac{6 r_0}{D} \right) \, .
\end{align}
These were obtained by extremizing the action over $r$ and plugging in the leading order 
$r \sim r_0/\Delta$.
Further expanding the bounce radius $r \sim (r_0 + \Delta^2 r_2)/\Delta$ to include the second order 
correction, we get
\begin{align}
  \mathcal T_2 &\ni \frac{\Omega v^{4-D}}{\lambda^{D/2-1} \Delta^{D-1}} \frac{r_0^{D-1}}{3}
  \left(1 + \Delta^2 \frac{r_2}{r_0} \right)^{D-1}
  \simeq \mathcal T_0 \left(1 + 3 \Delta^2 r_2 \right) \, ,
  \\
  \mathcal V_2 &\ni \mathcal V_0  - \mathcal T_0 3 \Delta^2 r_2 \, ,
\end{align}
where $\mathcal T_2$ and $\mathcal V_2$ stand for the kinetic and potential parts of the action
up to second order.
As expected, the action does not get corrected by $r_2$, however individual kinetic and potential
contributions are modified.
Keeping all the terms lead to a non-trivial validation of extremization via Derrick's theorem 
in~\eqref{eqDerrick}, which has to hold at all orders of $\Delta$ once all the terms in the action 
are obtained.

Further corrections come from the inclusion of friction when expanding the $\Delta z$ polynomial 
in~\eqref{eqSdn0}.
We will explicitly attribute the various contributions to the kinetic or potential
part, and check the validity of Derrick’s theorem at the end of the calculation. 
We start with the corrections that concern only the effects of $r_2$ and the damping of $\varphi_0$.
The $\Delta$ corrections vanish since the integrand is odd in $z$, while the $\Delta^2$ corrections
give us the following higher-order terms in the action (in the following few lines we omit the overall
constant $\Omega v^{4-D} \lambda^{1-D/2} \Delta^{1-D}$ in front of the action 
$S_2$, not to overburden with notation), proportional to
\begin{align}
\begin{split} \label{eqS2zero}
  S_2 &\ni 2 \, r_0^{D-3} \Delta^2 \frac{(D-1)(D-2)}{4} \int_{-\infty}^\infty \text{d} z \, z^2 \left(\frac{\text{d} \varphi_0}{\text{d} z}\right)^2 
  \\
  &- \frac{1}{D} r_0^{D-2} \Delta^2 \frac{D(D-1)}{2} \int_{-\infty}^\infty \text{d} z \, z^2 \frac{\text{d} \varphi_0}{\text{d} z}
\end{split}
  \\
  &= \frac{3}{2} \Delta^2  r_0^{D-1} \left( 3 \frac{D-2}{D-1} \int_{-\infty}^\infty \text{d} z \, z^2 \left(\frac{\text{d} \varphi_0}{\text{d} z}\right)^2 
  -  \int_{-\infty}^\infty \text{d} z \, z^2 \frac{\text{d} \varphi_0}{\text{d} z} \right)
  \\
  &
  = - \Delta^2 \left( \frac{\pi^2 + 6(D - 2)}{D-1} \right) r_0^{D-1} \, .
\end{align}
We emphasize a factor of 2 in the first line of~\eqref{eqS2zero} as we have two equal contributions; 
one belonging to the kintetic part and one to the potential part.
The second line belongs to the potential part only.
The addition of $\varphi_1 = -1$ brings in the following terms, proportional to
\begin{align} \label{eqS2one}
  S_2 \ni & \int_{-r}^\infty \text{d} z \,
  \left( r_0 + \Delta z \right)^{D-1} \left( \frac{3}{4} \Delta^2 \left(\varphi_0^2 - 1 \right)
  - \frac{1}{2} \Delta \varphi_0 \left(\varphi_0^2 - 1 \right)  
  - \frac{1}{2} \Delta^3 \left( \varphi_0 - 1 \right)  \right)  
  \\
  &= \int_{-r}^\infty \text{d} z \,
  \left( r_0 + \Delta z \right)^{D-1} \left( -\frac{3}{2} \Delta^2 \frac{\text{d} \varphi_0}{\text{d} z}
  + \Delta \varphi_0 \frac{\text{d} \varphi_0}{\text{d} z} 
  - \frac{1}{2} \Delta^3 \left( \varphi_0 - 1 \right)  \right) 
  \\ \label{eqS21nl}
  &= -\frac{3}{2} \Delta^2 r_0^{D-1} \int_{-1}^1 \text{d} \varphi_0
  + r_0^{D-2} (D-1) \Delta^2 \int_{-\infty}^\infty \text{d} z \, z \varphi_0 \frac{\text{d} \varphi_0}{\text{d} z}
  +\frac{D - 1}{3 D} \Delta^2 r_0^{D - 1}
  \\
  &= \Delta^2 r_0^{D-1} \left( 3 + \frac{r_0}{D} \right) \, .
\end{align}
where all the terms in~\eqref{eqS2one} belong to the potential part.
The final remaining pieces, depending on $\varphi_2$, are found to be
\begin{align} \label{eqStld2}
  S_2 \ni &\Delta^2 \int_{-r}^\infty \text{d} z \, \left( r_0 + \Delta z \right)^{D-1} 
  \left( 
    \frac{\text{d} \varphi_0}{\text{d} z} \frac{\text{d} \varphi_2}{\text{d} z} - \varphi_0 \frac{\text{d} \varphi_0}{\text{d} z} \varphi_2
    + 3 \Delta \frac{\text{d} \varphi_0}{\text{d} z} \varphi_2 \right) = 0 \, .
\end{align}
This can be understood from the following identity
\begin{align}
  &\int_{-\infty}^\infty \text{d} z \, \frac{\text{d}}{\text{d} z} \left( \frac{\text{d} \varphi_0}{\text{d} z} \varphi_2 \right)
  = \frac{\text{d} \varphi_0}{\text{d} z} \varphi_2 \biggr |_{-\infty}^\infty = 0
  = \int_{-\infty}^\infty \text{d} z \, \left( \frac{\text{d} \varphi_0}{\text{d} z} \frac{\text{d} \varphi_2}{\text{d} z} 
  + \frac{\text{d}^2 \varphi_0}{\text{d} z^2} \varphi_2 \right)
  \\
  &= \int_{-\infty}^\infty \text{d} z \, \left( \frac{\text{d} \varphi_0}{\text{d} z} \frac{\text{d} \varphi_2}{\text{d} z} 
  + \frac{1}{2} \varphi_0 \left( \varphi_0^2 - 1 \right) \varphi_2 \right)
  =  \int_{-\infty}^\infty \text{d} z \, \left( \frac{\text{d} \varphi_0}{\text{d} z} \frac{\text{d} \varphi_2}{\text{d} z} 
  - \varphi_0 \frac{\text{d} \varphi_0}{\text{d} z} \varphi_2 \right) \, ,
\end{align}
which shows that the first two terms in~\eqref{eqStld2} sum up to zero.
The last term in~\eqref{eqStld2} disappears, because $\varphi_2$ is odd and $\text{d} \varphi_0/\text{d} z$ is even and 
vanishes at the boundaries, allowing the limits of integration to be extened to $z \to \pm \infty$.

The final result is thus the Euclidean action, which can be written as
\begin{align} \label{eqS02}
  S_2 = S_0 \left(1 + \Delta^2 \left(\frac{1 + D \left(25 - 8 D - 3 \pi^2 \right)}{2 (D-1)} \right)\right) 
  + \mathcal O\left( \Delta^4 \right)\, .
\end{align}
This result can be used to get an improved precision on the bounce part of the action in~\eqref{eqS0}
for any $D$.
The $\Delta^3$ corrections vanish altogether, so this result is precise up to $\mathcal O(\Delta^4)$.
Moreover, it serves an estimate on the upper bound on $\Delta$, when the higher-order corrections
become relevant and thus the entire expansion becomes unreliable.
One should keep in mind that we have constructed such a TW setup, where higher orders of $\Delta$
simply describe the TV and FV more precisely and the bounce interpolates between those two.
To go beyond this ansatz, one would have to construct a bounce solution with 
$\varphi_0 \neq \varphi_{\text{TV}}$ already at leading order, which might be feasible but is certainly
beyond the scope of our paper.

It is useful to collect all of the corrections above and carefully attribute them to kinetic
and potential parts: we get the integrated kinetic and potential parts of the action
\begin{align}
  \mathcal T_2 &= \mathcal T_0 + \Delta^2 \mathcal T_0 \left( 
  \frac{1 + D(25 - 8 D - 3 \pi^2)}{2 (D - 1)} \right) \, ,
  \\
  \mathcal V_2 &= \mathcal V_0 - \Delta^2 \mathcal T_0 \left(\frac{D - 2}{D} \right) 
  \left( \frac{1 + D (25 - 8 D - 3 \pi^2)}{2 (D - 1)} \right) \, .
\end{align}
which demonstrates that Derrick's theorem in~\eqref{eqDerrick} holds at second-order in 
$\Delta$ and therefore all of the terms at this order have been properly included.

{\bf Counter-term integration.}
When calculating the counterterm action in~\eqref{eqSct}, we encountered the following two integrals
\begin{align} \label{eqPhi2Phi4}
  &\int_D \left( \phi^2 - \phi^2_\text{FV} \right) \, ,  &\text{and} & &
  &\int_D \left( \phi^4 - \phi^4_\text{FV} \right) \, .
\end{align}
These are evaluated in the TW limit using the same approach as above.
To get the leading order term proportional to $1/\Delta^{D-1}$ correctly, we must include the bounce up to $n=1$,
that is $\varphi = \varphi_0 + \Delta \varphi_1 = \varphi_0 - \Delta$.
The first integral in \eqref{eqPhi2Phi4} is
\begin{align}
  &\frac{\Omega v^{2-D}}{\lambda^{D/2} \Delta^{D-1}} \int_{-r}^\infty \text{d} z \,
  \left( r_0 + \Delta z \right)^{D-1} \left( (\varphi_0 - \Delta)^2 - (1 - \Delta)^2 \right)
  \\
  &= \frac{\Omega v^{2-D}}{\lambda^{D/2} \Delta^{D-1}} \left(   
  \int_{-\infty}^\infty \text{d} z \, r_0^{D-1} \left( \varphi_0^2 - 1 \right) - 2 
  \int_{-r}^\infty \text{d} z \, \left( r_0 + \Delta z \right)^{D-1} \Delta \left( \varphi_0 - 1 \right) \right)
  \\ \label{eqPhi2ct}
  &= \frac{\Omega v^{2-D}}{\lambda^{D/2} \Delta^{D-1}} \left( -4 r_0^{D-1} + \frac{4}{D} r_0^D \right)
  = -4 \frac{\Omega v^{2-D}}{\lambda^{D/2} \Delta^{D-1}} r_0^{D-1} \left( 1 - \frac{r_0}{D} \right) \, ,
\end{align}
where we used the same integration by parts as in~\eqref{eqSEn1}.
Proceeding along the same lines, the second term in~\eqref{eqPhi2Phi4} integrates into
\begin{align}
  &\frac{\Omega v^{4-D}}{\lambda^{D/2} \Delta^{D-1}} \int_{-r}^\infty \text{d} z \,
  \left( r_0 + \Delta z \right)^{D-1} \left( (\varphi_0 - \Delta)^4 - (1 - \Delta)^4 \right)
  \\
  &= \frac{\Omega v^{4-D}}{\lambda^{D/2} \Delta^{D-1}} \left(   
  \int_{-\infty}^\infty \text{d} z \, r_0^{D-1} \left( \varphi_0^4 - 1 \right) - 4 
  \int_{-r}^\infty \text{d} z \, \left( r_0 + \Delta z \right)^{D-1} \Delta \left( \varphi_0^3 - 1 \right) \right)
  \\ \label{eqPhi4ct}
  &= \frac{\Omega v^{4-D}}{\lambda^{D/2} \Delta^{D-1}} \left( -\frac{16}{3} r_0^{D-1} + \frac{8}{D} r_0^D \right)
  = -\frac{16}{3} \frac{\Omega v^{4-D}}{\lambda^{D/2} \Delta^{D-1}} r_0^{D-1} \left( 1 - \frac{3 r_0}{2 D} \right) \, .
\end{align}

%
%
\section{Calculation of the fluctuations} \label{appFluct}
In this appendix, we provide the details of the calculation of the fluctuations discussed in~\ref{secFluctuations}. 
Starting from \eqref{detratio}, We use the Gel'fand-Yaglom theorem, which states that
\begin{equation} \label{GelYag}
  \frac{\det {\cal O}_l}{\det {\cal O}_{l \rm{FV}}} = \lim_{\rho \to \infty} \left(  
  \frac{\psi_l(\rho)}{\psi_{l \rm{FV}}(\rho)} \right) ^{d_l} \, .
\end{equation}
Recall that the radial operator
\begin{equation} \label{radialop}
  {\mathcal O}_l = -\frac{\text{d}^2}{\text{d} \rho^2} - \frac{D-1}{\rho} \frac{\text{d}}{\text{d} \rho} + 
  \frac{l \left( l + D - 2 \right)}{\rho^2} + \ddV \, ,
\end{equation}
has degeneracy
\begin{align} \label{degl}
  d_l &= \frac{(2l + D-2)(l+D-3)!}{l! (D-2)!} \, ,
\end{align}
and the functions $\psi_l$ are solutions of
\begin{equation}  \label{eqspsi}
  {\cal O}_l \psi_l = 0 \, , \qquad \qquad  {\cal O}_{l\rm{FV}} \psi_{l\rm{FV}} = 0 \, ,
\end{equation}
with only one boundary condition fixed at $\rho \to 0$:
\begin{equation} \label{psibounds}
  \psi_l (\rho \to 0) \sim \rho^l \, \qquad \quad \psi_{l\rm{FV}} (\rho \to 0) \sim \rho^l \, .
\end{equation}

It is convenient to introduce
\begin{equation} \label{nudef}
 \nu = l + \frac{D}{2} - 1 \, ,
\end{equation}
and 
\begin{equation} \label{psinu}
  \psi_\nu(\rho) = \rho^{\frac{D-1}{2}} \psi_l (\rho) \, .
\end{equation}
Then from ${\cal O}_l \psi_l = 0$ we obtain ${\cal O}_\nu \psi_\nu = 0$, with
\begin{align} \label{Onu}
  \mathcal O_\nu = -\frac{\text{d}^2}{\text{d} \rho^2} + 
  \frac{\nu^2 - \frac{1}{4}}{\rho^2} + \ddV \,  .
 \end{align}
Thanks to the rescaling in \eqref{psinu}, ${\cal O}_\nu$ has no friction term. 
Defining
\begin{equation} \label{Rnudefin}
  R_\nu (\rho) \equiv \frac{\psi_\nu (\rho)}{\psi_{\nu\rm{FV}}(\rho)} = \frac{\psi_l (\rho)}{\psi_{l\rm{FV}}(\rho)}  \, ,
\end{equation}
we can rewrite the ratio of the determinants in equation \eqref{GelYag} as
\begin{equation} \label{funcdetRrho}
  \frac{\det {\cal O}_l}{\det {\cal O}_{l \rm{FV}}} = \frac{\det {\cal O}_\nu}{\det {\cal O}_{\nu \rm{FV}}} 
  = \left( \lim_{\rho \to \infty}  R_\nu(\rho) \right)^{d_\nu} \, .
\end{equation}
Here the degeneracy factor
\begin{equation} \label{degnu}
  d_\nu = \frac{2\nu \left( \nu + \frac{D}{2} -2 \right)!}{(D-2)! \left( \nu - \frac{D}{2} + 1 \right)!} \, ,
\end{equation}
is obtained from \eqref{degl} using \eqref{nudef}.
Taking the logarithm of equation \eqref{detratio} and using the Gel'fand-Yaglom theorem \eqref{funcdetRrho},
we can finally write\footnote{
We drop the prime from the determinant at the numerator here, we will comment on the removal of the zero modes
later on in the appendix.
}
\begin{equation} \label{logdetRnu} 
  \ln \left\vert  \frac{\det{\cal O}}{\det{\cal O}_{\rm FV}}  \right\vert^{- \frac{1}{2}}  =  
  -\frac{1}{2} \sum_{\nu = \frac{D}{2}-1}^\infty  d_\nu  \ln  \left( \lim_{x \to \infty}  R_\nu(x) \right) \, . 
\end{equation}
The parameter $\nu$ is an integer in even dimensions $D\geq 2$, a half-integer in odd dimensions $D\geq 3$. 
Our goal is to compute the quantity in \eqref{logdetRnu}, which implies eventually performing an 
infinite sum in $\nu$.
In this appendix we focus on computing the function $R_\nu$. To begin with, we need an equation for it.

We can trade ${\cal O}_\nu \psi_\nu = 0$ for an equation with $R_\nu$, which we derive as follows. 
From \eqref{Rnudefin} we have
\begin{align}
  \frac{1}{R_\nu} \drho R_\nu & = \frac{1}{\psi_\nu} \drho \psi_\nu - \frac{1}{\psi_{\nu{\rm FV}}} \drho \psi_{\nu{\rm FV}} \, , \label{dRnu}
  \\ \label{ddRnu}
  \begin{split}
  \frac{1}{R_\nu} \ddrho R_\nu - \left( \frac{1}{R_\nu} \drho R_\nu \right)^2 & = 
  \frac{1}{\psi_\nu} \ddrho \psi_\nu - \left( \frac{1}{\psi_\nu} \drho \psi_\nu \right)^2 
  \\
  &- 
  \frac{1}{\psi_{\nu{\rm FV}}} \ddrho \psi_{\nu {\rm FV}} + 
  \left( \frac{1}{\psi_{\nu{\rm FV}}} \drho \psi_{\nu{\rm FV}} \right)^2 \, , 
  \end{split}
\end{align}
with \eqref{ddRnu} obtained by taking the derivative of \eqref{dRnu}. From \eqref{dRnu} we have
\begin{equation}
  \frac{1}{\psi_\nu} \drho \psi_\nu = \frac{1}{R_\nu} \drho R_\nu + \frac{1}{\psi_{\nu{\rm FV}}} \drho \psi_{\nu{\rm FV}} \, ,
\end{equation}
which plugged into \eqref{ddRnu} gives
\begin{equation}
  \frac{1}{R_\nu} \ddrho R_\nu + 2 \frac{1}{R_\nu} \drho R_\nu  \frac{1}{\psi_{\nu{\rm FV}}} \drho \psi_{\nu{\rm FV}} =  
  \frac{1}{\psi_\nu} \ddrho \psi_\nu - \frac{1}{\psi_{\nu{\rm FV}}} \ddrho \psi_{\nu{\rm FV}} \, .
\end{equation}
The terms on the right-hand side, using \eqref{Onu}, are
\begin{equation}
 \frac{1}{\psi_\nu} \ddrho \psi_\nu =   \frac{\nu^2 - \frac{1}{4}}{\rho^2} + \ddV \, .
\end{equation}
Thus, we arrive at the following system of equations:
\begin{align}
  &  \left( -\frac{\text{d}^2}{\text{d} \rho^2} + 
  \frac{\nu^2 - \frac{1}{4}}{\rho^2} + \ddV_\text{FV} \right) \psi_{\nu \rm{FV}} = 0 \, , 
  \label{chiFVeq} 
  \\
  & \frac{1}{R_\nu}  \frac{\rm{d}^2 R_\nu}{\rm{d}\rho^2} + 2 \frac{1}{R_\nu} \frac{\rm{d} R_\nu}{\rm{d}\rho} 
  \left( \frac{\rm{d} \psi_{\nu\text{FV}} / \rm{d}\rho}{\psi_{\nu\text{FV}}} \right) -
  \left(\ddV - 
  \ddV_\text{FV} \right)  = 0  \, , 
  \label{Rlequ}
\end{align}
with boundary conditions
\begin{align} \label{bcondpsi}
  & \psi_{\nu \rm{FV}} (\rho \to 0)  \sim \rho^{\nu + \frac{1}{2}} \, , 
  \\
  & R_\nu(\rho = 0) = 1 \, , \qquad \quad \frac{\rm{d} R_\nu}{\rm{d}\rho}(\rho = 0) = 0 \, . 
\label{bconds}
\end{align}
\eqref{bconds} is obtained from the definition \eqref{Rnudefin} and the conditions \eqref{psibounds}.
Note that in the second term of \eqref{Rlequ} the behavior
\begin{equation} \label{dpsirho0}
  \frac{\rm{d} \psi_{\nu\text{FV}} / \rm{d}\rho}{\psi_{\nu\text{FV}}} \xrightarrow[\rho \to 0]{} \frac{1}{\rho} \, ,
\end{equation}
is compensated by the derivative condition in \eqref{bconds}.

Equation \eqref{chiFVeq} can be solved exactly:
\begin{equation} \label{chiFVexact}
  \psi_{\nu\text{FV}}(\rho) = 
  c_{\rm FV} \sqrt{\rho} \, I_\nu \left(\rho \sqrt{V^{(2)}_\text{FV}} \right) \, ,
\end{equation}
with $I_\nu$ a Bessel function.
We have solved a second-order differential equation with one boundary condition on the derivative
at the origin, so we are left with one integration constant, $c_{\rm FV}$. 
There is no need to fix this constant, as it cancels out in the ratio 
$\frac{\rm{d} \psi_{\nu\rm{FV}}/\rm{d}\rho}{\psi_{\nu\rm{FV}}}$ that enters the $R_\nu$ equation, \eqref{Rlequ}.

Equation \eqref{Rlequ} cannot be solved analytically. 
We are going to develop a method to solve it perturbatively, using expansions in the small parameter $\Delta$. 
As we will see shortly, the parameter $\nu$ will enter the equations in the combination $\Delta \nu$.
The final answer for the determinant, see \eqref{logdetRnu}, involves values of $\nu$ up to infinity.
This implies that for $\nu > 1/ \Delta$ the perturbation series in $\Delta$ will be ill defined.
To fix it we will consider the combination $\Delta \nu$ as a leading order term in the expansion. 
First, however, we will go through the simpler procedure of solving the $R_\nu$ equation for $\nu \ll 1/\Delta$. 
The result, \eqref{Rlowl}, is more illuminating written in terms of $l$ rather than $\nu$, as the negative eigenvalue 
for $l=0$ and the zero eigenvalues for $l=1$ are immediately apparent. 
For this reason, in the next section, we will relabel our functions with the subscript $l$ instead of $\nu$, to 
indicate that we are working with small values of $l$.

We will use the dimensionless variables
\begin{align}
  z  & = \sqrt{\lambda} v \ \rho - r \, , 
  &
  \varphi & = \phi / v \, , 
  &
  \ddVt & = \frac{1}{\lambda v^2} \ddV \, . 
\end{align}
%
%
{\bf Perturbative solutions for low multipoles.}
%
%
%
Despite having the exact solution \eqref{chiFVexact} for $\psi_{l{\rm FV}}$ at hand,
 it is more useful to have an approximate, simpler solution 
to \eqref{chiFVeq} with an explicit $\Delta$ dependence.
We computed it in Section~\ref{subsecFluctLowL} and found
\begin{equation}
  \psi_{l\rm{FV}}(z) = c_{\rm FV} {\rm Exp}\left[ \left(1 - \frac{3}{2}\Delta + 
  \frac{1}{2} \left( \frac{\nu^2 - \frac{1}{4}}{r_0^2} - \frac{21}{4}  \right) \Delta^2 + {\cal O}(\Delta^3) \right) z \right] \, .
\end{equation}
From this, we get the ratio
\begin{align} \label{glownu}
  g_l \equiv \frac{{\rm d} \psi_{l\rm{FV}} / {\rm d} z}{\psi_{l\rm{FV}}} & = 
  1 - \frac{3}{2}\Delta + \frac{1}{2} \left( \frac{\nu^2 - \frac{1}{4}}{r_0^2} - \frac{21}{4}  \right) \Delta^2 + {\cal O}(\Delta^3) 
  \\
  & = g_{l0} + \Delta g_{l1} +\Delta^2 g_{l2} + {\cal O}(\Delta^3) \, .  
\end{align}
%
%
%
Introducing the variable
\begin{equation} \label{xdef}
  x = e^z \, ,
\end{equation}
the $R_\nu$ equation~\eqref{Rlequ} turns into
\begin{equation} \label{Rlxeq}
  \frac{1}{R_l} \ddx R_l + \frac{2 g_l + 1}{x} \frac{1}{R_l} \dx R_l - \frac{1}{x^2} \left( 
  \ddVt - \ddVt_{\rm FV}  \right) = 0 \, ,
\end{equation}
with $g_l$ given in \eqref{glownu}. 
We relabeled $R_\nu \to R_l$ to indicate the low $l$ regime.
The boundary conditions from \eqref{bconds} become
\begin{equation} \label{bcsR}
  R_l(x = e^{-r}) = 1 \, , \qquad  e^{-r} \dx R_l(x = e^{-r}) = 0 \, .
\end{equation}
Note that we are retaining the $e^{-r}$ in front of the derivative condition. 
The reason is that in our expansions we keep track of powers of $\Delta$ but always drop exponentials of the form 
$e^{-r} \sim e^{-1/\Delta}$. 
Hence, we keep $e^{-r}$ explicit in \eqref{bcsR} to indicate that even with $\dx R_l(x = e^{-r}) $ equal to a constant or 
proportional to powers of $r$, the boundary condition is satisfied. 

We take a multiplicative expansion for $R_l$, 
\begin{align}
  R_l  & = R_{l 0} R_{l 1}^\Delta R_{l 2}^{\Delta^2} R_{l 3}^{\Delta^3} \cdots  \label{Rlmult}  
  \\ \label{Rlmultlog}
  & = R_{l0} \left( 1 + \Delta \ln R_{l1}  +\Delta^2  \left( \frac{1}{2} \ln^2 R_{l1} + \ln R_{l2}  \right) + {\cal O}(\Delta^3) \right) \, ,
\end{align}
and use the following expression
\begin{align}
  \left( \ddVt  - \ddVt_{\rm FV}  \right) & 
  = \sum_{n=0}^{\infty} \Delta^n  \left( \ddVt - \ddVt_{\rm FV} \right)_n 
  \nonumber \\
  & =  \frac{3}{2}(\varphi_0^2 -1) + 3 \Delta  \left(1 + \varphi_0 \varphi_1  \right) + \frac{3}{2} \Delta^2  
  \left( 2 + \varphi_1^2 + 2 \varphi_0 \varphi_2 \right) + {\cal O}(\Delta^3) \, .
\end{align}

At the zeroth order, $R_l = R_{l0}$, from \eqref{Rlxeq} we have
\begin{equation}
  \frac{1}{R_{l0}} \ddx R_{l0} + \frac{3}{x} \frac{1}{R_{l0}} \dx R_{l0} - \frac{1}{x^2} 
  \left( \ddVt - \ddVt_{\rm FV} \right)_0 = 0 \, ,
\end{equation}
with
\begin{equation} \label{dV20}
  \left( \ddVt - \ddVt_{\rm FV} \right)_0 = \frac{3}{2} (\varphi_0^2 - 1) 
  = - \frac{6x}{(1+x)^2} \, .
\end{equation}
The solution is
\begin{align} \label{Rl0}
  R_{l0} = \frac{1}{(1+x)^2} \, .
\end{align}

For orders $\Delta$ and higher, it is convenient to introduce
\begin{align} \label{fdef}
  f_l & \equiv \frac{1}{R_l} \dx R_l \, ,
\end{align}
so that the second-order differential equation \eqref{Rlxeq} turns into a first-order one,
\begin{equation} \label{flfull}
  \dx f_l + f_l^2 +  \frac{2 g_l + 1}{x} f_l - \frac{1}{x^2} \left( \ddVt - \ddVt_{\rm FV} \right) = 0 \, ,
\end{equation}
subject to the boundary condition
\begin{equation} \label{bcf}
  e^{-r} f_{l}(e^{-r}) = 0 \, ,
\end{equation}
where again we mean that even with $f_l(e^{-r})$ equal to a constant or proportional to powers of $r$,
the boundary condition is satisfied (see the comment below \eqref{bcsR}).
We expand $f_l$ in $\Delta$, 
\begin{equation}
  f_l = f_{l0} + \Delta f_{l1} + \Delta^2 f_{l2} + {\cal O}(\Delta^3) \, ,
\end{equation}
and plug it into \eqref{flfull}, then we proceed to solve the equation order by order in $\Delta$. 
At the lowest order, from \eqref{Rl0} we have
\begin{equation} \label{fl0}
  f_{l0}  = -\frac{2}{1+x} \, .
\end{equation}

At the order $\Delta^n$, with $n \geq 1$ we have the equations
\begin{equation} \label{fnloweq}
  \dx f_{ln} + \left( 2 f_{l0} + \frac{3}{x} \right) f_{ln} = P_{f_{ln}} \, ,
\end{equation}
with the functions
\begin{equation} \label{Pflow}
  P_{f_{ln}}  =  \frac{1}{x^2} \left( \ddVt - \ddVt_{\rm FV} \right)_n - \frac{2}{x} 
  \sum_{j=1}^n g_{lj} \ f_{l(n-j)} - \sum_{j=1}^{n-1} f_{lj}  \ f_{l(n-j)}  \, ,
\end{equation}
growing at every iteration. 
The last sum in \eqref{Pflow} contributes only for $n \geq 2$. 
The solution to \eqref{fnloweq} is given by
\begin{equation} \label{fnint}
  f_{ln}(x) = \frac{(1+x)^4}{x^3} \int_{e^{-r}}^x {\rm d}t \frac{t^3}{(1+t)^4} P_{f_{ln}}(t) \, .
\end{equation}
Here the lower limit of integration is set formally to $e^{-r}$ to guarantee the boundary condition \eqref{bcf}. 
In practice, however, one can set such a lower limit to 0, and check at every order that this corresponds to dropping 
only terms exponentially suppressed in $r$, that means terms of the form $e^{-1/\Delta}$.

At the first order in $\Delta$ we have
\begin{align} \label{dV21} 
  & \left( \ddVt - \ddVt_{\rm FV} \right)_1  
  = 3(\varphi_0 \varphi_1 + 1) = \frac{6}{1+x} \, ,
  \\
  & P_{f_{l1}} = \frac{1}{x^2} \left( \ddVt - \ddVt_{\rm FV} \right)_1 + 
  \frac{3}{x} f_{l0} = \frac{6}{x^2} \frac{1-x}{1+x} \, , 
\end{align}
so that
\begin{equation} \label{fl1full}
  f_{l1} = \frac{(1+x)^4}{x^3} \int_{e^{-r}}^x {\rm d}t \frac{t^3}{(1+t)^4} P_{f_{l1}}(t) 
  = \frac{(1+x)^4}{x^3} \left( \frac{3 x^2}{(1+x)^4} - \frac{3 e^{-2r}}{(1+e^{-r})^4} \right) \, .
\end{equation}
We can drop the last term in the squared parentheses as it is exponentially suppressed in $r$. 
Then\footnote{
The careful, and possibly skeptical reader might be bothered that \eqref{fl1} does not satisfy the boundary condition \eqref{bcf}.
However, one can check that using \eqref{fl1full}, which does satisfy \eqref{bcf}, instead of \eqref{fl1} to continue the 
calculation at $n \geq 2$, produces the same results after longer and more tedious algebra. 
The lesson is that we can drop terms exponentially suppressed in $r$ as soon as we encounter them.
}
\begin{equation} \label{fl1}
  f_{l1} = \frac{3}{x} \, .
\end{equation}
At the second order we have
\begin{align}
  & \left( \ddVt - \ddVt_{\rm FV} \right)_2  
  = \frac{3}{2} (2 + \varphi_1^2 + 2 \varphi_0 \varphi_2) 
  \nonumber \\
  & = \frac{3}{2} \bigg\{ 3 - \frac{1}{r_0 x(1+x)^3} (1-x) \Big[ x  \big( -1 + \pi^2 x + x^2 - 3 r_0(x^2-1) 
  \nonumber \\ 
  & \qquad \qquad \qquad \qquad \qquad \qquad \qquad +x(-3 -6r_0 +8x +x^2)\ln(x) + 3x \ln^2 (x) \big) 
  \nonumber \\  
  & \qquad \qquad \qquad \qquad \qquad \qquad  - (-1-8x+8x^3 + x^4)\ln(1+x) +12 x^2 {\rm Li}_2(-x)   \Big]  \bigg\} \, ,
  \label{dV22}
\end{align}
and
\begin{equation}
  P_{f_{l2}} = \frac{1}{x^2} \left( \ddVt - \ddVt_{\rm FV} \right)_2 +
  \frac{3}{x} f_{l1} - \frac{1}{x}\left( \frac{\nu^2 - \frac{1}{4}}{r_0^2} - \frac{21}{4}  \right) f_{l0} - f_{l1}^2  \, .
\end{equation}
Here note that the terms $\frac{3}{x} f_{l1} - f_{l1}^2$ cancel, given \eqref{fl1}. 
We can compute the integral in \eqref{fnint} analytically with $P_{f_{l2}}$. 
Setting the lower limit of integration $e^{-r}$ to zero, which is again equivalent to neglecting exponentially suppressed 
terms, we find
\begin{align}
  f_{l2} = &-\frac{1}{24 r_0^2 x^3 (x+1)^2}  
  \bigg(144 r_0 x^3 \text{Li}_2(-x)+9 r_0^2 x^6+6 r_0 x^6+12 r_0 x^6 \ln(x+1) \nonumber
  \\
  & +54 r_0^2 x^5+24 r_0 x^5+36 r_0 x^5 \ln (x+1)+9 r_0^2 x^4+60 r_0 x^4 \nonumber 
  \\
  & -36 r_0 x^4 \ln (x+1)  -36 r_0^2 x^3+12 \pi ^2 r_0 x^3+60 r_0 x^3+36 r_0 x^3 \ln ^2(x) \nonumber 
  \\ 
  & +30 r_0 x^2+36 r_0 x^2 \ln (x+1) -12 r_0 x^3 \left(6r_0+x^3+3 x^2-3 x+5\right) \ln (x) \nonumber 
  \\ 
  & +12 r_0 x -36 r_0 x \ln (x+1)-12 r_0 \ln (x+1) \nonumber 
   \\
  &   -4 \nu ^2 x^6+x^6-24 \nu ^2 x^5+6 x^5-36 \nu ^2 x^4+9 x^4-16 \nu ^2 x^3+4 x^3 \bigg) \, . \label{fl2}
\end{align}
This function has the following asymptotic behavior:
\begin{align}
  f_{l2} &  \xrightarrow[x\to 0]{}  \frac{8 \nu ^2-18 r_0 \ln ^2(x)+6 r_0 \left(6 r_0+5\right) \ln (x)+18
  r_0^2-\left(55+6 \pi ^2\right) r_0-2}{12 r_0^2} \, , \label{fl2zero} 
  \\
  f_{l2} &  \xrightarrow[x\to \infty]{} -\frac{x \left(-4 \nu ^2+9 r_0^2+6 r_0+1\right)}{24 r_0^2} \, .    \label{fl2infty}
\end{align}
Note that $f_{l2}$ diverges linearly at large $x$.

Next, we compute $\ln R_{ln}$, which from the definition \eqref{fdef} is simply given by
\begin{equation} \label{lnRln}
  \ln R_{ln}(x) = \int_{e^{-r}}^x {\rm d}t f_{ln}(t) \, .
\end{equation}
Here, the lower limit of integration is to enforce the boundary condition $R_l(x = e^{-r}) = 1$. 
At $n=1$, using \eqref{fl1}, we have
\begin{align} \label{lnRl1Delta}
  \ln R_{l1}(x) & = 3\left( \ln(x) + r \right) \, , \\
  R_{l1}(x) & = e^{\frac{3r_0}{\Delta}} x^3 \, , &
  R_{l1}^\Delta(x) & = e^{3r_0} x^{3 \Delta} \, , \label{Rl1Delta}
\end{align}
where we have used $r = r_0 / \Delta$. 
At $n=2$ we have
\begin{equation} \label{intfl2}
  \ln R_{l2}(x) = \int_{0}^x {\rm d}t f_{l2}(t) \, .
\end{equation}
Here we can set the lower limit of integration to zero. 
Doing so, given the asymptotic behavior \eqref{fl2zero}, we are throwing away terms suppressed by $e^{-r}$. 
The integral \eqref{intfl2} can be computed analytically plugging in \eqref{fl2}. 
However, we only need its asymptotic behavior at large $x$, which is easily obtained by integrating \eqref{fl2infty}:
\begin{equation} \label{Rl2infty}
  \ln R_{l2} (x \to \infty) = \int^x {\rm d}t f_{l2}(t \to \infty) =  -\frac{x^2 \left(-4 \nu ^2+9 r_0^2+6 r_0+1\right)}{48 r_0^2} \, .
\end{equation}  

We can now put together our low $l$ result for $R_l$ up to $\Delta^2$. 
Starting from \eqref{Rlmult} and \eqref{Rlmultlog}, and using \eqref{Rl1Delta}, we can factor out $e^{3r_0}$ and write
\begin{equation} \label{RlowUBA}
  R_l(x) = e^{3r_0} R_{l0}(x) 
  \left( 1 + 3\Delta \ln(x) + \Delta^2 \left( \frac{1}{2} 9 \ln^2(x) + \ln R_{l2}(x)  \right) + {\cal O}(\Delta^3) \right) \, .
\end{equation}
We are interested in the $x\to \infty$ limit.
From \eqref{Rl0} we have $R_{l0}(x \to \infty) = x^{-2}$, and using \eqref{Rl2infty} we obtain
\begin{equation} \label{Rlownu}
  R_l(x \to \infty) = - \Delta^2 e^{3r_0}\frac{-4 \nu ^2 + 9 r_0^2 + 6 r_0 + 1}{48 r_0^2} + {\cal O}(\Delta^3) \, .
\end{equation}
Substituting $\nu = l + D/2 -1$ and $r_0 = (D-1)/3$ we can rewrite it as
%
\begin{equation} \label{Rlowl}
  \lim_{\rho \to \infty} R_l(\rho) = R_l(x\to\infty) = \Delta^2 \frac{3}{4} e^{D-1} \frac{(l -1)(l + D -1)}{(D-1)^2} + {\cal O}(\Delta^3) \, ,
\end{equation}
which corresponds to~\eqref{eqRlinfLowl}.
Some comments are in order.
\begin{itemize}
  \item 
  The leading term in this low $l$ result is of order $\Delta^2$, with the ${\cal O}(\Delta^0)$ and ${\cal O}(\Delta)$ 
  terms vanishing. 
  It has the structure we expected: it is negative for $l=0$, it vanishes for $l=1$. 
  The latter signals the zero eigenvalues related to the translational invariance of the bounce. 
  We will soon go through a procedure to remove them.
  \item
  The $e^{3r_0}$ is an overall factor in \eqref{RlowUBA} and as such it ends up contributing to the first term, that is zeroth 
  order in the $\Delta$ expansion. 
  Note we got $e^{3r_0}$ at $n=1$ [see \eqref{lnRl1Delta} and \eqref{Rl1Delta}], rather than $n=0$, as a consequence 
  of the boundary condition at $x = e^{-r}$ (corresponding to $\rho = 0$). 
  In other words, we set up our equations based on the $\Delta^n$ power counting, but then found a solution at $n=1$
  that fed back into $n=0$.
  %
  %
  Let us refer to it as an order-breaking term.
  While this feature might be puzzling at first sight, there is no reason to panic.
  The calculation is sensible and the procedure well defined:
  at each order one computes $f_{ln}$ 
  from~\eqref{fnint}, and then checks from \eqref{lnRln} if there are contributions to $\ln R_{ln}$
   from the boundary at $x = e^{-r}$, with $r = r_0 / \Delta$, that feed back into lower orders. 
  In the low $l$ calculation this happens only at $n=1$. 
  In the next section, we will study the problem systematically for generic values of $l$ (or, better, of $\nu$) and show 
  that order-breaking terms will appear for every $n \geq 1$, but that indeed at low $l$ only the one from $n=1$ survives.
  \item
  There is at least another way of obtaining the $e^{3r_0}$ factor. 
  Instead of using \eqref{Rlmult}, one can approach the problem with an additive expansion, 
  $R_l = R_{l0} + \Delta R_{l1} + \Delta^2 R_{l2} + \cdots$, and solve the second order differential equation for $R_l$ 
  order by order. 
  The calculation is more involved, compared to the one we have presented here, and one finds at each order 
  boundary terms of the form $\Delta^n r^n$, which therefore feed back into $n = 0$. 
  Summing all such terms, up to $n = \infty$, one recovers exactly the exponential $e^{3r_0}$.
\end{itemize}

%
%
{\bf Zero removal at low multipoles.}
With the explicit solution valid up to second-order corrections in $\Delta^2$, one can proceed to
perturbatively remove the zero eigenvalue, as discussed in section~\ref{subsecFluctZero}.
The orbital mode dependence comes in at $\mathcal O(\Delta^2)$.
To remove the $l=1$ modes it is then sufficient to off-set the potential at this order with an
infinitesimal dimensional parameter
\begin{align} \label{eqV2tldeps}
  V^{(2)} \to V^{(2)} + \Delta^2 \mu_\varepsilon^2 \, ,
  \quad \text{ or } \quad 
  \tilde V^{(2)} \to \tilde V^{(2)} + \frac{\Delta^2 \mu_\varepsilon^2}{\lambda v^2} \, .
\end{align}
This offset does not affect the zero and first-order solutions, once $\mu_\varepsilon \to 0$,
 thus $R_0$ and $R_1$ remain the same.
On the other hand, the second-order solution with the above off-set is easily obtained starting from~\eqref{fnint},
\begin{align}
  f_{l2}^\varepsilon(x) &= \frac{(1+x)^4}{x^3} \int_{e^{-r}}^x {\rm d}t \frac{t^3}{(1+t)^4} \left(
  P_{f_{l2}}(t) + \frac{ \mu_\varepsilon^2}{\lambda v^2} \frac{1}{t^2} \right) 
  \\
  &= f_{l2} + \frac{\mu_\varepsilon^2}{\lambda v^2} \frac{(1+x)^4}{x^3} 
  \int_0^\infty \text{d} t \frac{t}{(1+t)^4} \, .
  \\
  &= f_{l2} + \frac{ \mu_\varepsilon^2}{\lambda v^2} \frac{(1+x)^4}{6 x^3}  \, .  
\end{align}
From here, we plug the $f_{l2}^\varepsilon(x)$ into~\eqref{lnRln}, such that
\begin{align} 
  \ln R_{2 1}^\varepsilon &= \int^x \text{d} t \, f_{l2}^\varepsilon(t) 
  \\
  &= \int^x \text{d} t \left( f_{l2} + \frac{ \mu_\varepsilon^2}{\lambda v^2} \frac{(1+t)^4}{6 t^3} \right)
  \\ \label{eqR21primeDer}
  &=  \ln R_{2 1} +  \frac{\mu_\varepsilon^2}{\lambda v^2} \frac{x^2}{12} 
  =  \frac{\mu_\varepsilon^2}{\lambda v^2} \frac{x^2}{12} \, ,
\end{align}
which is precisely what was given in~\eqref{eqR21prime0}.

%
%
{\bf Perturbative solutions for generic multipoles.}
In the previous section we computed $\lim_{x \to \infty}  R_\nu(x)$, but our result was only valid for small 
values of $\nu$.
In this section we are going to revisit our calculation, generalizing it to generic values of $\nu$. The starting point
is to consider the combination $\Delta \nu$ as on order one parameter instead of a $\Delta$ suppressed one.
Then the $\psi_{\nu{\rm FV}}$ equation, see~\eqref{eqPsiFVLowL}, has the leading order solution
\begin{equation} \label{psinulead}
\psi_{\nu{\rm FV}}^{\rm lead} = c_{\rm FV} e^{k_\nu z} \, ,
\end{equation}
with
\begin{equation} \label{knuApp}
k_\nu = \sqrt{1 + \frac{\Delta^2 \nu^2}{r_0^2}} \, ,
\end{equation}
which counts as an ${\cal O}(\Delta^0)$ parameter. We define
\begin{equation}  \label{gFVexpApp}
 \frac{1}{\psi_{\nu{\rm FV}}} \dz \psi_{\nu{\rm FV}} \equiv g_\nu =  \sum_{n=0}^\infty \Delta^n g_{\nu n} \, . 
\end{equation}
Given \eqref{psinulead}, we immediately have that $g_{\nu 0} = k_\nu$.
Note, as a rather trivial check, that using \eqref{knuApp}, expanding in $\Delta$ and keeping only the zeroth order, 
that is $k_\nu = 1$, we match the low $l$ result: $g_{\nu 0}(k_\nu = 1) = g_{l 0} = 1$.
With~\eqref{gFVexpApp} the differential FV equation \eqref{chiFVzeq} turns into
\begin{equation} \label{gFVeqApp}
  \dz g_\nu + g_\nu^2 = \frac{\nu^2 - \frac{1}{4}}{(z+r)^2} + \ddVt_\text{FV} \, .
\end{equation}
With the counting stated above the right hand side has the expansion
\begin{align}
\frac{\nu^2 - \frac{1}{4}}{(z+r)^2} + \ddVt_\text{FV} & =  k_\nu^2 + \left( -3 - 2 (k_\nu^2 -1) \frac{z}{r_0}  \right) \Delta
+ \left( -3 -\frac{1}{4 r_0^2} + 3 (k_\nu^2 -1) \frac{z^2}{r_0^2} \right) \Delta^2 \nonumber \\
& + \left( -\frac{15}{2} + \frac{z}{2 r_0^3} - 4 (k_\nu^2 - 1) \frac{z^3}{r_0^3}  \right) \Delta^3 + {\cal O}(\Delta^4) \, .
\label{DelExp3}
\end{align}
Using this and pugging~\eqref{gFVexpApp} into~\eqref{gFVeqApp}, we get the $n=1$ equation
\begin{equation}
  \dz g_{\nu 1} + 2 k_\nu g_{\nu 1} =  -3 - 2 (k_\nu^2 -1) \frac{z}{r_0} \, ,
\end{equation}
which has the solution
\begin{equation} \label{gnu1}
  g_{\nu 1} = \frac{ -2 (k_\nu^2 - 1)}{2 k_\nu}  \frac{z}{r_0} + \frac{k_\nu^2 - 1 - 3 k_\nu r_0}{2 k_\nu^2 r_0}\, .
\end{equation}
Note the first term is linear in $z$, while the second is $z$ independent.
With this we can solve~\eqref{gFVeqApp} at $n=2$, where we find
\begin{equation}
g_{\nu 2} = \frac{1}{2 k_\nu} \left( 3 (k_\nu^2 -1) -  4 (k_\nu^2 -1)^2  \frac{1}{(2k_\nu)^2}  \right) \left( \frac{z}{r_0} \right)^2 + {\rm terms \ with \ lower \ powers \ of} \ z \, . \label{gnu2}
\end{equation}
Here we don't bother writing explicitly the terms with lower powers of $z$
 because they are not needed for the rest of the calculation,
as we are going to explain.
Let us keep the higher orders $g_{\nu n}$ on hold for a moment, and turn to $R_\nu$.

The $R_\nu$ equation to solve for generic multipoles,
\begin{equation} \label{Rnuxeq}
  \frac{1}{R_\nu} \ddx R_\nu + \frac{2 g_{\nu} + 1}{x} \frac{1}{R_\nu} \dx R_\nu - \frac{1}{x^2} \left( 
  \ddVt - \ddVt_{\rm FV} \right) = 0 \, ,
\end{equation}
reads the same as \eqref{Rlxeq}, but $g_l$ from~\eqref{glownu} is replaced with $g_\nu$ from~\eqref{gFVexpApp}. 
At order zero we have
\begin{equation}
  \frac{1}{R_{\nu 0}} \ddx R_{\nu 0} + \frac{2k_\nu + 1}{x} \frac{1}{R_{\nu 0}} \dx R_{\nu 0} - \frac{1}{x^2} \left( 
  \ddVt - \ddVt_{\rm FV} \right)_0 = 0 \, ,
\end{equation}
with the last term given by \eqref{dV20}. The solution is
\begin{equation} \label{Rnu0}
  R_{\nu 0}(x) = \frac{1- 4x + x^2 + 2k_\nu^2(1+x)^2 + 3k_\nu(1 - x^2)}{(1+k_\nu)(1 + 2k_\nu)(1+x)^2} \, .
\end{equation}
This matches the boundary condition $R_{\nu 0}(x \to 0) = 1$, reduces to \eqref{Rl0} for $k_\nu =1$, and has 
the asymptotic behavior
\begin{equation} \label{Rnu0infty}
  R_{\nu 0} (\infty) = \frac{(1-k_\nu) (1 - 2k_\nu)}{(1 + k_\nu) (1+2k_\nu)} \, .
\end{equation}
Then we introduce
\begin{equation} \label{fnudef} 
  f_{\nu}  \equiv \frac{1}{R_\nu} \dx R_\nu \, , 
\end{equation}
in terms of which \eqref{Rnuxeq} becomes
\begin{equation} \label{fnufull}
  \dx f_\nu + f_\nu^2 +  \frac{2 g_\nu + 1}{x} f_\nu - \frac{1}{x^2} \left( \ddVt - \ddVt_{\rm FV} \right) = 0 \, ,
\end{equation}
with the boundary condition
\begin{equation} \label{fnubc}
  e^{-r} f_\nu (e^{-r}) = 0 \, .
\end{equation}
We expand
\begin{equation}
  f_\nu = \sum_{n=0}^\infty \Delta^n f_{\nu n} \, ,
\end{equation}
and $g_\nu$ with \eqref{gFVexpApp}, then solve \eqref{fnufull} order by order in $\Delta$. 
The $f_{\nu 0}$ can be immediately obtained from \eqref{Rnu0},
\begin{equation} \label{fnu0}
  f_{\nu 0}(x) = \frac{-6(1+k_\nu - x + k_\nu x)}{(1+x)\left(1 - 4x + x^2 + 2k_\nu^2(1+x)^2 + 3k_\nu(1 - x^2) \right) } \, .
\end{equation}
For order $n \geq 1$, from \eqref{fnufull} we get
\begin{equation} \label{fnuperteq}
  \dx f_{\nu n} + \left( 2 f_{\nu 0} + \frac{2k_\nu + 1}{x} \right) f_{\nu n} = P_{f_{\nu n}} \, ,
\end{equation}
with 
\begin{equation} \label{Pfnu}
  P_{f_{\nu n}}  =  \frac{1}{x^2} \left( \ddVt - \ddVt_{\rm FV} \right)_n - 
  \frac{2}{x} \sum_{j=1}^n g_{\nu j} \ f_{\nu(n-j)} - \sum_{j=1}^{n-1} f_{\nu j}  \ f_{\nu(n-j)}  \, .
\end{equation}
Here the last sum contributes only for $n \geq 2$.
The solution to \eqref{fnuperteq} is
\begin{align} \label{fnusol}
\begin{split}
  f_{\nu n}(x) = & \frac{x^{-(2k_\nu + 1)} (1 + x)^4}{
  \left( 1 - 4x + x^2 + 2k_\nu^2(1 + x)^2 + 3k_\nu (1 - x^2) \right)^2} \,  
  \\
  & \int_{e^{-r}}^x {\rm d} t \frac{t^{2k_\nu + 1}}{(1 + t)^4} 
  \left( 1 - 4t + t^2 + 2k_\nu^2(1 + t)^2 + 3k_\nu (1 - t^2) \right)^2 P_{f_{\nu n}} (t) \, ,
\end{split}
\end{align}
and from the definition \eqref{fnudef} we have
\begin{equation} \label{logRnu}
  \ln R_{\nu n}(x) = \int_{e^{-r}}^x {\rm d} t \ f_{\nu n}(t) \, .
\end{equation}
Analogously to the low $l$ case, in \eqref{fnusol} we can always set the lower limit of integration to zero, 
dropping only terms exponentially suppressed in $r$ in doing so. 
On the contrary, in \eqref{logRnu} it is crucial to maintain $e^{-r}$ as the lower limit of integration, which implements 
the boundary condition $R_{\nu}(\rho = 0) = R_{\nu}(x = e^{-r}) = 1$. Note that, as $R_{\nu 0} (x = e^{-r} \to 0) = 1$,
we must require $\ln R_{\nu n}(x = e^{-r}) = 0$ for $n \geq 1$.

Formally we have the recipe to compute $\ln R_{\nu n}$ at any $n$ and get the result
\begin{align} \label{Rnumultexp}  
  R_{\nu} & = R_{\nu 0} R_{\nu 1}^\Delta R_{\nu 2}^{\Delta^2} R_{\nu 3}^{\Delta^3} \cdots
  \\
  & =  R_{\nu 0} \left( 1 + \Delta \ln R_{\nu 1}  + 
  \Delta^2  \left( \frac{1}{2} \ln^2 R_{\nu 1} + \ln R_{\nu 2}  \right) + {\cal O}(\Delta^3) \right) \, ,
\end{align}
up to the desired order in $\Delta$.
In practice, we can solve the integral in \eqref{fnusol} analytically only for $n=1$, the result containing hypergeometric 
functions, but already fail when trying to integrate it to compute the full form of $\ln R_{\nu 1}$.
We have already found that $R_{\nu 0}(x \to \infty)$ in~\eqref{Rnu0infty} does not vanish, in contrast with 
$R_{l0}$ in the low-$l$ regime, which did vanish at large $x$. Thus, it seems like we already have
the leading order term in the expansion~\eqref{Rnumultexp}. But is that the full contribution at the leading order?
In the low-$l$ calculation we found an order-breaking term at $n=1$ that fed back into $n=0$.
It came from the lower boundary of integration of $\ln R_{l1}$. We should then check if we also get
order-breaking terms in the current calculation for generic multipoles. 
To do so, we must study the lower boundary of the integral~\eqref{logRnu}, for which 
it is sufficient to know the $x \to 0$ limit of $f_{\nu n}$. That is obtained from~\eqref{fnusol}:
\begin{equation} \label{fnunx0}
  f_{\nu n}(x \to 0) = x^{-(2k_\nu + 1)} \int_0^x {\rm d}t \ t^{2k_\nu +1} P_{f_{\nu n}}(t \to 0) \, .
\end{equation}

At $n=1$, from~\eqref{Pfnu} we have
\begin{equation} \label{Pf1}
  P_{f_{\nu 1}}  =  \frac{1}{x^2} \left( \ddVt - \ddVt_{\text{FV}} \right)_1 - 
  \frac{2}{x}  g_{\nu1} \ f_{\nu 0} \, ,
\end{equation}
with [see \eqref{dV21} and \eqref{gnu1}]
\begin{align}
  & \left( \ddVt - \ddVt_{\text{FV}} \right)_1  = \frac{6}{1+x} 
  &
  & g_{\nu 1} (x) = -2 \frac{k_\nu^2 - 1}{2k_\nu} \frac{\ln(x)}{r_0} \, ,
\end{align}
and $f_{\nu 0}$ given in \eqref{fnu0}. Here we have kept only the highest power of $z = \ln(x)$
in $g_{\nu 1}$. 
In the limit $x\to 0$:
\begin{align}
  \frac{1}{x^2}  \left( \ddVt - \ddVt_{\text{FV}} \right)_1 & 
  \xrightarrow[x\to 0]{} \frac{6}{x^2} \, , \label{dV21x0} 
  \\
  \frac{2}{x} g_{\nu 1} \ f_{\nu 0} & \xrightarrow[x\to 0]{} 2 \frac{\ln(x)}{x} \left( 
   -2 \frac{k_\nu^2 - 1}{2k_\nu} \frac{1}{r_0} \right) \left( \frac{-6}{1+2k_\nu}  \right) \, . \label{g1f0x0}
\end{align}
%
%
From the first term, \eqref{dV21x0}, using \eqref{fnunx0}, we find
\begin{align} \label{fnu1x0}
  P_{f_{\nu 1}} (x \to 0) & = \frac{6}{x^2} \, , 
  & 
  f_{\nu 1}(x \to 0) & = \frac{3}{k_\nu x} \, ,
\end{align}
which plugged into~\eqref{logRnu} gives
\begin{equation} \label{U1}
  \Delta \ln R_{\nu 1} \ni \Delta \int_{e^{-r}} {\rm d}x \ f_{\nu 1}(x \to 0)  = 3 r_0 \frac{1}{k_\nu} \equiv U_1 \, .
\end{equation}
This is an order-breaking term which feeds back into the $n=0$ result.
When $k_\nu \to 1$ it matches the factor in \eqref{Rl1Delta} we obtained in the low-$l$ calculation. 
Note that $U_1$ originates from the second derivative of the potential, \eqref{dV21x0}, calculated at $n=1$,
which involves the bounce up to first order in $\Delta$, that is $\varphi_1$.
It is easy to check that the other term, \eqref{g1f0x0}, gives $f_{\nu 1}(x \to0) \propto \ln(x)$, does not
produce a boundary term in $\ln R_{\nu 1}$, and so does not feed back into $n=0$. Recall that in~\eqref{g1f0x0}
we have dropped a term coming from the $x$-independent piece in $g_{\nu 1}$, see~\eqref{gnu1}.
Clearly, that cannot produce a boundary term in $\ln R_{\nu 1}$ either.

At $n=2$ we have
\begin{equation}
  P_{f_{\nu 2}}  =  \frac{1}{x^2} \left( \ddVt - \ddVt_{\text{FV}} \right)_2 - 
  \frac{2}{x} \left(  g_{\nu 1} \ f_{\nu 1} + g_{\nu 2} \ f_{\nu 0} \right) - f_{\nu 1}^2 \, .
\end{equation}
Using \eqref{dV22}, \eqref{fnu0}, \eqref{gnu1}, \eqref{gnu2}, \eqref{fnu1x0}, for $x\to 0$ we get:
\begin{align}
  \frac{1}{x^2} \left( \ddVt - \ddVt_{\text{FV}} \right)_2 & 
  \xrightarrow[x\to 0]{} \ \frac{3 \left(12 r_0 (\ln (x)+2)-6 \ln ^2(x)+6 \ln (x)-2 \pi ^2-15\right)}{4 r_0 x} \, , \label{dV22x0} 
  \\
  \frac{2}{x} g_{\nu 1} \ f_{\nu 1} &  \xrightarrow[x\to 0]{} \  2 \cdot  \frac{ -2 (k_\nu^2 - 1)}{2 k_\nu}  \frac{1}{r_0} 
  \frac{3}{k_\nu} \frac{\ln(x)}{x^2} \, ,  \label{g1f1x0} 
  \\ \label{g2f0x0} 
  \frac{2}{x} g_{\nu 2} \ f_{\nu 0} &  \xrightarrow[x\to 0]{} \ 2  \frac{k_\nu^2 - 1}{2 k_\nu}   \left( 3 -  4 (k_\nu^2 -1)  
  \frac{1}{(2k_\nu)^2}  \right)  \frac{1}{r_0^2} \frac{-6}{1+2k_\nu} \frac{\ln^2(x)}{x} \, ,  
  \\ \label{f1sqx0}
  f_{\nu 1}^2  &  \xrightarrow[x\to 0]{} \ \frac{9}{k_\nu^2 x^2} \, . 
\end{align}
Here there are two contributions proportional to $1/x^2$: \eqref{f1sqx0} and \eqref{g1f1x0}, which also
contains $\ln(x)$ from $g_{\nu 1}$. 
From the latter we find
\begin{align}
  P_{f_{\nu 2}}(x\to 0) & = -\frac{2}{x} g_{\nu 1} f_{\nu 1}(x\to 0) = \frac{6}{r_0} \frac{k_\nu^2 - 1}{k_\nu^2} \frac{\ln(x)}{x^2} \, , 
  \\
  f_{\nu 2}(x \to 0) & = \frac{6}{r_0} \frac{1}{2 k_\nu}  \frac{k_\nu^2-1}{k_\nu^2}  \frac{\ln(x)}{x} \, , \label{fnu2x0}
\end{align}
which gives the order-breaking contribution
\begin{equation} \label{U2}
  \Delta^2 \ln R_{\nu 2} \ni \Delta^2 \int_{e^{-r}} {\rm d}x \ f_{\nu 2}(x \to 0)  = 
  3 r_0 \left( -\frac{1}{2 k_\nu} + \frac{1}{2k_\nu^3} \right) \equiv U_2 \, ,
\end{equation}
feeding back into $n=0$. In the limit $k_\nu \to 1$ we get $U_2 \to 0$. This confirms the absence
of such a contribution in the low-multipole limit, in agreement with our low-$l$ result. 
From~\eqref{f1sqx0}, one ends up with a boundary term 
$\Delta^2 \ln R_{\nu 2} \propto \Delta^2 r = {\cal O}(\Delta)$, 
which is also order-breaking, but feeds back into $n=1$ rather than $n=0$,
so it is not a leading order contribution.
From the other two terms,  \eqref{dV22x0} and \eqref{g2f0x0}, both proportional to $1/x$,
 we get terms in $\Delta^2 \ln R_{\nu 2}$ which are genuinely of order $\Delta^2$.
%
%
%

At $n=3$ we have
\begin{equation} \label{Pfnu3full}
  P_{f_{\nu 3}}  =  \frac{1}{x^2} \left( \ddVt - \ddVt_{\text{FV}} \right)_3 - 
  \frac{2}{x} \left(  g_{\nu 1} \ f_{\nu 2} + g_{\nu 2} \ f_{\nu 1} + g_{\nu 3} \ f_{\nu 0}    \right) - 2 f_{\nu 1} \ f_{\nu 2} \, .
\end{equation}
Here, the term
\begin{equation}
  \frac{1}{x^2} \left( \ddVt - \ddVt_{\text{FV}} \right)_3 = 
  \frac{1}{x^2} \left( \frac{15}{2} + 3 \varphi_1 \varphi_2 + 3 \varphi_0 \varphi_3 \right) 
  \xrightarrow[x \to 0]{} \frac{1}{x^2} \left( 15 + {\cal O}(x) \right) \, ,
\end{equation}
gives a contribution $\Delta^3 \ln R_{\nu 3} \propto {\cal O}(\Delta^2)$, which feeds back into $n=2$, but not 
into the leading order.
The following pieces
\begin{align}
  P_{f_{\nu 3}} (x\to 0) & = -\frac{2}{x}( g_{\nu 1} \ f_{\nu 2} + g_{\nu 2} \ f_{\nu 1} )(x\to 0) = 
  -\frac{9}{r_0^2} \frac{k_\nu^2 -1}{k_\nu^4}\frac{\ln^2(x)}{x^2} \, , \\
    f_{\nu 3}(x \to 0) & = -\frac{9}{r_0^2} \frac{1}{2k_\nu} \frac{k_\nu^2 -1}{k_\nu^4}\frac{\ln^2(x)}{x} \, ,
\end{align}
give the contribution
\begin{equation} \label{U3}
  \Delta^3 \ln R_{\nu 3} \ni \Delta^3 \int_{e^{-r}} {\rm d}x \ f_{\nu 3}(x \to 0)  = 
  3 r_0 \left( -\frac{1}{2 k_\nu^3} + \frac{1}{2k_\nu^5} \right) \equiv U_3 \, ,
\end{equation}
which feeds back into $n=0$.
Here we have used~\eqref{fnu1x0}, \eqref{fnu2x0}, and we have kept only the terms with the leading 
power of $z = \ln(x)$ in $g_{\nu 1}$ and $g_{\nu 2}$ from~\eqref{gnu1} and~\eqref{gnu2}; 
sub-leading powers of $z$ do not contribute to \eqref{U3}. Likewise, one can check that the remaining terms
in~\eqref{Pfnu3full} don't produce a contribution that feeds back into $n=0$.


At $n=4$ we have
\begin{align} \label{Pfnu4full}
  P_{f_{\nu 4}} & = \frac{1}{x^2} \left( \ddVt - \ddVt_{\text{FV}} \right)_4  - \frac{2}{x} \left(  g_{\nu 1} \ f_{\nu 3} + g_{\nu 2} \ f_{\nu 2} + g_{\nu 3} \ f_{\nu 1} + g_{\nu 4} \ f_{\nu 0}    \right) - 
  2 f_{\nu 1} \ f_{\nu 3} - f_{\nu 2}^2 \, ,
\end{align}
%
with
\begin{equation}
  \frac{1}{x^2} \left( \ddVt - \ddVt_{\text{FV}} \right)_4 
  = \frac{1}{x^2} \frac{3}{2}(16 + 2\varphi_0 \varphi_4 + 2 \varphi_1 \varphi_3 + \varphi_2^2) 
  \xrightarrow[x \to 0]{} {\cal O}\left( \frac{1}{x} \right) \, .
\end{equation} 
The order-breaking contribution feeding back into $n=0$, comes from
\begin{align}
\label{Pfnu4select}
P_{f_{\nu 4}} (x\to 0) & = -\frac{2}{x}( g_{\nu 1} \ f_{\nu 3} + g_{\nu 2} \ f_{\nu 2} + g_{\nu 3} \ f_{\nu 1} )(x\to 0) 
= -\frac{3}{r_0^3} \frac{5 - 6k_\nu^2  + k_\nu^4}{k_\nu^6}\frac{\ln^3(x)}{x^2} \, , \\
  f_{\nu 4}(x \to 0) & =  -\frac{3}{r_0^3} \frac{1}{2k_\nu} \frac{5 - 6k_\nu^2  + k_\nu^4}{k_\nu^6}\frac{\ln^3(x)}{x}\, ,
 \label{fnu4x0}
\end{align}
and is given by
\begin{equation} \label{U4}
  \Delta^4 \ln R_{\nu 4} \ni \Delta^4 \int_{e^{-r}} {\rm d}x \ f_{\nu 4}(x \to 0)  = 
  3 r_0 \left( \frac{1}{8 k_\nu^3} - \frac{6}{8k_\nu^5} + \frac{5}{8k_\nu^7} \right) \equiv U_4 \, .
\end{equation}
Here we have used $g_{\nu 3} \propto z^3 = \ln^3(x)$, see the derivation below.
The terms in~\eqref{Pfnu4full} not included in~\eqref{Pfnu4select} do not feed back into $n=0$.

The pattern should be clear at this point. We are after order-breaking contributions of the form
$\Delta^n \ln R_{\nu n} = {\cal O}(\Delta^0)$. They come from the lower limit of the integral~\eqref{logRnu},
when $f_{\nu n}$ has the form
\begin{equation} \label{fnulog}
  f_{\nu n}(x \to 0) \propto \frac{\ln^{n-1}(x)}{x} \, .
\end{equation}
This, in turn, is obtained from
\begin{equation} \label{Pisolate}
  P_{f_{\nu n}}(x \to 0) \propto \frac{\ln^{n-1}(x)}{x^2} \, ,
\end{equation}
as one can check by plugging it into \eqref{fnunx0},
\begin{equation} \label{Pisolateintegral}
  x^{-(2k_\nu + 1)} \int_0^x {\rm d}t \ t^{2k_\nu +1} \frac{\ln^{n-1}(t)}{t^2} = \frac{1}{2k_\nu} \frac{\ln^{n-1}(x) + \cdots}{x} \, ,
\end{equation}
with the dots denoting terms with powers of $\ln(x)$ lower than $n-1$ in the numerator, which we are dropping. 
Let us look into the general form of $P_{f_{\nu n}}$ in~\eqref{Pfnu}, which has the following behavior. 
For the odd orders
\begin{equation} \label{dV2odd}
  \frac{1}{x^2} \left( \ddVt - \ddVt_{\text{FV}} \right)_{2 j + 1}
  \xrightarrow[x \to 0]{} \frac{1}{x^2}[{\rm constant} + {\cal O}(x)] \, , 
\end{equation}
while for the even orders
\begin{equation}
  \frac{1}{x^2} \left( \ddVt - \ddVt_{\text{FV}} \right)_{2 j}
  \xrightarrow[x \to 0]{} \frac{1}{x^2}[{\cal O}(x)] \, .
\end{equation}
Such a behavior is a direct consequence of the boundary condition $\varphi(x\to 0) = \varphi(\rho = 0) = \varphi_{\rm TV}$
of the bounce solution. 
Then we see that only for $n=1$ such a term is of the form \eqref{Pisolate}, and from it we get first $f_{\nu 1}$, then $U_1$. 
Proceeding to higher orders, $n \geq 2$, we can forget about the second-derivative-potential term in \eqref{Pfnu}, as at most 
the odd orders \eqref{dV2odd} will give contributions feeding back into the previous order, not back into $n=0$. 
We can also drop the last sum, $-\sum_{j=1}^{n-1} f_{\nu j} f_{\nu (n-j)}$, which never produces terms like \eqref{Pisolate}.
We are left with the sum $-\frac{2}{x} \sum_{j=1}^n g_{\nu j} f_{\nu(n-j)}$.
We have seen that only the term with the highest power of $z$ in $g_{\nu n}$, that is $z^n$, gives~\eqref{Pisolate}. 
Let us then compute the coefficients of such terms. To do so, we have to get back to~\eqref{gFVeqApp}.
From~\eqref{DelExp3} we see that we can write the expansion as
\begin{align} \label{k0zexp}
 \frac{\nu^2 - \frac{1}{4}}{(z+r)^2} + \ddVt_{\text{FV}} & = k_\nu^2 + (k_\nu^2 - 1) \sum_{n=1}^\infty 
 (-)^n (n+1) \frac{\Delta^n z^n}{r_0^n}  \\
 & \quad + \sum_{n > q \geq 0} c_{n,q} \Delta^n z^q \, . \label{uselessk0z}
\end{align}
Then we can write~\eqref{gFVeqApp}, using~\eqref{gFVexpApp}, for $n \geq 1$ as
\begin{equation} \label{gneq}
  \dz g_{\nu n} + 2 k_\nu g_{\nu n} = P_{g_n} \, ,
\end{equation}
with the function
\begin{equation} \label{Pgneq}
  P_{g_n} =  (-)^n (n+1) (k_\nu^2 - 1)\frac{z^n}{r_0^n} - \sum_{j=1}^{n-1} g_{\nu j} \ g_{\nu (n-j)} \, ,
\end{equation}
built iteratively. 
Here the last sum is present only for $n \geq 2$. In $P_{g_n}$ 
we have dropped the terms from~\eqref{uselessk0z},
because they contain sub-leading powers of $z$, which we do not need.
The solution to \eqref{gneq} is
\begin{equation} \label{gnsol}
  g_{\nu n}(z) = e^{-2k_\nu z} \int_{-r}^z {\rm d}t \ e^{2 k_\nu t} P_{g_n}(t) \, .
\end{equation}
The lower limit of integration can be understood as follows. 
From \eqref{dpsirho0}, switching from $\rho$ to $z$, we get the boundary condition
\begin{equation}
  g_\nu(z \to -r) = \frac{1}{z+r} = \frac{\Delta}{r_0} \sum_{n = 0}^{\infty} (-)^n \frac{\Delta^n z^n}{r_0^n} \, .
\end{equation}
The sum with alternating signs on the right hand side, obtained after expanding in $\Delta$, implies that 
$g_{\nu n} (z = -r = -r_0/ \Delta) = 0$, as implemented in \eqref{gnsol}.
We can infer from \eqref{Pgneq} that $P_{g_n}$ is a polynomial in 
$z$ of degree $n$. 
Hence, we can set $-r \to -\infty$ in the lower limit of integration of \eqref{gnsol}, dropping terms 
exponentially suppressed in $r$. 
%
At $n=1$, from \eqref{Pgneq} we get 
\begin{equation} \label{Pg1}
  P_{g_1} = -2 (k_\nu^2 - 1) \frac{z}{r_0} \, .
\end{equation}
To get $g_{\nu 1}$ we must do the integral in \eqref{gnsol}. 
The following result is useful:
\begin{equation}
  \frac{1}{r_0^n} e^{-2k_\nu z} \int_{-\infty}^z {\rm d}t \ e^{2 k_\nu t} t^n 
  = \frac{1}{2k_\nu}  \left( \frac{z}{r_0} \right)^n + \cdots \, , \label{gnpoly}
\end{equation}
with the dots indicating terms with powers of $z$ lower than $n$, which we will be dropping systematically 
in what follows.
With this we reproduce the first term of $g_{\nu 1}$ in~\eqref{gnu1}.
Using it into \eqref{Pgneq}, at $n=2$ we have
\begin{equation}
  P_{g_2} = \left( 3 (k_\nu^2 - 1) - 4 (k_\nu^2 -1)^2  \frac{1}{(2k_\nu)^2} \right) \left( \frac{z}{r_0} \right)^2 \, ,
\end{equation}
which upon integration with~\eqref{gnpoly} gives $g_{\nu 2}$ in \eqref{gnu2}.
We can systematize the procedure by writing 
\begin{equation}
  P_{g_n}  = c_{g_n}(k_\nu) \left( \frac{z}{r_0} \right)^n \, , \label{Pgnpoly} 
\end{equation}
and using \eqref{gnpoly} to retain only the highest power of $z$ at any order $n$
\begin{equation} \label{gnunhigh}
  g_{\nu n}  = \frac{c_{g_n}(k_\nu)}{2k_\nu}  \left( \frac{z}{r_0} \right)^n \, .
\end{equation}
The coefficients $c_{g_n}$, which are functions of $k_\nu$, are built iteratively from \eqref{Pgneq}:
\begin{equation} \label{cgn}
  c_{g_n}(k_\nu) = (-)^n (n+1)(k_\nu^2 - 1) - \frac{1}{(2k_\nu)^2} \sum_{j=1}^{n-1} c_{g_j}(k_\nu) c_{g_{n-j}}(k_\nu) \, .
\end{equation}
Here the last sum is present only for $n \geq 2$. The explicit form of $g_{\nu 3}$, which we used to 
compute $f_{\nu 4}$ above, is 
\begin{align}
  g_{\nu 3} & = \frac{k_\nu^2 -1}{2k_\nu}\left( -16\frac{\left(k_\nu^2-1\right){}^2}{(2k_\nu)^4}+
  12\frac{\left(k_\nu^2-1\right)}{(2k_\nu)^2}-4 \right)  \left( \frac{z}{r_0} \right)^3   \, , \label{eqgnu3} 
\end{align}
%

With $g_{\nu n}$ at hand, we can return to the expression for $P_{f_{\nu n}}$ and focus on 
$-\frac{2}{x} \sum_{j=1}^n g_{\nu j} f_{\nu(n-j)}$. 
The last term in this sum, $\frac{2}{x}g_{\nu n} f_{\nu 0} \propto \frac{\ln^n(x)}{x}$, 
does not have the behavior of~\eqref{Pisolate}, and indeed does not feed back into $n=0$, so we drop it.
We are left with
\begin{equation} \label{Pisolateiter}
  P_{f_{\nu n}}(x \to 0) = -\frac{2}{x} \sum_{j=1}^{n-1} g_{\nu j} \ f_{\nu(n-j)} = \frac{3}{r_0^{n-1}} c_{f_{\nu n}}(k_\nu) 
  \frac{\ln^{n-1}(x)}{x^2} \, , \qquad n \geq 2 \, ,
 \end{equation}
where we have introduced the coefficients $c_{f_{\nu n}}(k_\nu)$, which are functions only of $k_\nu$. 
Integrating with~\eqref{Pisolateintegral} we get
\begin{equation} \label{fnuiter}
  f_{\nu n}(x \to 0) =\frac{3}{r_0^{n-1}}  \frac{c_{f_{\nu n}}(k_\nu)}{2 k_\nu} \frac{\ln^{n-1}(x)}{x} \, , \qquad n \geq 2 \, .
\end{equation}
Now we can proceed iteratively. Note~\eqref{Pisolateiter} and~\eqref{fnuiter} start at $n=2$, so the iteration
requires
\begin{equation} \label{cf1}
  c_{f_{\nu 1}} = 2 \, , \qquad  f_{\nu 1} = \frac{3}{k_\nu x} \, , 
\end{equation}
which we computed explicitly above, as the starting point.
Plugging \eqref{gnunhigh}, with $z = \ln(x)$, and \eqref{fnuiter} into \eqref{Pisolateiter} we get the recursive relation
\begin{equation} \label{cfniter}
  c_{f_{\nu n}}(k_\nu) = - \frac{1}{k_\nu^2} \left( c_{g_{n-1}}(k_\nu) +
   \frac{1}{2} \sum_{j=1}^{n-2} c_{g_j}(k_\nu) \ c_{f_{\nu(n-j)}}(k_\nu) \right)\, , 
   \qquad n \geq 2 \, ,
\end{equation}
where we have used \eqref{cf1}. Here the $c_{g_n}$ coefficients are given by \eqref{cgn}.
We have thus managed to formally isolate all the contributions $\Delta^n \ln R_{\nu n} = {\cal O}(\Delta^0)$
which feed back into the leading order. They are given by
\begin{equation}
  U_n \equiv  \Delta^n \frac{3}{r_0^{n-1}}  \frac{c_{f_{\nu n}}(k_\nu)}{2 k_\nu} \int_{e^{-r_0/\Delta}} {\rm d}x \frac{\ln^{n-1}(x)}{x} 
  = 3 r_0 \left(- \frac{(-)^n}{n}  \frac{c_{f_{\nu n}}(k_\nu)}{2 k_\nu} \right) \, .
\end{equation}
The $U$ factor is obtained by summing\footnote{
What we have to perform is a complicated nested sum, given~\eqref{cfniter}, which in turn contains $c_{g_n}$, given
by~\eqref{cgn}. When we include enough orders in $c_{f_{\nu n}}$ for a fixed $n$, the sum turns into
\begin{equation} \label{aleksum}
  \sum_{n=1}^\infty  \left(- \frac{(-)^n}{n}  \frac{c_{f_{\nu n}}(k_\nu)}{2 k_\nu} \right) =
  \frac{1}{k_\nu} \sum_{j = 0}^\infty \frac{(2 j)!}{2^{2 j + 1} (j + 1)! j!} \frac{1}{k_\nu^{2j}} = 
  k_\nu - \sqrt{k_\nu^2 -1} \, .
\end{equation}
} over all of them:
\begin{equation} \label{Ufinal}
  U = \sum_{n=1}^\infty U_n = 3r_0 \left( k_\nu - \sqrt{k_\nu^2 -1} \right) \, .
\end{equation}
With this, we have the full contribution at the leading order,
\begin{align}
  \lim_{x \to \infty} \ln R_\nu(x)  &  = \ln R_{\nu 0}(x\to \infty) + U + {\cal O}(\Delta) \nonumber 
  \\ \label{lnRnuinfty}
  & = \ln \frac{(1-k_\nu) (1-2k_\nu)}{(1+k_\nu) (1+2k_\nu)} + 3r_0 \left(k_\nu - \sqrt{k_\nu^2 - 1}\right) + {\cal O}(\Delta) \, ,
\end{align}
as in~\eqref{lnRnuU}. 
After this long derivation, the result in~\eqref{lnRnuinfty} deserves some comments.
\begin{itemize}
  \item 
  Expression \eqref{lnRnuinfty} holds in any number of dimensions $D$. 
  \item 
  Recall the definition of $k_\nu$ from \eqref{knuApp}
   and remember we are treating $k_\nu$ as a leading order parameter in the expansion in $\Delta$. 
 One could be tempted to expand $R_{\nu}(\infty)$
  for $\Delta \nu \ll 1$,
  \begin{equation}
    e^U  \frac{(1-k_\nu) (1-2k_\nu)}{(1+k_\nu) (1+2k_\nu)} \xrightarrow[\Delta\nu \ll 1]{}  \Delta^2 e^{3r_0} \frac{\nu^2}{12 r_0^2} \, ,
  \end{equation}
  and compare to the result \eqref{Rlownu}. 
  The two do not quite match, and indeed we should not expect so. 
  The point is that the calculation we performed in this section is valid up to ${\cal O}(\Delta)$, as written explicitly 
  in~\eqref{lnRnuinfty}: we cannot expect to get the $\Delta^2$ order correctly and match the 
  result at low $\nu$, \eqref{Rlownu}, which was derived consistently up to that order in the earlier section.
  In other words, there are terms of order $\Delta^2$ in ${\cal O}(\Delta)$ in \eqref{lnRnuinfty} which we have not computed, 
  and do not know how to compute for generic values of $\nu$. Such terms have the form~\eqref{uselessk0z},
  and we have dropped them in our calculation.
  As we argued in section~\ref{secRenorm}, when we perform the sum in $\nu$
   our result \eqref{lnRnuinfty} is sufficient to compute the log of 
  the determinant at the leading order in $\Delta$.
  \item 
  When $\nu \ll 1/\Delta$, $k_\nu = 1 + {\cal O}(\Delta^2)$, and $U = U_1 = 3 r_0$, with $U_n = 0$ for $n \geq 2$.
  This confirms what we anticipated at the end of the low $l$ calculation: at low $l$ (low $\nu$) the only order-breaking 
  contribution comes from $n=1$ and results in the factor $e^{3r_0}$ in $R_l$. 
  For generic values of $\nu$, we have found instead that there are order-breaking contributions for any $n \geq 1$ and 
  we have managed to organize them systematically. 
  \item 
  The contribution to $U$ from $n=1$ is, in some sense, special. It is the only one that survives at low $\nu$, and the only 
  one that originates from the potential, that is from the term
  \begin{equation}
    \frac{1}{x^2} \left( \ddVt - \ddVt_{\text{FV}} \right)_1 = \frac{3}{x^2} \left(1 + \varphi_0 \varphi_1 \right) \, ,
  \end{equation}
  in $P_{f_{\nu 1}}$, see \eqref{Pf1}. 
  From this we got $f_{\nu 1}(x\to 0) = 3/(k_\nu x)$ and $U_1 = 3r_0/k_\nu$. 
  Then, for $n \geq 2$ we just had to iterate \eqref{Pisolateiter} and \eqref{fnuiter}, and we could do it to arbitrary 
  order $n$. 
  The knowledge of $f_{\nu 1}$ was the crucial starting point, then the iteration was automatic. 
  This implies that our result~\eqref{lnRnuinfty} is constructed only based on knowing the bounce up to $n=1$.
  \item
  What is often done in the literature, to compute analytically the fluctuations in the TW limit, is to set $z + r \simeq r$
  in the denominator on the left hand side of~\eqref{k0zexp}, based on the observation that the bounce radius $r$ is large.
  For us, on the contrary, it is crucial to maintain the $z$ dependence in that denominator, which then enters
  in the expansion on the right hand side of~\eqref{k0zexp}. It is from such a $z$ dependence that we build the $U$
  factor, which is crucial to get the full correct result for the fluctuations at the leading order.
  
\end{itemize}


\section{UV Integrals} \label{appUVint}
In the process of renormalization, we encounter integrals of the generic form
\begin{align} \label{eqIn}
  I_m = \int_0^\infty \text{d} \rho \, \rho^{2 m - 1} \left( V^{(2) m} - V^{(2) m}_\text{FV} \right) \, .
\end{align}
This section of the appendix deals with the evaluation of these integrals, starting from the two
explicit cases of $I_{1,2}$ and then with the generic expression, valid for any $m$.
After the usual substitutions for $\rho$ and the bounce, the calculation of $I_1$ goes as
\begin{align}
  I_1 &= \int_{-r}^\infty \text{d} z 
  \left(r + z \right) \left(\frac{3}{2} \left( \varphi^2 - 1\right) - 3 \Delta \right) \,.
\end{align}
The second term in the bracket is of higher-order, so we drop it here to get the leading order
\begin{align}
  I_1^{n=0} = -6 r \int_{0}^\infty \frac{\text{d} x}{x} \frac{x}{\left( 1 + x \right)^2} = -6 r \, .
\end{align}
Here we are allowed to extend the limits of integration from $x = e^{-r}$ to $0$.
Moving on to the higher-order and including $\varphi_1$, we get
\begin{align}
  I_1^{n=1,2} = \Delta \int_{e^{-r}}^\infty \frac{\text{d} x}{x} \left( r + \ln x \right) \left( \frac{6}{1 + x} \right) \, ,
\end{align}
where the $\ln x$ term is $\Delta$ suppressed with respect to $r$ and is thus counted as $n=2$.
These two higher orders are dominated by the lower boundary, which cannot be extended to $x=0$. 
However, we may expand the integrand for small $x$ and simplify the integration
\begin{align}
  I_1^{n=1} &\simeq 6 \Delta r \int_{e^{-r}} \frac{\text{d} x}{x} = 6 \Delta r^2 \, ,
  &
  I_1^{n=2} &= 6 \Delta \int_{e^{-r}} \frac{\text{d} x}{x} \ln x
  = 6 \Delta \int_{-r} \text{d} z \, z = -3 \Delta r^2 \, .
\end{align}
Further including $\varphi_2$ does not bring any additional $1/\Delta$ terms to $I_1$,
they are $\Delta^2$ suppressed.
Combining all of the terms, we get
\begin{align}
  I_1 = -6 r + 3 \Delta r^2 \simeq - 3 \left( 2 - r_0 \right) \left(\frac{r_0}{\Delta}\right) \, ,
\end{align} 
which agrees with the expression in~\eqref{eqI1}.
Note that the $n=0$ part of $I_1$ cancels the divergence of $\ln R_{\nu 0}$, while combining $n = 1, 2$ orders
comes with an additional $r_0$, just like the $U$-factor and it precisely cancels its dependence.
In other words $n=0$ part of $I_1$ constitutes the asymptotic $\ln R_{\nu 0}^a$ and the $n \geq 1$ go into
the $U^a$ part of the subtraction.
Now that we have the complete expression for $I_1$ collected in~\eqref{eqI1}, we can move on.

At the leading order, the $I_2$ UV integral (after all the substitutions) can be written as
\begin{align}
  I_2^{n=0} = -12 r^3 \int_{0}^\infty \frac{\text{d} x}{x} \frac{x \left(1 - x + x^2 \right)}{\left( 1 + x \right)^4} = -6 r^3 \, .
\end{align}
To get to the sub-leading orders, we add $\varphi_1$ into $V^{(2) 2} - V^{(2) 2}_\text{FV} = 12 \Delta (1 - x + 4 x^2)/(1 + x)^3$, which brings in one power of $\Delta$.
We then expand the integrand at small $x$, because that is what dominates the integral, expand 
$\rho^3 = (r + z)^3$ and count each power of $z$ as one order higher, to end up with
\begin{align}
  \begin{split}
    I_2^{n=1} &= \int_{e^{-r}} \frac{\text{d} x}{x} r^3 12 \Delta \left( \frac{1 - x + 4 x^2}{\left( 1 + x \right)^3} \right)
    \\
    & \simeq 12 \Delta r^3 \int_{-r} \text{d} z = 12 \Delta r^4 \, ,
  \end{split}
  \\
  I_2^{n=2} &= 12 \Delta \int_{-r} \text{d} z \, 3 r^2 z  = -18 \Delta r^4 \, ,
  \\
  I_2^{n=3} &\simeq 12 \Delta \int_{-r} \text{d} z \, 3 r z^2 = 12 \Delta r^4 \, ,
  \\
  I_2^{n=4} &\simeq 12 \Delta \int_{-r} \text{d} z \, z^3 = -3 \Delta r^4 \, .
\end{align}
We dropped all the terms sub-leading in $\Delta$ and collected the leading $\Delta^0$ powers.
We now have five orders in $n = 0, \ldots 4$ and again the $n=0$ matches the $\ln R_{\nu 0}^a$, while
all the higher orders combine and give an extra factor of $r_0$ that matches to $U^a$;
altogether, we have
\begin{align}
  I_2^{\sum n} = -6 r^3 + 3 \Delta r^4 \simeq - 3 \left( 2 - r_0 \right) \left(\frac{r_0}{\Delta}\right)^3 \, ,
\end{align}
which is what is given in~\eqref{eqI2}.

The $\tilde I_2$ integral is
\begin{align}
\tilde I_2 & =   \int_0^\infty \text{d} \rho  \ \rho^3 \left( V^{(2)2} - V_{\rm FV}^{(2)2} \right) \left( \frac{1}{\epsilon} + \gamma_E + 1 + \ln \left( \frac{\mu \rho}{2} \right) \right)  \label{I2tdef} \\
& = I_2 \left( \frac{1}{\epsilon} + \gamma_E + 1 + \ln \left( \frac{\mu R}{2} \right) \right) + 
 \int_0^\infty \text{d} \rho  \ \rho^3 \left( V^{(2)2} - V_{\rm FV}^{(2)2} \right) \ln \frac{\rho}{R} \, . \label{I2t}
\end{align}
Here, $R = \frac{r}{\sqrt{\lambda}v} \simeq \frac{r_0}{\sqrt{\lambda}v \Delta}$ is the bubble radius. To perform the integral
with $\ln \rho/R$ we switch to the dimensionless variables,
\begin{equation}
r^3 \int_{-r}^\infty \text{d} z  \left( 1 + \frac{z}{r} \right)^3  \left( \tilde V^{(2)2} - \tilde V_{\rm FV}^{(2)2} \right) \ln\left(1 + \frac{z}{r} \right) \, ,  \label{deltaI2}
\end{equation}
and expand the log term as
\begin{equation}
 \ln\left(1 + \frac{z}{r} \right) = \sum_{n=1}^\infty \frac{(-1)^{n-1}}{n} \left( \frac{z}{r} \right)^n \, .
\end{equation}
It is useful to expand in $\Delta$ the term with the second derivative of the potential squared:
\begin{equation}
\left( \tilde V^{(2)2} - \tilde V_{\rm FV}^{(2)2} \right) = \left( \tilde V^{(2)2} - \tilde V_{\rm FV}^{(2)2} \right)_0 + \Delta \left( \tilde V^{(2)2} - \tilde V_{\rm FV}^{(2)2} \right)_1 + {\cal O}(\Delta^2) \, ,
\end{equation}
where
\begin{align}
\left( \tilde V^{(2)2} - \tilde V_{\rm FV}^{(2)2} \right)_0 & = \frac{3}{4} \left(3 \tanh ^4\left(\frac{z}{2}\right)-2 \tanh
   ^2\left(\frac{z}{2}\right)-1\right) \, , \label{V20} \\
\left( \tilde V^{(2)2} - \tilde V_{\rm FV}^{(2)2} \right)_1 & = -9 \tanh ^3\left(\frac{z}{2}\right)+3 \tanh \left(\frac{z}{2}\right)+6 \equiv h_1(z) \, .   \label{V21}
\end{align}
The zeroth order, in \eqref{V20}, is an even function of $z$ which vanishes exponentially at $z \to \pm \infty$. 
In that case we can then extend the lower limit of integration to $-\infty$ in \eqref{deltaI2}, and the final contribution
is $\Delta^2$ suppressed. 

The first order, in \eqref{V21}, vanishes exponentially at $z \to \infty$, but goes to a constant at $z \to - \infty$:
\begin{equation}
h_1(z \to -\infty) = 12 \, .
\end{equation} 
This gives a contribution at the lower boundary of integration in \eqref{deltaI2}. To compute it we integrate by parts:
\begin{align}
& {}  \Delta r^3  \sum_{n=1}^\infty \frac{(-1)^{n-1}}{n} \int_{-r}^\infty \text{d} z  \left( \frac{z}{r} \right)^n  \left( 1 + \frac{z}{r} \right)^3 h_1(z) \\
& =  \Delta r^3  \sum_{n=1}^\infty \frac{(-1)^{n-1}}{n} \left[ \left. z \left(\frac{z}{r} \right)^n \left( \frac{1}{n+1} + 3\frac{z}{r}\frac{1}{n+2} + 3\frac{z^2}{r^2} \frac{1}{n+3} + \frac{z^3}{r^3} \frac{1}{n+4} \right) h_1(z) \right\vert_{-r}^\infty   \right. \label{line1} \\
& \left.  \qquad \qquad \qquad - \int_{-\infty}^\infty \text{d} z \ z \left(\frac{z}{r} \right)^n \left( \frac{1}{n+1} + 3\frac{z}{r}\frac{1}{n+2} + 3\frac{z^2}{r^2} \frac{1}{n+3} + \frac{z^3}{r^3} \frac{1}{n+4} \right) \frac{\text{d}}{\text{d}z} h_1(z)  \right] \, . \label{line2}
\end{align}
In \eqref{line2} we can extend the lower limit of integration to $-\infty$, as $\frac{\text{d}}{\text{d}z}h_1(z)$ vanishes exponentially at
$z\to \pm \infty$. We can drop the contribution from that line, as it is also $\Delta^2$ suppressed. In \eqref{line1}, the contribution at $\infty$ vanishes because 
of $h_1(z)$, while at $-r$ we have $h_1(-r) \simeq h_1(z\to -\infty) = 12$, so we have
\begin{align}
 &   \Delta r^3  \sum_{n=1}^\infty \frac{(-1)^{n-1}}{n} 
\left[-12  (-r) (-1)^n \left( \frac{1}{n+1} - 3 \frac{1}{n+2} +3 \frac{1}{n+3} - \frac{1}{n+4}  \right)    \right] \\
& = -12 \Delta r^4  \sum_{n=1}^\infty \frac{1}{n}  \left( \frac{1}{n+1} - 3 \frac{1}{n+2} +3 \frac{1}{n+3} - \frac{1}{n+4}  \right) 
 = -\frac{3}{4} r_0 \left( \frac{r_0}{\Delta} \right)^3  \, .
\end{align}
To summarize, we have found the leading order contribution to the second term in \eqref{I2t}:
\begin{equation}
 \int_0^\infty \text{d}\rho  \ \rho^3 \left( V^{(2)2} - V_{\rm FV}^{(2)2} \right) \ln \frac{\rho}{R} =  -\frac{3}{4} r_0 \left( \frac{r_0}{\Delta} \right)^3 \left( 1+ {\cal O}(\Delta) \right).
\end{equation}
In $D=4$, where $r_0 = 1$, this contribution is equal to $I_2/4$, so in the end we get \eqref{eqI2tld}:
\begin{equation}
\tilde I_2 =  I_2 \left( \frac{1}{\epsilon} + \gamma_E + \frac{5}{4} + \ln \left( \frac{\mu }{2 \sqrt{\lambda}v \Delta} \right) \right) \, .
\qquad \qquad (D = 4)
\end{equation}

{\bf General order.} 
We can carry out the above analysis for any order.
First, expand the integrand up to $\Delta$
\begin{align} \label{eqVppm}
  V^{(2) m} - V^{(2) m}_\text{FV} \propto \frac{1}{2^m} \left( 3 \varphi_0^2 - 1 \right)^m - 1 - 3 \Delta m \left(
  \frac{1}{2^{m-1}} \left( 3 \varphi_0^2 - 1 \right)^{m-1} \varphi_0 - 1 \right) \, .
\end{align}
In the first piece, coming from the $\mathcal O(\Delta^0)$ term in~\eqref{eqVppm}, we can extend the lower
limit of integration to zero.
Furthermore, we can expand the power of $\rho^{2m-1} \simeq r^{2m-1}$, the rest is subdominant in $\Delta$
\begin{align}
  I_m^{n=0} &= r^{2 m - 1} \int_0^\infty \frac{\text{d} x}{x} \, \left( \frac{1}{2^m} \left( 3 \varphi_0^2 - 1 \right)^m - 1 \right) \, ,
  &
  \varphi_0 &= \frac{x - 1}{x + 1} \, .
\end{align}
This integral is given as a finite sum of incomplete Beta functions and Harmonic numbers
\begin{align}
  I_m^{n=0} &= \frac{r^{2 m - 1}}{2^{m-1}} \sum_{i=0}^{m} \binom{m}{i} (-)^{i+1} 3^{m - i} \biggl(
  B\left(-1, 2 (m-i), 0 \right) + \log 2 + H_{2(m-i)-1} \biggr) \, ,
\end{align}
here we list some of them from $m = 1 \ldots 10$
\begin{align}
  \frac{I_m^{n=0}}{r^{2 m - 1}}  &= -\left \{ 6, 6, \frac{36}{5}, \frac{264}{35}, \frac{282}{35}, \frac{3222}{385}, \frac{43476}{5005},
  \frac{44736}{5005}, \frac{780462}{85085}, \frac{1378158}{146965} , \ldots \right \} \, .
\end{align}
For the linear term in~\eqref{eqVppm}, we have to keep the lower limit of integration, but we can set 
$x = 0$, or equivalently $\varphi_0 = -1$, in the integrand, such that
\begin{align}
  I_m^{n \geq 1} &= 6 \Delta m \int_{-r} \text{d} z \, \left( r + z \right)^{2 m - 1}
  = 6 \Delta m \sum_{i=0}^{2 m - 1} \binom{2m - 1}{i} r^{2m-1-i} \int_{-r} \text{d} z \, z^i \, 
  \\&
  = 6 \Delta m r^{2m} \sum_{i=0}^{2 m - 1} \binom{2 m - 1}{i} \frac{(-1)^i}{i+1}
  = 6 \Delta m r^{2m} \frac{1}{2 m}\simeq 3 r_0 \left(\frac{r_0}{\Delta} \right)^{2 m - 1} \, .
\end{align}


\section{Quartic functional determinant} \label{appQuartic}
We discuss the calculation of the functional determinant of an (unstable) quartic potential, specified
by $V = \lambda/4 \phi^4$, where $\lambda < 0$. 
It can be computed either by use of Feynman diagrams or via the $\zeta$/WKB formalism.
The bounce is given by $\phi = \sqrt{8/(-\lambda)} R/(R^2 + \rho^2)$ and the ratio of multipoles $R_l$ can be 
calculated using Gel'fand-Yaglom~\cite{Isidori:2001bm, Andreassen:2017rzq, Guada:2020ihz} with $R_l = l(l-1)/(l+2)/(l+3)$.
To compute the determinant, we have to regulate the sum of $d_l \ln R_l$, such that we first subtract the divergent
growth at large $l$ and then renormalize it.

The Feynman diagrammatic approach~\cite{Isidori:2001bm, Andreassen:2017rzq} specifies the subtraction of $\ln R_l$ 
in powers of $V^{(2)} - V^{(2)}_\text{FV}$ insertions.
The renormalization in $D = 4$ is performed by calculating the Feynman diagrams with one and two insertions
in momentum space, using the $\overline{\text{MS}}$ scheme.
After performing the finite sum and adding back the renormalized determinant, one gets
\begin{align} \label{eqFuncDetQFeyn}
  -\frac{1}{2} \ln \left(\frac{\det^{\prime} \mathcal O}{\det \mathcal O_{\text{FV}}} \right) =
  \frac{3}{\epsilon} - \frac{5}{4} \left(1 - 2 \ln \left(\frac{5}{6} \right) \right) + 3 \ln \left(\frac{\mu  R}{2} \right) + 6 \zeta '(-1) \, ,
\end{align}
as stated e.g. in Eq.~(5.6) of~\cite{Andreassen:2017rzq} (here we translated to our convention $\varepsilon = 4 - D$).

According to the $\zeta$/WKB prescription, we use~\eqref{eqLnDetRenD4} with the UV integrals coming from integrating 
$V^{(2)} = 3 \lambda \phi^2$ using the quartic bounce given above, such that
\begin{align}
  I_1 &= -12 \, , &  I_2 &= -48 \, ,
  &
  \tilde I_2 &= 48 \left( \frac{1}{\varepsilon} + 1 + \gamma_E + \ln \left(\frac{\mu  R}{2} \right) \right) \, .
\end{align}
The finite sum is performed by starting from $l = 2$ (or $\nu = 3$)
\begin{align}
  S_{\text{fin}} = \sum_{\nu=3}^\infty \nu^2 \left( \ln R_l + \frac{6}{\nu} + \frac{6}{\nu^3} \right) = 
  - \frac{37}{2} + 9 \ln 2 + 5 \ln 3 - 12 \zeta'(-1) \, .
\end{align}
To complete the calculation, we have to add the $l=0$ and $l=1$ terms.
This includes the $\ln R_l$ with zeroes removed, given by $R^{\prime}_0 = -1/5$ and $R^{\prime}_1 = 1/10$ (see e.g~(4.37) 
of~\cite{Andreassen:2017rzq}), as well as the asymptotic subtractions, such that
\begin{align}
  -\frac{1}{2} \ln \left(\frac{\det^{\prime} \mathcal O}{\det \mathcal O_{\text{FV}}} \right) =
  -\frac{1}{2} \left( \ln |R^\prime_0| + 4 \ln R^\prime_1 + \sum_{\nu=1}^2 \left( 6 \nu + \frac{6}{\nu} \right) + 
  S_{\text{fin}} - \frac{1}{8} \tilde I_2 \right) \, .
\end{align}
Although the intermediate steps in these two approaches differ significantly, the final result precisely agrees 
with~\eqref{eqFuncDetQFeyn}.
This further confirms that a single integration over $\rho$ in~\eqref{eqLnDetRenD4} is equivalent to using
Feynman diagrams in the $\overline{\text{MS}}$ scheme.

%
%
 \section{Finite determinants in generic dimensions} \label{appFinSumGenD}
The procedure that we discussed for the specific $D = 3, 4$ above, deriving the asymptotics, subtracting 
the infinities and applying the EuMac approximation, can be generalized for generic $D$.
We first focus on the asymptotic behavior of the determinant that needs to be subtracted to regulate the 
sum, then on the sums in generic $D$.
We treat odd and even dimensions in separate sub-sections, starting with the easier odd ones.
Some parts of the summation are valid for any $D$, so we will keep the discussion general for as long
as possible and reduce to a specific choice when this becomes unavoidable.

{\bf Asymptotics of determinants.}
We will derive the asymptotic behavior of the determinant for high multipoles in inverse powers of $\nu$.
To compute the finite sum in~\eqref{eqDefSigmaD}, we need to obtain $\ln R_\nu^a$ that regulates the 
infinites with a minimal number of $\nu^{-j}$ terms.
The order depends on the dimension, because the degeneracy factor at high $\nu$ scales as 
$d_\nu \propto \nu^{D-2}$.
We thus have to expand to high enough powers of $\nu^{-j}$,  to get to the lowest 
one, which is $d_\nu \ln R_\nu \sim \nu^{-1}$.
This defines the notion of minimal subtraction.

To keep track of the calculation, we will separate the asymptotic into two pieces: the logarithmic
one $\ln R_{\nu 0}^a$, and the $U$ factor, such that $\ln R_\nu^a = \ln R_{\nu 0}^a + U^a$.
In what follows we will be evaluating the integrals by performing a large $k_\nu$ expansion of
$\ln R_\nu$, integrating the pieces of the sums and summing them back up.
For this purpose, it is sensible to first expand in large $k_\nu$, which is by definition larger than
one, thus the sums with $k_\nu^{-j}$ converge, and then further expand in powers
of $\nu^{-j}$ or better $y^{-j}$.
The last step brings us to the well defined $1/\nu$ expansions, which we get during the process 
of renormalization.

Let us start with the logarithmic piece and rewrite the log of~\eqref{eqRnu0inf} as:
\begin{align} \label{eqLnRn0Expk0}
  \ln R_{\nu 0} &
  = -2 \sum_{j=1}^\infty \frac{1 + 2^{- 2 j + 1}}{2 j - 1}k_\nu^{- 2 j + 1} \, ,
\end{align}
where the powers of $k_\nu$ are going down by factors of 2.
Further expanding the powers of $k_\nu$ in various roots of large $y \gg 1$, using
\begin{align}
  k_\nu^{- 2 j + 1} = y^{- 2 j + 1} \left( 1 + y^{-2} \right)^{- j + \frac{1}{2}} 
  = y^{- 2 j + 1} \sum_{p=0}^\infty \binom{- j + \frac{1}{2}}{p} y^{-2p} \, ,
\end{align}
we then express it as a double infinite sum
\begin{align} \label{eqLnRn0Expy}
  \ln R_{\nu 0} & = -2 \sum_{j=1}^\infty \frac{1 + 2^{-2 j + 1}}{2 j - 1} \sum_{p=0}^\infty \binom{-j +\frac{1}{2}}{p} y^{-2(j+p)+1} \, .
\end{align}
As discussed above, the powers in the asymptotic functions have to stop when we reach $\nu^{-1}$.
To derive the upper bound in the sum, note that the degeneracy factor goes as $d_\nu \propto y^{D-2}$ 
in the UV and we need to remove all the negative $y$ powers, including the highest one with $y^{-1}$.
Comparing the powers leads to the maximal one at $j = m$, for both odd and even dimensions, 
where $D = 2 m + 1$ for the odd ones and $D = 2 m$ for the even, thus
\begin{align} \label{eqLnRnu0Asm}
  \ln R_{\nu 0}^a & = -2 \sum_{j=1}^m \frac{1 + 2^{-2 j + 1}}{2 j - 1} \sum_{p=0}^{m - j} \binom{-j +\frac{1}{2}}{p} y^{-2(j+p)+1} \, .
\end{align}
We proceed along similar lines to obtain the asymptotics of the $U$ factor
\begin{align} \label{eqUExpk0}
  U &= - 3 r_0 \sum_{j=1}^\infty \binom{\frac{1}{2}}{j}(-)^j k_\nu^{-2 j + 1}
  =-3 r_0 \sum_{j = 1}^\infty \binom{\frac{1}{2}}{j} (-)^j y^{-2j+1}\sum_{p = 0}^\infty \binom{-j+\frac{1}{2}}{p}y^{-2p} \, ,
\end{align}
and truncating the series at the appropriate maximal power
\begin{align} \label{eqUAsm}
  U^a &= -3 r_0 \sum_{j=1}^m \binom{\frac{1}{2}}{j} (-)^j y^{-2j+1}\sum_{p=0}^{m-j}\binom{-j+\frac{1}{2}}{p}y^{-2p} \, .
\end{align}  
With these minimal subtractions in powers of $y$, the sum in~\eqref{eqDefSigmaD} becomes convergent 
and can be evaluated using the EuMac approximation in~\eqref{eqDefEM}.
Plugging the appropriate boundaries in~\eqref{eqLnRnu0Asm} and~\eqref{eqUAsm}, we get precisely
the asymptotic functions $\ln R_\nu^a$, which were calculated in $D=3,4$ via the integrals 
$I_{1,2}$ in~\eqref{eqI1} and~\eqref{eqI2}.

From the generic behavior of the asymptotics, we already see that the odd and even $D$ will
behave differently.
Because the degeneracy factor grows as $y^{D-2}$ and the UV expansion of $\ln R_{\nu0} + U$ behave as 
$y^{-2j+1}$, we get $y^{-1}$ terms only for even $D$.
After the summation in $\nu$, these become $\ln y$ divergent terms.
Conversely, there are no log divergencies in odd dimensions, which makes them
generally easier to handle.

%
%
{\bf Generic odd $D$.} \label{appFinSumOddD}
Let us generalize the summation procedure to generic $D$, beginning with an integral part of the 
$U$-factor in the EuMac approximation
\begin{align} \label{eqSigmaOdd}
  \Sigma_D^{\int U} &= \int_{\nu_0}^\infty \text{d} \nu \, d_\nu \left( U - U^a \right) \, .
\end{align}  
We found out that in $D=3$ the EuMac integral was insensitive to the lower bound/multipoles.
We shall argue that this is true for any odd $D$.
The dominant contribution to the integral comes from $\nu \sim 1/\Delta$ terms, where
the degeneracy factor is approximated as
\begin{align} \label{eqDlApprxOdd}
  d_\nu &= \frac{2^{4-D} \nu}{(D-2)!} \prod_{j=1}^{\frac{D-3}{2}} \left( (2\nu)^2 - (2 j - 1)^2 \right) 
  \simeq \frac{2}{(D-2)! } \nu^{D-2} \, .
\end{align}
In the $\nu \sim 1/\Delta$ range, the $2 j -1$ terms above are $\Delta^2$ suppressed with respect to $\nu^2$.
Using this approximation of $d_\nu$, we get the following EuMac integral
\begin{align} \label{eqSigmaUOddDef}
  \Sigma_D^{\int U} &= \frac{2}{(D-2)!}
  \int_{\nu_0}^\infty \text{d} \nu \, \nu^{D-2} \left( U - U^a \right)
  \\
  &= \frac{2}{(D-2)!}  \left(\frac{r_0}{\Delta} \right)^{D-1} 
  \int_{\sqrt{1 + y_0^2}}^\infty \text{d} k_\nu \, k_\nu(k_\nu^2-1)^{\frac{D - 3}{2}} \left( U - U^a \right) \, ,
\end{align}  
where we will use the two expansions for $U$ in~\eqref{eqUExpk0} and the asymptotic $U^a$ in~\eqref{eqUAsm}.

After recasting the integrands $U, U^a$ into sums over $j$, it becomes clear that we can 
divide~\eqref{eqSigmaUOddDef} into a lower part with $j \in [1, m]$ that contains the 
subtraction of $U^a$ and the high region $j \in[m+1, \infty]$, which remains unsubtracted.
The high region is given by
\begin{align} \label{eqUnSubgenD}
    \Sigma_{D \text{ uns}}^{\int U}  &= -\frac{6 r_0}{(D-2)!}  \left(\frac{r_0}{\Delta} \right)^{D-1} \sum_{j = m + 1}^\infty(-)^j \binom{\frac{1}{2}}{j}
  \int_{\sqrt{1+y_0^2}}^\infty \text{d} k_\nu \, k_\nu^{-2j+2}(k_\nu^2-1)^{\frac{D - 3}{2}} 
  \\
  &= \frac{6 r_0}{(D-2)!}  \left(\frac{r_0}{\Delta} \right)^{D-1}  \frac{\left( D - 2 \right)! \left( m - \frac{D}{2} \right)!}{2^D \left(\frac{D}{2}\right)! m!} \, .
\end{align}
This result is already valid for any $D$ and does not depend on the lower boundary $y_0$.
The subtracted part contains a double sum coming from the lower part of 
the sum over $j$ with $j \in [1, m]$.
When we take the difference $U - U_a$, the lower part of the sum over $p$ is removed, and we are left 
with the task of having to evaluate
\begin{align} \label{eqUSubgenD}
\begin{split}
  \Sigma_{D \text{ sub}}^{\int U} 
  &= -\frac{6 r_0}{(D-2)!}  \left(\frac{r_0}{\Delta} \right)^{D-1} 
  \sum_{j=1}^m \binom{\frac{1}{2}}{j} (-)^j \sum_{p = m - j + 1}^\infty\binom{-j+\frac{1}{2}}{p} \times
  \\
  &\int_{\sqrt{1 + y_0^2}}^\infty \text{d} k_\nu \, k_\nu \left(k_\nu^2 - 1 \right)^{\frac{D}{2} - j - p - 1} \, .
\end{split}
\end{align}
Let us focus first on the integral and perform the sum over $p$
\begin{align} \label{eqIntSump}
  &\sum_{p = m - j + 1}^\infty\binom{-j+\frac{1}{2}}{p} 
  \int_{\sqrt{1 + y_0^2}}^\infty \text{d} k_\nu \, k_\nu \left(k_\nu^2 - 1 \right)^{\frac{D}{2} - j - p - 1}
  \\
  &= \sum_{p = m - j + 1}^\infty\binom{-j+\frac{1}{2}}{p} \frac{y_0^{D - 2 (j+p)}}{D - 2(j + p)} 
  \\ \label{eqIntSumpGenD}
  \begin{split} 
  & =\binom{\frac{1}{2}-j}{-j+m+1} \frac{-y_0^{D - 2 m - 2}}{D-2 m-2} \times
  \\
  &\quad \,_3F_2\left(1,m+\frac{1}{2},m+1-\frac{D}{2};m+2-\frac{D}{2},m+2-j;-\frac{1}{y_0^2}\right) \, .
  \end{split}
\end{align}
The result is also valid for any $D$, however, the dependence on $y_0$ differs from odd to even.
Therefore, we need to specify that we are working in odd dimensions, so we set $D = 2 m + 1$ and 
expand for small $y_0$ to end up with a simple coefficient
\begin{align} \label{eqSumOverpUOddD}
  \eqref{eqIntSumpGenD} \xrightarrow[y_0 \to 0]{D=2m+1} \frac{(-)^m}{m \binom{m-j+\frac{1}{2}}{m}} \, .
\end{align}
Now we come back to evaluate the remaining outer sum over $j$, which gives
\begin{align}
  \Sigma_{2m + 1 \text{ sub}}^{\int U} &= -\frac{6 r_0}{(D-2)!}  \left(\frac{r_0}{\Delta} \right)^{D-1} 
  \sum_{j = 1}^m \binom{\frac{1}{2}}{j} \frac{(-)^{j+m}}{m \binom{m-j+\frac{1}{2}}{m}} 
  \\
  &= -\frac{6 r_0}{(D-2)!}  \left(\frac{r_0}{\Delta} \right)^{D-1}
  \frac{(-)^m}{m \binom{m+\frac{1}{2}}{\frac{1}{2}}} \left( (m+1) \binom{\frac{1}{2}}{m+1} - \frac{1}{2} \right) \, .
\end{align}
Combining the subtracted sum with the unsubtracted part in~\eqref{eqUnSubgenD}, we get an elegant 
expression for the $U$ part in odd dimensions
\begin{align} \label{eqSigOddIntUFin}
  \Sigma_{2m+1}^{\int U} = \frac{2}{(D-2)!} \left(\frac{r_0}{\Delta} \right)^{D-1}
  \frac{(-)^m 3 r_0}{2 m \binom{m+\frac{1}{2}}{\frac{1}{2}}} \, .
\end{align}

Now we understand why the integral is dominated by $\nu \sim 1/\Delta$: the lowest multipoles come with 
higher orders, expanded in $y^2$ and would be $y_0^2$ or $\Delta^2$ suppressed.
This justifies the approximation in~\eqref{eqDlApprxOdd}. 
If we were to keep the additional constant terms
next to $\nu^2$ in~\eqref{eqDlApprxOdd}, they would be $\Delta^2$ suppressed once we reach the 
intermediate multipoles with $\nu \sim 1/\Delta$.

Let us show that this is already the final answer and discuss why the additional EuMac terms in~\eqref{eqDefEM} 
do not contribute.
Inserting the lower boundary with $\nu_0, d_\nu \sim \mathcal O(1)$ and $y_0 \sim \mathcal O(\Delta)$, we get 
the following orders in $\Delta$
\begin{align} \label{eqSigmaLowOdd}
  \sigma_{2m+1}(\nu_0) \propto \mathcal O(1) \left( \sqrt{\Delta^2 + 1} - \Delta - 
  \sum_{j=1}^m \binom{\frac{1}{2}}{j} \Delta^{-2 j + 1} \right) \, .
\end{align}
Remember that the leading order result in~\eqref{eqSigmaOdd} goes as $\Delta^{-D+1}$ or equivalently 
$\Delta^{-2m}$.
On the other hand, the term with the most negative power of $\Delta$ in~\eqref{eqSigmaLowOdd} 
is still $\Delta^{-2m+1}$, which is suppressed compared to the leading one.
The same argument goes through for the higher derivatives of $\sigma_{2m+1}(\nu_0)$:
when we multiply the last term with the degeneracy factor, we will at most get a term $\propto \nu$.
Taking higher derivatives of $\nu$ just makes it less important and therefore the Bernoulli terms are negligible.

Let us proceed with the evaluation of the $\ln R_{\nu 0}$ term using the same line of reasoning.
Using the expansions in~\eqref{eqLnRn0Expk0},~\eqref{eqLnRn0Expy} and~\eqref{eqLnRnu0Asm}, we again 
split the sum into two pieces,
\begin{align}
  \Sigma_{2 m + 1}^{\int \ln} &= 
  \frac{2}{(D - 2)!} \left(\frac{r_0}{\Delta} \right)^{D - 1} \int_{y_0}^\infty \text{d} y \, 
  y^{D - 2} \left( \ln R_{\nu 0} - \ln R_{\nu 0}^a \right) 
  \\
  &=  \frac{2}{(D - 2)!} \left(\frac{r_0}{\Delta} \right)^{D - 1}
  \int_{y_0}^\infty \text{d} y \, y^{D-2} \left( \sum_{j = 1}^m \ldots + \sum_{j = m + 1}^\infty \ldots \right) \, ,
\end{align}
where the first part in the parenthesis is the subtracted one and the second part is the unsubtracted one,
independent of $\ln R_{\nu 0}^a$.
The latter part is again simpler, valid for any $D$ and independent of $y_0$
\begin{align} \label{eqSigmaOddLogUns}
\begin{split}
  \Sigma_{D \text{ uns}}^{\int \ln} &= \frac{2}{(D-2)!} \left(\frac{r_0}{\Delta} \right)^{D-1}
  \frac{1}{(D-1)(2m-1) \binom{m-\frac{D}{2}}{m-\frac{3}{2}}}
  \times
  \\
  &\left(1+ \frac{D - 1}{4^{m+1} (m+\frac{1}{2})} 
  \,_2 F_1\left(1, m + 1 - \frac{D}{2}; m+\frac{3}{2}; \frac{1}{4} \right) \right) \, . 
\end{split}
\end{align}  
The subtracted piece is slightly more involved.
Taking the difference of $\ln R_{\nu0}$ and $\ln R_{\nu0}^a$ in the subtracted region, where $j \in [1, m]$, gives us
\begin{align}
  \Sigma_{D \text{ sub}}^{\int \ln} &= -\frac{4}{(D-2)!} \left(\frac{r_0}{\Delta} \right)^{D-1}
  \sum_{j=1}^{m} \frac{1 + 2^{-2 j + 1}}{2 j - 1}  \sum_{p = m - j + 1}^\infty\binom{-j+\frac{1}{2}}{p} \times 
  \\
  & \qquad \qquad \qquad \qquad \qquad \qquad \int_{\sqrt{1+y_0^2}}^\infty \text{d} k_\nu \, k_\nu(k_\nu^2-1)^{\frac{D}{2} - j - p - 1} \, ,
\end{align}
which is very similar to the subtracted part of the $U - U^a$ integration in~\eqref{eqUSubgenD}.
Indeed, the integral and the sum over $p$ are exactly the same as in~\eqref{eqIntSump} and once again
we have to have to separate the odd ones by setting $D = 2 m + 1$ and expanding \eqref{eqIntSump} for 
small $y_0$ to get~\eqref{eqSumOverpUOddD}.
What remains is a slightly different sum over the $j$ index
\begin{align}
  \Sigma_{2m + 1 \text{ sub}}^{\int \ln} &= -\frac{4}{(D-2)!} \left(\frac{r_0}{\Delta} \right)^{D-1} \frac{(-)^{m}}{m}
  \sum_{j = 1}^m \frac{1 + 2^{-2 j + 1}}{(2 j - 1) \binom{m-j+\frac{1}{2}}{m}}
  \\ \label{eqSigmaOddLogSub}
  \begin{split}
  &= \frac{4}{(D-2)!} \left(\frac{r_0}{\Delta} \right)^{D - 1}
  \frac{(-)^{m+1}}{2 m^2} \biggl( 
  \frac{1 + m \,_2 F_1 \left(1, m - \frac{1}{2}; \frac{3}{2}; \frac{1}{4}\right)}{\binom{m}{m-\frac{1}{2}}} 
  \\
  & \qquad \qquad \qquad \qquad \qquad \qquad \qquad
  - \frac{1 + 4^{-m} \frac{m}{2m+1} \,_2 F_1 \left( \frac{1}{2}, 1; m+ \frac{3}{2}; \frac{1}{4} \right)}{\binom{-\frac{1}{2}}{m}} \biggr) \, .
  \end{split}
\end{align}
To wrap up the $\ln R_\nu$ integration, we combine the subtracted part with the unsubtracted piece 
from~\eqref{eqSigmaOddLogUns}, where we set $D = 2 m + 1$, and end up with the following result
\begin{align}
  \Sigma_{2m+1}^{\int \ln} &= \Sigma_{2m + 1 \text{ uns}}^{\int \ln} + \Sigma_{2m + 1 \text{ sub}}^{\int \ln}
  \\ \label{eqSigmaOddLog}
  &=\frac{2}{(D-2)!} \left(\frac{r_0}{\Delta} \right)^{D-1} \frac{(-)^{m+1} \left(1 + m 
  \,_2 F_1 \left(1, \frac{1}{2} - m; \frac{3}{2}; \frac{1}{4} \right)  \right) }{m^2 \binom{m-\frac{1}{2}}{m}} \, .
\end{align}

Before moving on to even dimensions, let us discuss the EuMac corrections for the log terms.
The reasoning here goes along the same lines as for the $U$-factor above.
The lowest power of $\Delta$ comes from taking the ultimate term with $p = m - j$ in the asymptotic 
$\ln R_{\nu 0}^a$, given in~\eqref{eqLnRnu0Asm}, such that
\begin{align}
  \ln R_{\nu 0}^a & \ni \frac{-2}{y^{2 m - 1}} \sum_{j=1}^m \frac{1 + 2^{-2 j + 1}}{2 j - 1} \binom{-j +\frac{1}{2}}{- j + m}
  \\ \label{eqLnRnuMaxAsmp}
  &=  -\frac{2}{y^{2 m - 1}} \frac{m}{m - \frac{1}{2}} \binom{\frac{1}{2}}{m} \left( 1 + \left(m - \frac{1}{2} \right) 
  \,_2 F_1 \left(1, 1 - m; \frac{3}{2}; \frac{1}{4} \right) \right) \, .
\end{align}
Now we plug in the low values of $\nu_0$ and perform the same $\Delta$ counting as in~\eqref{eqSigmaLowOdd},
to realize again that the term with the most negative power of $\Delta$ goes as $\Delta^{-2m+1}$.
This is suppressed compared to the leading one with $\Delta^{-2m}$ and the higher derivatives even more so,
thus all the EuMac corrections are irrelevant here.

To summarize the situation for the odd $D$, we combine the integral of the log given by $\Sigma_{2 m + 1}^{\int \ln}$ 
in~\eqref{eqSigmaOddLog} with the integral of the $U$-factor given by $\Sigma_{2m+1}^{\int U}$ 
from~\eqref{eqSigOddIntUFin}.
Perhaps surprisingly, the final expression comes out to be remarkably simple
\begin{align}
  \Sigma_{2 m + 1} &= \Sigma_{2 m + 1}^{\int U} + \Sigma_{2 m + 1}^{\int \ln} 
  \\
  &= \frac{2}{(D - 2)!} \left(\frac{r_0}{\Delta} \right)^{D - 1} \frac{(-)^{m + 1}}{m \binom{m - \frac{1}{2}}{m}}
  \left(\frac{1}{m} + \! \,_2 F_1 \left(1, \frac{1}{2} - m; \frac{3}{2}; \frac{1}{4} \right) - \frac{3}{2} r_0 \right) \, ,
\end{align}
Evaluating the hypergeometric function gives us the following numbers for a couple of low odd $D$s
\begin{align}
  \Sigma_D = \frac{2}{(D-2)!} \left(\frac{r_0}{\Delta} \right)^{D-1} 
  \begin{cases}
  \frac{5}{6}  +\frac{3}{8}  \ln 3  \, , 					& D = 3 \, ,
  \\
  -\frac{37}{240} - \frac{9}{64} \ln 3   \, , 		& D = 5 \, ,
  \\
 -\frac{3}{224} + \frac{9}{128} \ln 3  \, , 	& D = 7 \, .
  \end{cases}
\end{align}

%
%
{\bf Generic even $D$.} \label{appFinSumEvenD}
The main difference for generic even dimensions is the appearance of a log term on the lower boundary 
and thus the importance of low multipoles in $\ln R_\nu^a$.
This translates into the fact that the EuMac sum over the Bernoulli numbers diverges if we apply the EuMac 
approximation na\"ively on the entire interval from $\nu_0 = \mathcal O(1)$ to $\infty$, as we explained
in section~\ref{secRenorm}.

Let us discuss how these issues appear and how to resolve them in general $D = 2m$.
In even dimensions, the asymptotic parts will always contain a term that goes as $1/\nu$.
When we take higher derivatives in the last term of~\eqref{eqDefEM} to sum over the Bernoulli numbers,
the $1/\nu$ will turn into a term that goes as $\nu^{-(p>1)}$.
If we plug in the lower boundary with $\nu = D/2-1$, the sum over the Bernoulli numbers will eventually diverge.

A way out of this impasse is the same as in $D = 4$: split up the sum over $\nu$ into two parts, the first 
coming from $\nu_0 = \mathcal O(1)$ up to $\nu_1 \lesssim r_0/\Delta$ and the other from $\nu_1$ to $\infty$.
As we shall see, the precise position of this separation is irrelevant, as long as $\nu_1 \gg 1$.
Let us now consider the low part of the sum in more concrete detail.
Here we need to extract the leading powers of $\Delta$ from the asymptotic terms $\ln R_{\nu0}^a$ and $U^a$,
which means we have to set $p = m - j$ and perform the sum over $j$, just as in~\eqref{eqLnRnuMaxAsmp}.
This gives us the most relevant low multipole terms, which can be written as
\begin{align} \label{eqSigmaULow}
  \sigma_{2m} &\simeq \frac{d_\nu}{y^{2 m - 1}} C_m \, ,
\end{align}
where the $\nu$-independent coefficient $C_m$ is given by
\begin{align} \label{eqDefCm}
  C_m &= \frac{2 m}{m - \frac{1}{2}} \binom{\frac{1}{2}}{m} \left( 1 + \left(m - \frac{1}{2} \right) 
  \,_2 F_1 \left(1, 1 - m; \frac{3}{2}; \frac{1}{4} \right) \right) + 3 r_0 \, \binom{\frac{1}{2}}{m} \, .
\end{align}
The sum over low multipoles gives
\begin{align}
  \Sigma_{2 m}^{\text{low} U} &= 
  \sum_{\nu = m - 1}^{\nu_1} \sigma_{2m} = 
  \frac{2 C_m}{(D-2)!} \left(\frac{r_0}{\Delta} \right)^{D-1}
  \prod_{j = 1}^{2 m - 3} \frac{\nu - m + 1 + j}{\nu^{2 m - 2}} \, .
\end{align}
We can further rework it into a product of pairs and ultimately into a sum
\begin{align}
  \prod_{j = 1}^{2m-3} \frac{\nu - m + 1 + j}{\nu^{2 m - 2}} 
  = \frac{1}{\nu^{2 m - 3}}\prod_{j = 1}^{m - 2} \left( \nu^2 - j^2 \right)
  = \frac{1}{\nu} \sum_{j = 0}^{m - 2} \anynom{m - 2}{j} (-)^j \nu^{-2 j} \, ,
\end{align}
where the curly binomial stands for the multiply nested sum
\begin{align}
  \anynom{m-2}{j} &\equiv \sum_{i_1 = 1}^{m - 2} \sum_{i_2 = i_1 + 1}^{m - 2} \ldots 
  \sum_{i_j = i_{j - 1}+1}^{m - 2} \left(i_1 \ldots i_j \right)^2 \, .
\end{align}
This is just a convenient way to write the numerical coefficients, which are independent of $\nu$.
The finite sums of powers of $1/\nu^{2p+1}$, on the other hand, are given by
\begin{align}
  \sum_{\nu = m-1}^{\nu_1} \frac{1}{\nu^n} = 
  \begin{cases}
    \ln \nu_1 - \psi(m-1) \, ,  & n = 1 \, ,
    \\
    - \frac{1}{(n-1)!} \psi^{(n-1)}(m-1) \, , & n \geq 1 \, ,
  \end{cases}
\end{align}
where $\psi^{(n)}$ is the $n$-th derivative of the digamma function.
We thus end up with
\begin{align} \label{eqLowSum}
\begin{split}
  \Sigma_{2 m}^{\text{low}} = &\frac{2 C_m }{(D-2)!} \left(\frac{r_0}{\Delta} \right)^{D - 1} 
  \biggl(\ln \nu_1 - \psi(m - 1) +
  \sum_{j = 1}^{m - 2} \anynom{m - 2}{j} \frac{(-)^{j + 1}}{( 2 j )!} \psi^{(2 j)}(m - 1) \biggr) \, .
\end{split}  
\end{align}
%
%

With the low-$\nu$ summation at hand, we can proceed to the second part of the sum over high multipoles
$\nu \in [\nu_1, \infty]$ via the usual EuMac approximation in~\eqref{eqDefEM}.
Because we delayed the lower end of the sum to $\nu_1 \gg 1$, we do not have to worry 
about EuMac corrections in~\eqref{eqDefEM}.
All of the terms from the lower boundary at $\nu_1$ are $\Delta$ suppressed and we can drop them safely.

Most of the work needed for the high-multipole sum has already been done in the previous section, 
so we can easily handle the EuMac integrals by specifying the even $D = 2 m$ in the generic 
terms derived above.
Let us begin with the simpler $U$-factor and augment the unsubtracted part in~\eqref{eqUnSubgenD} 
to even dimensions, such that
\begin{align} \label{eqUnSubEvenD}
    \Sigma_{2m \text{ uns}}^{\int U} &= \frac{6 r_0}{(D-2)!} \left(\frac{r_0}{\Delta} \right)^{D-1}  
    \frac{\left(2 m - 2 \right)! }{4^m m!^2} \, .
\end{align}
Moving on to the subtracted part, we take the expression from~\eqref{eqIntSumpGenD}, specify 
$D = 2 m$, shift the lower bound from $y_0 \to y_1$ and expand for smallish $y_1$
\begin{align} \label{eqSumOverpUEvenD}
  \eqref{eqIntSumpGenD} \xrightarrow[y_1 \ll 1]{D = 2 m} 
  \frac{j - m - 1}{2 m - 1} \binom{\frac{1}{2}-j}{m + 1 - j} \left(H_{m-j}-H_{m-\frac{3}{2}}+ 2 \ln y_1 \right) \, ,
\end{align}
which leaves us with the following finite sum
\begin{align}
\begin{split}
  \Sigma_{2 m \text{ sub}}^{\int U} = \frac{6 r_0}{(D-2)!}  &\left(\frac{r_0}{\Delta} \right)^{D-1} 
  \sum_{j = 1}^m \binom{\frac{1}{2}}{j} \frac{(-)^j (m + 1 - j)}{2 m - 1} \times
  \\
  & \binom{\frac{1}{2} - j}{m + 1 - j} \left(H_{m - j} - H_{m - \frac{3}{2}}+ 2 \ln y_1 \right) \, .
\end{split}
\end{align}
While we do not get a closed-form expression when summing the $H_{m-j}$ term, 
the two remaining terms, which are independent of $j$, nicely sum into
\begin{align}
  \Sigma_{2 m \text{ sub}}^{\int U} \ni - \frac{6 r_0}{(D-2)!} \left(\frac{r_0}{\Delta} \right)^{D - 1} 
  \binom{\frac{1}{2}}{m} \left(\frac{1}{2} H_{m-\frac{3}{2}} - \ln y_1 \right) \, .
\end{align}
This again shows that when we combine with the low part of the sum in~\eqref{eqLowSum}, 
the $\nu_1$ parameter cancels out (i.e. the part that comes from the $U$-factor of $C_m$),
so that the final result does not depend on the precise position where we split the sum.

The two final remaining pieces come from the integration of the log terms.
The unsubtracted one is
\begin{align} \label{eqSigmaEvenLogUns}
  \Sigma_{2m \text{ uns}}^{\int \ln} &= -\frac{1}{(D-2)!} \left(\frac{r_0}{\Delta} \right)^{D-1} \left(
    \frac{1}{\left(m - \frac{1}{2} \right)^2} + \frac{\,_2 F_1 \left(1, 1; m + \frac{3}{2}; \frac{1}{4} \right)}{
    4^{m + \frac{1}{2}} \left(m + \frac{3}{2} \right) \left(m + \frac{1}{2} \right)} \right) \, .
\end{align}  
The subtracted part is very similar to the $U$-factor above, where the sum over $p$
was given in~\eqref{eqSumOverpUEvenD}.
Here the coefficient enters into the sum defined by $\ln R_{\nu 0}^a$ in~\eqref{eqLnRnu0Asm}
\begin{align}
\begin{split}
  \Sigma_{2 m \text{ sub}}^{\int \ln} = \frac{4}{(D-2)!}  &\left(\frac{r_0}{\Delta} \right)^{D-1} 
  \sum_{j = 1}^m \frac{1 + 2^{-2 j + 1}}{2 j - 1} \frac{m + 1 - j}{2 m - 1} \times
  \\
  & \binom{\frac{1}{2}-j}{m+1-j} \left(H_{m-j} - H_{m-\frac{3}{2}}+ 2 \ln y_1 \right) \, .
\end{split}
\end{align}
Just as in the case of the $U$-factor, we do not find a closed-form expression when summing over the $H_{m-j}$
term, while the last two terms sum into
\begin{align}
\begin{split}
  \Sigma_{2 m \text{ sub}}^{\int \ln} \ni -\frac{4}{(D-2)!} & \left(\frac{r_0}{\Delta} \right)^{D - 1} 
  \frac{m}{m - \frac{1}{2}} \binom{\frac{1}{2}}{m} \left(\frac{1}{2} H_{m-\frac{3}{2}} - \ln y_1 \right)  \times
  \\
  &\left( 1 + \left(m - \frac{1}{2} \right) \,_2 F_1 \left(1, 1 - m; \frac{3}{2}; \frac{1}{4} \right) \right) \, .
\end{split}
\end{align}
Let us compare this to the first part of the $C_m$ coefficient in~\eqref{eqDefCm} that multiplies
the $\ln \nu_1$ term in~\eqref{eqLowSum}.
Again, we see how the arbitrary splitting point defined by $\nu_1$ cancels out and
we are left with an unambiguous result for any even $D$.

To summarize this arduous section, the final result for the finite sum is obtained by collecting all
of the five terms above into
\begin{align}
  \Sigma_{2 m} = \Sigma_{2m}^\text{low} + \Sigma_{2 m}^{\int} = 
 \Sigma_{2m}^\text{low}  + \Sigma_{2 m \text{ uns}}^{\int U}  + \Sigma_{2 m \text{ sub}}^{\int U}
  + \Sigma_{2 m \text{ uns}}^{\int \ln} + \Sigma_{2 m \text{ sub}}^{\int \ln} \, .
\end{align}
In this final result, the $\ln \nu_1$ terms partially cancel with $\ln y_1$ ones, leaving us with 
a residual dependence on $\ln \Delta/r_0$.
Combining the closed-form expressions obtained above does not lead to 
a simple expression, as the one we found for odd $D$s.
Still, plugging in the integers leads to rather compact expressions for the first couple of even $D$'s:
\begin{align} \label{eqSigEvenLowFinNum}
  \Sigma_D^\text{low} = \frac{2}{(D-2)!} \left(\frac{r_0}{\Delta} \right)^{D-1} 
  \begin{cases}
    \frac{5}{2} \left( \gamma_E + \ln \nu_1 \right) 				\, , & D = 2 \, ,
    \\
    - \frac{3}{8} \left( \gamma_E + \ln \nu_1 \right) 		\, , & D = 4 \, ,
    \\
    \frac{1}{10} \left( \gamma_E - \zeta(3) + \ln \nu_1 \right)	\, , & D = 6 \, ,
  \end{cases}
\end{align}
and
\begin{align} \label{eqSigEvenFinNum}
  \Sigma_D^{\int} = \frac{2}{(D-2)!} \left(\frac{r_0}{\Delta} \right)^{D-1}
  \begin{cases}
    \frac{\pi}{2 \sqrt{3}} - \frac{11}{4} - \frac{5}{2} \ln 2 y_1 			\, , & D = 2 \, ,
    \\
    - \frac{\pi}{8 \sqrt{3}} + \frac{3}{32} + \frac{3}{8} \ln 2 y_1 			\, , & D = 4 \, ,
    \\
    \frac{3 \sqrt{3} \pi}{160}-\frac{211}{4800}-\frac{1}{10} \ln 2 y_1		\, , & D = 6 \, .
  \end{cases}
\end{align}
Together with the result in generic odd $D$s, these finite sums provide us with
the finite log of the determinant in any dimension.
To complete the calculation of the decay rate one
has to perform the renormalization procedure, as described in section~\ref{secRenorm}
for even $D$s.
This might be a useful exercise if there is enough physical
motivation, but is beyond the scope of this paper.
We point out that what enters in this calculation will be integrals with
a certain number of insertions of $V^{(2)}$ and the accompanying powers of $\rho$ to compensate
for the dimensions.
These types of integrals, as shown in~\eqref{eqIn}, are considered and evaluated in closed form
in the appendix~\ref{appUVint}.

%

\def\arxiv#1[#2]{\href{http://arxiv.org/abs/#1}{[#2]}}
\def\Arxiv#1[#2]{\href{http://arxiv.org/abs/#1}{#2}}

\end{document}